\documentclass[twocolumn,twocolappendix]{aastex701}

\usepackage{siunitx}
\usepackage{amsmath}
\usepackage{CJK}
\usepackage[utf8]{inputenc}

\newcommand{\rsun}[1]{$\mathrm{R_{\odot}}$}
\newcommand{\lsun}[1]{$\mathrm{L_{\odot}}$}
\newcommand{\Halpha}[1]{$\mathrm{H\alpha}$}
\newcommand{\Hbeta}[1]{$\mathrm{H\beta}$}

\begin{document}
\title{JWST observations of SN 2024abup: First Detection of CO in a broad-lined Type Ic Supernova and Constraints on r-process Nucleosynthesis}
\correspondingauthor{Manisha Shrestha}
\email{manisha.shrestha@monash.edu}

\newcommand{\LCO}{\affiliation{Las Cumbres Observatory, 6740 Cortona Drive, Suite 102, Goleta, CA 93117-5575, USA}}
\newcommand{\UCSB}{\affiliation{Department of Physics, University of California, Santa Barbara, CA 93106-9530, USA}}
\newcommand{\KITP}{\affiliation{Kavli Institute for Theoretical Physics, University of California, Santa Barbara, CA 93106-4030, USA}}
\newcommand{\UCD}{\affiliation{Department of Physics and Astronomy, University of California, Davis, 1 Shields Avenue, Davis, CA 95616-5270, USA}}
\newcommand{\WIS}{\affiliation{Department of Particle Physics and Astrophysics, Weizmann Institute of Science, 76100 Rehovot, Israel}}
\newcommand{\OKC}{\affiliation{Oskar Klein Centre, Department of Astronomy, Stockholm University, Albanova University Centre, SE-106 91 Stockholm, Sweden}}
\newcommand{\OAPD}{\affiliation{INAF-Osservatorio Astronomico di Padova, Vicolo dell'Osservatorio 5, I-35122 Padova, Italy}}
\newcommand{\Caltech}{\affiliation{Cahill Center for Astronomy and Astrophysics, California Institute of Technology, Mail Code 249-17, Pasadena, CA 91125, USA}}
\newcommand{\GSFC}{\affiliation{Astrophysics Science Division, NASA Goddard Space Flight Center, Mail Code 661, Greenbelt, MD 20771, USA}}
\newcommand{\UMD}{\affiliation{Joint Space-Science Institute, University of Maryland, College Park, MD 20742, USA}}
\newcommand{\UCB}{\affiliation{Department of Astronomy, University of California, Berkeley, CA 94720-3411, USA}}
\newcommand{\UCBMMA}{\affiliation{Berkeley Center for Multi-messenger Research on Astrophysical Transients and Outreach (Multi-RAPTOR), University of California, Berkeley, CA 94720-3411, USA}}
\newcommand{\TTU}{\affiliation{Department of Physics, Texas Tech University, Box 41051, Lubbock, TX 79409-1051, USA}}
\newcommand{\STScI}{\affiliation{Space Telescope Science Institute, 3700 San Martin Drive, Baltimore, MD 21218-2410, USA}}
\newcommand{\UT}{\affiliation{University of Texas at Austin, 1 University Station C1400, Austin, TX 78712-0259, USA}}
\newcommand{\IoA}{\affiliation{Institute of Astronomy, University of Cambridge, Madingley Road, Cambridge CB3 0HA, UK}}
\newcommand{\QUB}{\affiliation{Astrophysics Research Centre, School of Mathematics and Physics, Queen's University Belfast, Belfast BT7 1NN, UK}}
\newcommand{\IPACSSC}{\affiliation{Spitzer Science Center, California Institute of Technology, Pasadena, CA 91125, USA}}
\newcommand{\IPAC}{\affiliation{IPAC, California Institute of Technology, 1200 East California Boulevard, Pasadena, CA 91125, USA}}
\newcommand{\JPL}{\affiliation{Jet Propulsion Laboratory, California Institute of Technology, 4800 Oak Grove Dr, Pasadena, CA 91109, USA}}
\newcommand{\Southampton}{\affiliation{Department of Physics and Astronomy, University of Southampton, Southampton SO17 1BJ, UK}}
\newcommand{\LANL}{\affiliation{Space and Remote Sensing, MS B244, Los Alamos National Laboratory, Los Alamos, NM 87545, USA}}
\newcommand{\Tsinghua}{\affiliation{Physics Department and Tsinghua Center for Astrophysics, Tsinghua University, Beijing, 100084, People's Republic of China}}
\newcommand{\NAOC}{\affiliation{National Astronomical Observatory of China, Chinese Academy of Sciences, Beijing, 100012, People's Republic of China}}
\newcommand{\YNAO}{\affiliation{Yunnan Observatories (YNAO), Chinese Academy of Sciences (CAS), Kunming, 650216, People's Republic of China}}
\newcommand{\ICEY}{\affiliation{International Centre of Supernovae, Yunnan Key Laboratory, Kunming 650216, People's Republic of China}}
\newcommand{\Itagaki}{\affiliation{Itagaki Astronomical Observatory, Yamagata 990-2492, Japan}}
\newcommand{\Einstein}{\altaffiliation{Einstein Fellow}}
\newcommand{\Hubble}{\altaffiliation{Hubble Fellow}}
\newcommand{\CfA}{\affiliation{Center for Astrophysics \textbar{} Harvard \& Smithsonian, 60 Garden Street, Cambridge, MA 02138-1516, USA}}
\newcommand{\UA}{\affiliation{Steward Observatory, University of Arizona, 933 North Cherry Avenue, Tucson, AZ 85721-0065, USA}}
\newcommand{\MPIA}{\affiliation{Max-Planck-Institut f\"ur Astrophysik, Karl-Schwarzschild-Stra\ss{}e 1, D-85748 Garching, Germany}}
\newcommand{\DSFP}{\altaffiliation{LSSTC Data Science Fellow}}
\newcommand{\HCO}{\affiliation{Harvard College Observatory, 60 Garden Street, Cambridge, MA 02138-1516, USA}}
\newcommand{\Carnegie}{\affiliation{Observatories of the Carnegie Institute for Science, 813 Santa Barbara Street, Pasadena, CA 91101-1232, USA}}
\newcommand{\TAU}{\affiliation{School of Physics and Astronomy, Tel Aviv University, Tel Aviv 69978, Israel}}
\newcommand{\Edinburgh}{\affiliation{Institute for Astronomy, University of Edinburgh, Royal Observatory, Blackford Hill EH9 3HJ, UK}}
\newcommand{\Birmingham}{\affiliation{Birmingham Institute for Gravitational Wave Astronomy and School of Physics and Astronomy, University of Birmingham, Birmingham B15 2TT, UK}}
\newcommand{\Bath}{\affiliation{Department of Physics, University of Bath, Claverton Down, Bath BA2 7AY, UK}}
\newcommand{\CTIO}{\affiliation{Cerro Tololo Inter-American Observatory, National Optical Astronomy Observatory, Casilla 603, La Serena, Chile}}
\newcommand{\Potsdam}{\affiliation{Institut f\"ur Physik und Astronomie, Universit\"at Potsdam, Haus 28, Karl-Liebknecht-Str. 24/25, D-14476 Potsdam-Golm, Germany}}
\newcommand{\INPE}{\affiliation{Instituto Nacional de Pesquisas Espaciais, Avenida dos Astronautas 1758, 12227-010, S\~ao Jos\'e dos Campos -- SP, Brazil}}
\newcommand{\UNC}{\affiliation{Department of Physics and Astronomy, University of North Carolina, 120 East Cameron Avenue, Chapel Hill, NC 27599, USA}}
\newcommand{\Ohio}{\affiliation{Astrophysical Institute, Department of Physics and Astronomy, 251B Clippinger Lab, Ohio University, Athens, OH 45701-2942, USA}}
\newcommand{\AAS}{\affiliation{American Astronomical Society, 1667 K~Street NW, Suite 800, Washington, DC 20006-1681, USA}}
\newcommand{\MMT}{\affiliation{MMT and Steward Observatories, University of Arizona, 933 North Cherry Avenue, Tucson, AZ 85721-0065, USA}}
\newcommand{\Geneva}{\affiliation{ISDC, Department of Astronomy, University of Geneva, Chemin d'\'Ecogia, 16 CH-1290 Versoix, Switzerland}}
\newcommand{\IUCAA}{\affiliation{Inter-University Center for Astronomy and Astrophysics, Post Bag 4, Ganeshkhind, Pune, Maharashtra 411007, India}}
\newcommand{\CMU}{\affiliation{Department of Physics, Carnegie Mellon University, 5000 Forbes Avenue, Pittsburgh, PA 15213-3815, USA}}
\newcommand{\NAOJ}{\affiliation{Division of Science, National Astronomical Observatory of Japan, 2-21-1 Osawa, Mitaka, Tokyo 181-8588, Japan}}
\newcommand{\IfA}{\affiliation{Institute for Astronomy, University of Hawai`i, 2680 Woodlawn Drive, Honolulu, HI 96822-1839, USA}}
\newcommand{\UCSC}{\affiliation{Department of Astronomy and Astrophysics, University of California, Santa Cruz, CA 95064-1077, USA}}
\newcommand{\Purdue}{\affiliation{Department of Physics and Astronomy, Purdue University, 525 Northwestern Avenue, West Lafayette, IN 47907-2036, USA}}
\newcommand{\Princeton}{\affiliation{Department of Astrophysical Sciences, Princeton University, 4 Ivy Lane, Princeton, NJ 08540-7219, USA}}
\newcommand{\Moore}{\affiliation{Gordon and Betty Moore Foundation, 1661 Page Mill Road, Palo Alto, CA 94304-1209, USA}}
\newcommand{\Durham}{\affiliation{Department of Physics, Durham University, South Road, Durham, DH1 3LE, UK}}
\newcommand{\JHU}{\affiliation{Department of Physics and Astronomy, The Johns Hopkins University, 3400 North Charles Street, Baltimore, MD 21218, USA}}
\newcommand{\Toronto}{\affiliation{David A.\ Dunlap Department of Astronomy and Astrophysics, University of Toronto,\\ 50 St.\ George Street, Toronto, Ontario, M5S 3H4 Canada}}
\newcommand{\Duke}{\affiliation{Department of Physics, Duke University, Campus Box 90305, Durham, NC 27708, USA}}
\newcommand{\NCU}{\affiliation{Graduate Institute of Astronomy, National Central University, 300 Jhongda Road, 32001 Jhongli, Taiwan}}
\newcommand{\Columbia}{\affiliation{Department of Physics and Columbia Astrophysics Laboratory, Columbia University, Pupin Hall, New York, NY 10027, USA}}
\newcommand{\Flatiron}{\affiliation{Center for Computational Astrophysics, Flatiron Institute, 162 5th Avenue, New York, NY 10010-5902, USA}}
\newcommand{\CIERA}{\affiliation{Center for Interdisciplinary Exploration and Research in Astrophysics and Department of Physics and Astronomy, \\Northwestern University, 1800 Sherman Avenue, 8th Floor, Evanston, IL 60201, USA}}
\newcommand{\GeminiNorth}{\affiliation{Gemini Observatory, 670 North A`ohoku Place, Hilo, HI 96720-2700, USA}}
\newcommand{\Keck}{\affiliation{W.~M.~Keck Observatory, 65-1120 M\=amalahoa Highway, Kamuela, HI 96743-8431, USA}}
\newcommand{\UW}{\affiliation{Department of Astronomy, University of Washington, 3910 15th Avenue NE, Seattle, WA 98195-0002, USA}}
\newcommand{\catalyst}{\altaffiliation{LSSTC Catalyst Fellow}}
\newcommand{\USask}{\affiliation{Department of Physics \& Engineering Physics, University of Saskatchewan, 116 Science Place, Saskatoon, SK S7N 5E2, Canada}}
\newcommand{\Thacher}{\affiliation{Thacher School, 5025 Thacher Road, Ojai, CA 93023-8304, USA}}
\newcommand{\Rutgers}{\affiliation{Department of Physics and Astronomy, Rutgers, the State University of New Jersey,\\136 Frelinghuysen Road, Piscataway, NJ 08854-8019, USA}}
\newcommand{\FSU}{\affiliation{Department of Physics, Florida State University, 77 Chieftan Way, Tallahassee, FL 32306-4350, USA}}
\newcommand{\Melbourne}{\affiliation{School of Physics, The University of Melbourne, Parkville, VIC 3010, Australia}}
\newcommand{\ASTROthreeD}{\affiliation{ARC Centre of Excellence for All Sky Astrophysics in 3 Dimensions (ASTRO 3D)}}
\newcommand{\Stromlo}{\affiliation{Mt.\ Stromlo Observatory, The Research School of Astronomy and Astrophysics, Australian National University, ACT 2601, Australia}}
\newcommand{\NCPAS}{\affiliation{National Centre for the Public Awareness of Science, Australian National University, Canberra, ACT 2611, Australia}}
\newcommand{\TAMU}{\affiliation{Department of Physics and Astronomy, Texas A\&M University, 4242 TAMU, College Station, TX 77843, USA}}
\newcommand{\Mitchell}{\affiliation{George P.\ and Cynthia Woods Mitchell Institute for Fundamental Physics \& Astronomy, College Station, TX 77843, USA}}
\newcommand{\ESO}{\affiliation{European Southern Observatory, Alonso de C\'ordova 3107, Casilla 19, Santiago, Chile}}
\newcommand{\ICE}{\affiliation{Institute of Space Sciences (ICE, CSIC), Campus UAB, Carrer
de Can Magrans, s/n, E-08193 Barcelona, Spain}}
\newcommand{\IEEC}{\affiliation{Institut d'Estudis Espacials de Catalunya (IEEC), Edifici RDIT, Campus UPC, 08860 Castelldefels (Barcelona), Spain}}
\newcommand{\Warwick}{\affiliation{Department of Physics, University of Warwick, Gibbet Hill Road, Coventry CV4 7AL, UK}}
\newcommand{\Macquarie}{\affiliation{School of Mathematical and Physical Sciences, Macquarie University, NSW 2109, Australia}}
\newcommand{\AAARC}{\affiliation{Astronomy, Astrophysics and Astrophotonics Research Centre, Macquarie University, Sydney, NSW 2109, Australia}}
\newcommand{\Capodimonte}{\affiliation{INAF - Capodimonte Astronomical Observatory, Salita Moiariello 16, I-80131 Napoli, Italy}}
\newcommand{\INFNNapoli}{\affiliation{INFN - Napoli, Strada Comunale Cinthia, I-80126 Napoli, Italy}}
\newcommand{\ICRANet}{\affiliation{ICRANet, Piazza della Repubblica 10, I-65122 Pescara, Italy}}
\newcommand{\MSU}{\affiliation{Center for Data Intensive and Time Domain Astronomy, Department of Physics and Astronomy,\\Michigan State University, East Lansing, MI 48824, USA}}
\newcommand{\SETI}{\affiliation{SETI Institute,
339 Bernardo Ave, Suite 200, Mountain View, CA 94043, USA}}
\newcommand{\IAIFI}{\affiliation{The NSF AI Institute for Artificial Intelligence and Fundamental Interactions}}
\newcommand{\ANUC}{\affiliation{Department of Astronomy, AlbaNova University Center, Stockholm University, SE-10691 Stockholm, Sweden}}

\newcommand{\Konkoly}{\affiliation{Konkoly Observatory,  CSFK, Konkoly-Thege M. \'ut 15-17, Budapest, 1121, Hungary}}
\newcommand{\ELTE}{\affiliation{ELTE E\"otv\"os Lor\'and University, Institute of Physics, P\'azm\'any P\'eter s\'et\'any 1/A, Budapest, 1117 Hungary}}
\newcommand{\SZTE}{\affiliation{Department of Experimental Physics, University of Szeged, D\'om t\'er 9, Szeged, 6720, Hungary}}
\newcommand{\IdAlta}{\affiliation{Instituto de Alta Investigaci\'on, Sede Esmeralda, Universidad de Tarapac\'a, Av. Luis Emilio Recabarren 2477, Iquique, Chile}}
\newcommand{\Kavli}{\affiliation{Kavli Institute for Cosmological Physics, University of Chicago, Chicago, IL 60637, USA}}
\newcommand{\UofChicago}{\affiliation{Department of Astronomy and Astrophysics, University of Chicago, Chicago, IL 60637, USA}}
\newcommand{\Fermi}{\affiliation{Fermi National Accelerator Laboratory, P.O.\ Box 500, Batavia, IL 60510, USA}}
\newcommand{\Dartmouth}{\affiliation{Department of Physics and Astronomy, Dartmouth College, Hanover, NH 03755, USA}}
\newcommand{\Surrey}{\affiliation{Department of Physics, University of Surrey, Guildford GU2 7XH, UK}}
\newcommand{\NU}{\affiliation{Center for Interdisciplinary Exploration and Research in Astrophysics (CIERA) and Department of Physics and Astronomy, Northwestern University, Evanston, IL 60208, USA}}
\newcommand{\itagaki}{\affiliation{Itagaki Astronomical Observatory, Yamagata 990-2492, Japan}}
\newcommand{\UdChile}{\affiliation{Departamento de Astronomia, Universidad de Chile, Camino El Observatorio 1515, Las Condes, Santiago, Chile}}
\newcommand{\UVA}{\affiliation{Department of Astronomy, University of Virginia, Charlottesville, VA 22904, USA}}
\newcommand{\IfK}{\affiliation{Institut f\"ur Kernphysik, Technische Universita\"t Darmstadt, Schlossgartenstr. 2, Darmstadt D-64289, Germany}}
\newcommand{\MPIK}{\affiliation{Max-Planck-Institut f\"ur Kernphysik, Saupfercheckweg 1, Heidelberg D-69117, Germany}}
\newcommand{\UCSD}{\affiliation{Department of Astronomy \& Astrophysics, University of California, San Diego, 9500 Gilman Drive, MC 0424, La Jolla, CA 92093-0424, USA}}
\newcommand{\Monash}{\affiliation{School of Physics and Astronomy, Monash University, Clayton, Victoria 3800, Australia}}
\newcommand{\OzGrav}{\affiliation{OzGrav: The ARC Centre of Excellence for Gravitational Wave Discovery, Clayton, Victoria 3800, Australia}}
\newcommand{\UCBPh}{\affiliation{Department of Physics, University of California, 366 Physics North MC 7300, Berkeley, CA 94720, USA}}
\author[0000-0002-4022-1874]{Manisha Shrestha}
\Monash
\OzGrav
\email[show]{manisha.shrestha@monash.edu}
\author[0000-0003-3108-1328]{Lindsey~A.~Kwok}
\CIERA \email{lindsey.kwok@northwestern.edu}

\author[0000-0003-4102-380X]{David J. Sand}
\UA\email{}

\author[0000-0003-4800-2737]{Stan Bartmentloo}
\OKC \email{stan.barmentloo@astro.su.se}

\author[orcid=0000-0003-0528-202X, gname=Collin, sname=Christy]{Collin Christy}
\UA \email{collinchristy@arizona.edu}

\author[0000-0001-8005-4030]{Anders Jerkstrand}
\OKC \email{anders.jerkstrand@astro.su.se}

\author[0000-0002-4924-444X]{K. Azalee Bostroem}
\catalyst\UA \email{}

\author[0000-0003-0123-0062]{Jennifer E. Andrews}
\GeminiNorth \email{}

\author[orcid=0000-0002-8297-2473, gname=Kate, sname=Alexander]{Kate D. Alexander}
\UA \email{kdalexander@arizona.edu}

\author[0000-0002-7937-6371]{Yize Dong \begin{CJK*}{UTF8}{gbsn}(董一泽)\end{CJK*}}
\CfA
\email{yize.dong@cfa.harvard.edu}

\author[]{Carl E. Fields}
\UA \email{}

\author[0000-0003-2744-4755]{Emily Hoang}
\UCD \email{}
\author[0000-0002-0832-2974]{Griffin Hosseinzadeh}
\UCSD \email{}
\author[0000-0002-9454-1742]{Brian Hsu}
\UA \email{}
\author[0000-0003-0549-3281]{Daryl Janzen}
\USask \email{}
\author[0000-0001-8738-6011]{Saurabh W.\ Jha}
\Rutgers \email{}

\author[0000-0001-5975-290X]{Joel Johansson}
\OKC \email{joeljo@fysik.su.se}

\author[0000-0002-0744-0047]{Jeniveve Pearson}
\UA \email{}

\author[0000-0001-9589-3793]{M.~J. Lundquist}

\Keck \email{}
\author[0009-0008-9693-4348]{Darshana Mehta}
\UCD \email{}

\author[0009-0001-3106-0917]{Aidan Martas}
\UCB \email{aidmart@berkeley.edu}
\author[0000-0001-7132-0333]{Maryam Modjaz}
\UVA \email{}

\author[0000-0002-4470-1277]{Bernhard M\"uller}
\Monash \email{}

\author[0000-0003-4175-4960]{Conor L. Ransome}
\UA\email{cransome@arizona.edu}

\author[0000-0002-7352-7845]{Aravind P.\ Ravi}
\UCD \email{}

\author[0000-0002-6718-9472]{Mathieu Renzo}
\UA\email{}

\author[0000-0002-7015-3446]{Nicol\'as Meza Retamal}
\UCD \email{}
\author[0000-0001-8073-8731]{Bhagya Subrayan}
\UA\email{bsubrayan@arizona.edu}

\author[0000-0001-5510-2424]{Nathan Smith}
\UA \email{}

\author[0000-0001-8818-0795]{Stefano Valenti}
\UCD \email{}
\author[0000-0002-4951-8762]{Sergiy Vasylyev}
\UCSD
\email{svasylyev@ucsd.edu}

\author[0000-0003-1626-1355]{Giacomo Ricigliano}
\MPIK\IfK \email{}
\author[0000-0001-6272-5507]{Peter J. Brown}
\TAMU \email{}

\author[0000-0002-1895-6639]{Moira Andrews}
\LCO \UCSB \email{}
\author[0000-0003-4914-5625]{Joseph Farah}
\LCO 
\UCSB \email{}

\author[0000-0003-4253-656X]{D.\ Andrew Howell}
\LCO\UCSB \email{}
\author[0000-0001-5807-7893]{Curtis McCully}
\LCO \email{}
\author[0000-0001-9570-0584]{Megan Newsome}
\LCO 
\UCSB \email{}

\author[orcid=0009-0006-7296-728X]{Kathryn Wynn}
\LCO \UCSB \email{}

\author[0000-0002-7706-5668]{Ryan Chornock}
\UCB\UCBPh
\UCBMMA \email{}

\author[0000-0002-2249-0595]{Natalie LeBaron}
\UCB \UCBPh
\UCBMMA \email{}

\author[0000-0003-4768-7586]{Raffaella Margutti}
\UCB \UCBPh
\UCBMMA \email{}

\author[0000-0002-9301-5302]{Melissa Shahbandeh}
\JHU
\STScI \email{}
\author[0000-0002-5221-7557]{Chris Ashall}
\IfA \email{}
%\author[0000-0002-9301-5302]{Ori D. Fox}
%\STScI \email{}
\author[0000-0002-4338-6586]{Peter Hoeflich}
\FSU \email{}

% \author[0000-0002-0810-5558]{Ricardo R. Munoz}
% \UdChile

% \author[0000-0003-0105-9576]{Gustavo E. Medina}
% \Toronto

% \author[0000-0002-9110-6163]{Ting S. Li}
% \Toronto

% \author{Paula Diaz}
% \UdChile

% \author[0000-0002-1125-9187]{Daichi Hiramatsu}
% \CfA
% \IAIFI

% \author{Brad E. Tucker}
% \Stromlo

% \author[0000-0003-1349-6538]{J. C. Wheeler}
% \UT

% \author[0000-0002-7334-2357]{Xiaofeng Wang}
% \Tsinghua

% \author{Qian Zhai}
% \YNAO

% \author[0000-0002-8296-2590]{Jujia Zhang}
% \YNAO
% \ICEY

% \author[0000-0002-3884-5637]{Anjasha Gangopadhyay}
% \OKC

% \author[0000-0002-6535-8500]{Yi Yang}
% \Tsinghua

% \author[0000-0003-2375-2064]{Claudia P. Guti\'{e}rrez}
% \IEEC\ICE



\begin{abstract}
SN~2024abup is a nearby broad-lined Type Ic supernova (SN\,Ic-bl) in NGC~0681 at a distance of $23.3 \pm 1.6$ Mpc. As energetic explosions of massive stars, SNe Ic-BL are considered a plausible site for rapid-neutron capture nucleosynthesis ($r$-process) and chemical enrichment from short-lived progenitors. They may also contribute to dust production in the early Universe. We present JWST near- to mid-infrared (NIR$+$MIR) observations (1--14 $\mathrm{\mu}$m) of SN Ic-bl~2024abup at $+$41 days after the V band maximum ($+$54\,days after explosion), the first-ever JWST+MIR observation of a SN Ic-bl along with radio and optical data. Using the spectral synthesis code \texttt{SUMO}, we identify the observed broad IR line features in SN~2024abup and find significant contributions from C, O, Mg, and carbon monoxide (CO) - the earliest detection of molecules in a core-collapse SN so far. 
%While there are some discrepancies in the strength and shape of lines, the model helps us provide an unambiguous detection of dust formation in a Type Ic-BL SN for the first time.
The spectrum shows continuum emission at wavelengths greater than 1.5 $\mathrm{\mu m}$, which could be explained by dust---preexisting, newly formed, or a combination---heated by the SN. We do not find compelling evidence for infrared signatures of $r$-process elements, though our search is hampered by the presence of many broad and blended features from the non-$r$-process elements. These new observations indicate that SNe Ic-BL could be a contributor to early-universe dust production, and suggest that if $r$-process elements are produced, revealing their presence from spectra requires very high-quality data and models to disentangle blends. 

\end{abstract}



\section{Introduction} \label{sec:intro}
The explosion of massive stars with $\mathrm{M_{ZAMS}}$ $\gtrsim$ 8 $\mathrm{M_{\odot}}$ are known as core-collapse supernovae (CCSNe). Among the zoo of CCSN types, only $\sim$1\% \citep{Li_2011,Shivvers_2017} do not exhibit any hydrogen (H) or helium (He) and exhibit broad features, classifying them as Type Ic broad-lined supernovae (SN Ic-bl)\citep[also see reviews by e.g.][]{Woosley_2006,Cano_2017,Modjaz_2019}. These explosions are characterized by ejecta velocities of 20--30,000 km s$^{-1}$ \citep[e.g.][]{Modjaz_2016}, well above the typical $\sim$10,000 km s$^{-1}$ for normal core-collapse SNe, along with inferred kinetic energies an order of magnitude higher. SN Ic-bl are the only type of SN that has been associated with normal long-duration gamma-ray bursts (LGRBs), however, not all SN Ic-bl are associated with LGRBs \citep[e.g.][]{Corsi_2016,Modjaz_2016,Japelj_2018,Taddia_2019,Corsi_2023,Srinivasaragavan_2024,Schroeder_2025}. Several competing progenitor scenarios for SN Ic-bl have been proposed to explain the stripping of H and He from the envelope: 1) mass loss via binary interaction in massive stars \citep[e.g.][]{Podsiadlowski_1993,Nomoto_1995}, 2) mass loss via strong stellar winds of single Wolf-Rayet stars \citep[e.g.][]{Gaskell_1986,Smartt_2009}, 3) eruptions that create dense shells \citep{Salas_2013,Chen_2026}, and 4) chemical homogeneous evolution \citep{Nicholl_2017, Renzo_2026}. In either case, the progenitor is expected to be a rapidly rotating massive star whose core collapse drives a hyperenergetic explosion. 
%Additionally, SN Ic-bl could play an important role in the formation of heavy elements via rapid neutron capture nucleosynthesis ($r$-process; \citet{Li_1998}) and could contribute to dust production in the early universe.

The first electromagnetic counterpart to a gravitational wave (GW) source was detected for GW170817 \citep[e.g.][]{ Alexander_2017,Andreoni_2017, Arcavi_2017, Covino_2017, Cowperthwaite_2017, Drout_2017, Evans_2017, Goldstein_2017,Haggard_2017, Kasliwal_2017, Margutti_2017, Pian_2017,Savchenko_2017,Smartt_2017, Soares-Santos_2017, Tanvir_2017, Troja_2017, Utsumi_2017,Valenti_2017}, in the form of short-duration GRB~170817A \citep{Goldstein_2017,Savchenko_2017} and subsequent kilonova emission in the form of AT~2017gfo \citep{Coulte_2017}. This discovery provided a breakthrough in our search for astrophysical mechanisms to produce $r$-process elements (see \citealt{Margutti_2021} for a review). While the production of such material in binary neutron star mergers is confirmed, there is emerging evidence that they are not the sole source of the $r$-process elements.  For instance, chemical abundance patterns in Milky Way halo stars and dwarf galaxies both argue for a prompt enrichment channel in the early universe, and $r$-process events that are substantially different from that expected from binary neutron stars \citep[e.g.,][]{Ji_2016,Yong21,Naidu22}. The rate of binary neutron star mergers is also highly uncertain, and recent analysis of GW data from the O4a Laser Interferometer Gravitational-Wave Observatory (LIGO), Virgo and KAGRA collaboration run finds that the rate could be two times lower than previously expected from the O2 run \citep{GW_2025_mergerrate,gwtc5_2026}. This lower rate implies that $r$-process abundances seen in the Universe could not be explained by only BNS mergers as thought previously \citep{Drout_2017}. Thus, another channel of r-process element production is likely needed to match the observed abundances \citep[see examples,][]{Papish_2015,Naidu22,Siegel_2022,Grichener_2025,Patel_2025}.  

SNe Ic-BL have been proposed as one of the many viable mechanisms for producing heavy elements via $ r$-process. Here, $r$-process material is most likely produced in stars that collapse to a black hole with an accretion disk (collapsar) but could also occur when the collapse produces a rapidly spinning, strongly magnetized neutron star \citep{MacFadyen99,Siegel_2019}. Recent 3D relativistic MHD simulations have found that collapsars could be the production site for a large fraction of all the $r$-process elements in the Universe \citep{Siegel_2019}. In the simulations by \citet{Siegel_2019}, they found that the production of $r$-process elements impacts the near-infrared (NIR) photometry and spectroscopy. \citet{Barnes_2022} expanded the \cite{Siegel_2019} model to include a wider parameter space to investigate the possibility of detecting $r$-process production in SN Ic-bl. \citet{Barnes_2022} find that $r$-process elements produce a near-infrared excess compared to regular SN Ic-bl light-curve that can be observed even during the photospheric phase (within $\sim$ 75 days). \citet{Barnes_2022} finds that $r$-process effects are most prominent in the NIR wavelength ranges. The strength of this signature is predicted to be more visible for typical or low-mass ejecta SN Ic-bl. Recently, \citet{Ricigliano_2025} modeled the nebular spectrum of SNe Ic-BL using a plasma model including the heavy element line list by \citet{Hotokezaka_2022}, and found that, if SNe Ic-BL are a relevant production site for $r$-process elements, the latter would be likely identifiable in the spectra. In particular, the imprints from these elements would be observable between $\sim$ 1 and 10 $\mathrm{\mu m}$, for a few $\times$0.1\% of the ejecta mass being constituted by heavy species. However, there are also several theoretical works predicting that typical collapsars do not produce significant neutron-rich ejecta, leading to $r$-process \citep[e.g.][]{Fujibayashi_2020,Issa_2025,Just_2015,Miller_2020}. Thus, there is an open question whether collapsars could be a site for $r$-process element production. These theoretical works underscore the need for IR observations of SN Ic-bl to answer this question of $r$-process productions via this alternative channel.

\cite{Anand2024} performed a systematic study of 25 SNe Ic-BL following the models of \cite{Barnes_2022}. Their analysis shows that the light curves of their sample better match $r$-process-free models.  Their data, however, is limited to low-cadence optical and near-infrared photometry. Additionally, \cite{Rastinejad_2024} analyzed four SN Ic-bl associated with LGRBs using optical-to-NIR photometric data from the Hubble Space Telescope (HST) and found no conclusive evidence for $r$-process elements. The four SNe Ic-BL have limited NIR data at different epochs. In addition, neither of these two studies utilizes spectroscopic analysis. Thus, it is still essential to search for $r$-process signatures in spectroscopic data, which may be able to distinguish subtle features that broadband photometry cannot. %Thus, a higher cadence and infrared observations in both photometry and spectroscopy are required to determine the production of these heavy elements in collapsars. 

In addition to heavy element production, SN Ic-bl could produce dust. A significant amount of dust has been observed in high-redshift galaxies (from z $\simeq$4 to z$\simeq$8) \citep[e.g.,][]{Laporte_2017}. These galaxies are too young for low-mass stars to evolve to the asymptotic giant branch (AGB) phase and produce dust (which is the major dust producer in our current universe; e.g. \citealt[][]{Gehrz_1989,Dellagli_2018,Ventura_2014}). Hence, other channels such as CCSNe have been proposed as a prompt channel for dust production \citep[e.g.,][]{Cherchneff_2026, Dwek_2011,Liljegren_2020_Molecules,Schneider_2024}. 
%Stripped-envelope SNe such as SN Ic and SN Ic-bl could be efficient in producing dust because the progenitor of these SNe have their envelope stripped, thus, there are a lot of C and O in the ejecta which could form CO. Also due to their low ejecta mass and high kinetic energy, the environment around these SNe could be efficient in cooling down faster than other CCSNe, making it easier to form CO molecules. 
There have been several detections of CO in SN Ic, where CO can efficiently cool SN ejecta for dust formation to take place \citep{Rho_2021, Ravi_2023,Tinyanont_2026}. These detections indicate that SN Ic could be contributing to the dust budget of the early universe; however, so far, we have not detected any CO in SN Ic-bl.  

\begin{figure}
    \centering
    \includegraphics[width=\columnwidth]{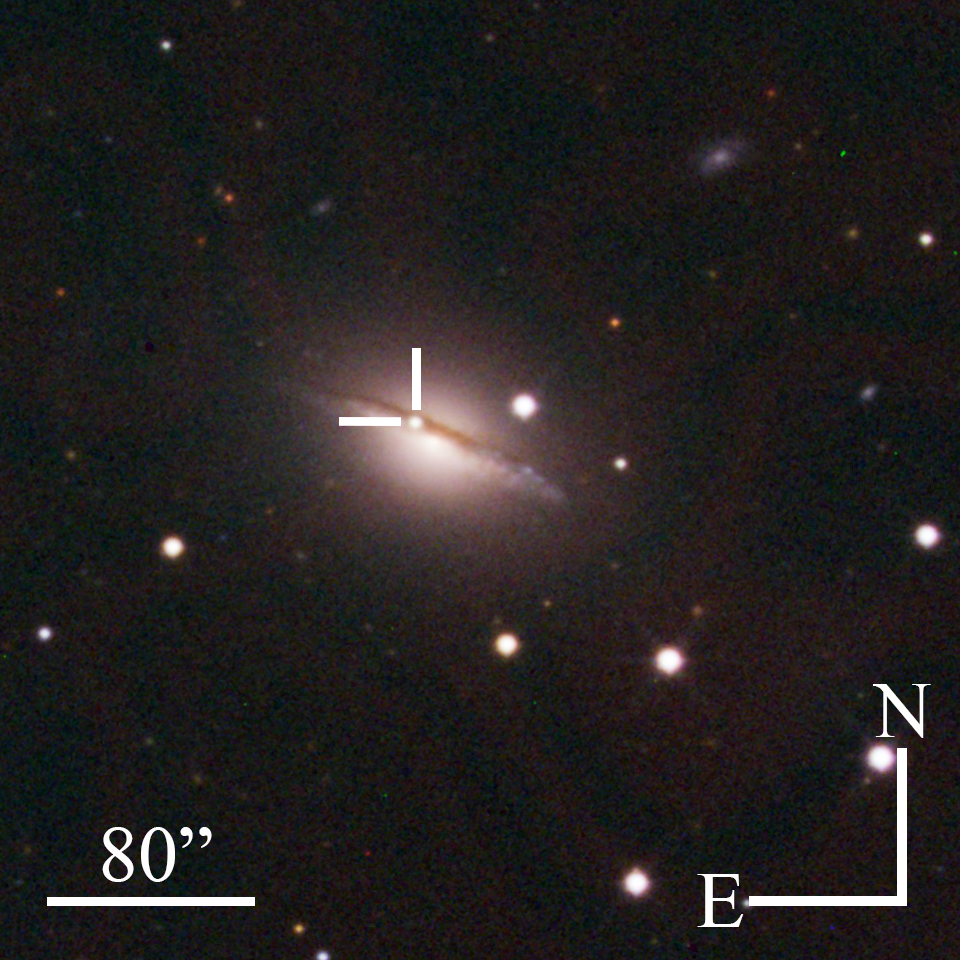}
    \caption{SN~2024abup in NGC\,0681 taken 25 days post explosion. The image is a composite of $g$, $r$, and $i$ filter data from the Las Cumbres Observatory.}
    \label{fig:hostgalaxy}
\end{figure}

\begin{figure}
    \centering
    \includegraphics[width=0.5\textwidth]{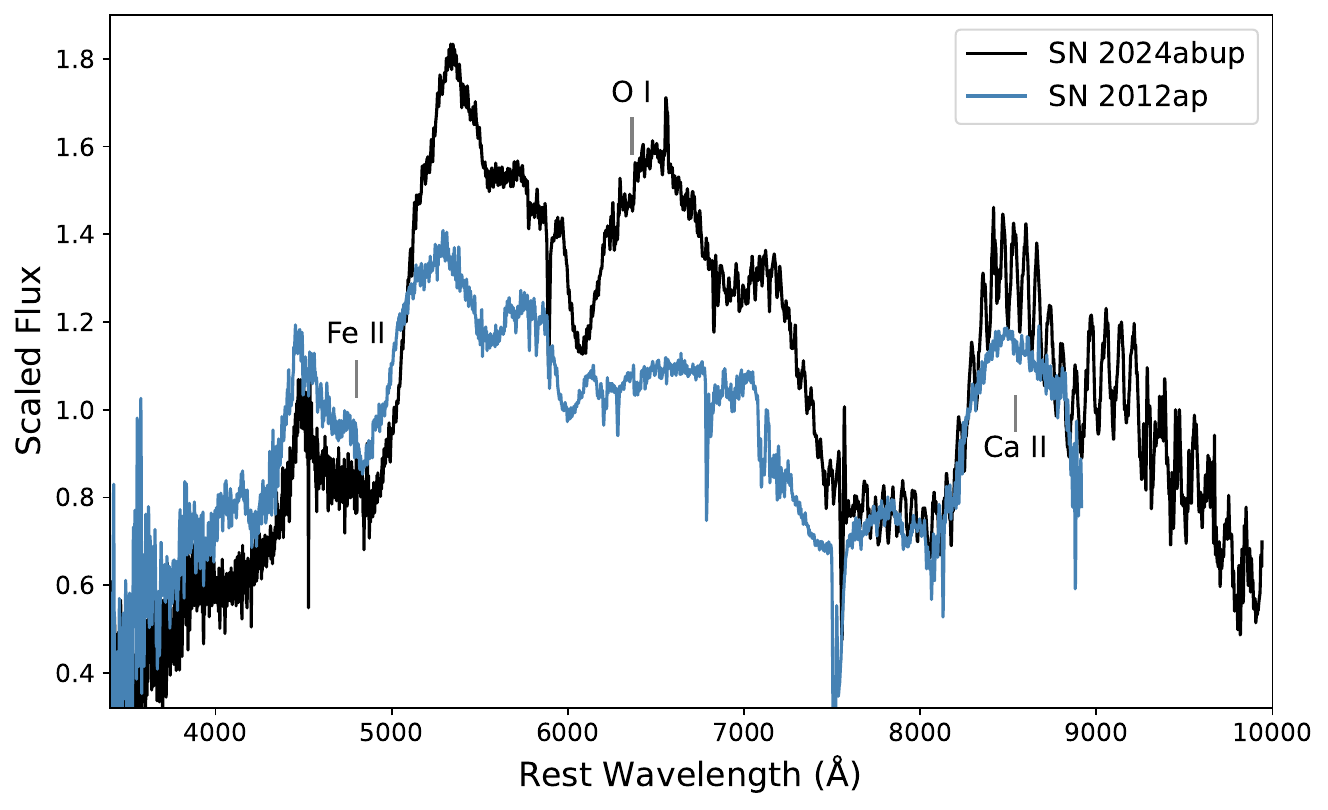}

    \caption{Extinction corrected optical spectrum of SN~2024abup from FLOYDS around V$_{\rm max}$ on 60649.40 MJD compared to SN~2012ap, the best fit SN\,Ic-bl \citep{Milisavljevic_2015} according to SNID-SAGE \citep{Stoppa2026}, near peak in V band. The similarity between the two objects provides confirmation of SN\,2024abup's classification as a SN Ic-bl \citep{Balcon_2024_classification}.}
    \label{fig:spec_max}
\end{figure}

The discovery of the nearby Type Ic-BL SN~2024abup provides a great opportunity to test whether these explosions can produce $r$-process elements and dust. SN~2024abup was discovered on 2024-11-22 08:25:17.760 UT (60636.35 MJD) by the Asteroid Terrestrial-Impact Last Alert System \citep[ATLAS;][]{Tonry_2011,Tonry_2018,Smith_2020,Fulton_2023} in NGC 0681 at a distance of 23.3 $\pm$1.6 Mpc using the Hubble flow distance method assuming $H_0=73 \pm 5~\mathrm{km/sec/Mpc}$\footnote{https://ned.ipac.caltech.edu} (see \autoref{fig:hostgalaxy}). It was classified as a stripped-envelope SN \citep{Balcon_2024_classification} and later confirmed as an SN Ic-bl \citep{Lidman_2024_classification_confirm} (see \autoref{fig:spec_max}). We present further explosion properties such as ejecta mass, expansion velocities, nickel mass, etc., of SN~2024abup in \autoref{tab:results}. 

%We present a JWST NIR$+$MIR ($\sim$1--14\,$\mathrm{\mu m}$) spectrum of SN~2024abup at $+$54\,days after explosion, the first-ever MIR spectrum of a SN Ic-bl. We search for the imprints of both $r$-process elements and CO. 
In \autoref{sec:obs} we detail the data reduction of our photometric and spectroscopic data, and in \autoref{sec:host} we analyze the properties of the host galaxy and calculate the line-of-sight extinction. We present explosion properties in \autoref{sec:photometry}. We present our spectral analysis in \autoref{sec:spectra}, including line identifications, model comparison, CO detection, comparison to other objects, and our search for r-process lines. The results from radio observations are presented in \autoref{sec:radio}.  %We compare the properties of SN~2024abup with a larger population of Ic and Ic-BL SNe in \autoref{sec:comp}.  
Finally, we discuss our results and summarize our conclusions in \autoref{sec:conclusions}.

\section{Observations \& data reduction} \label{sec:obs}
We performed multi-wavelength photometric and spectroscopic follow-up of SN~2024abup as described below. In this work, we focus on the analysis of the optical to MIR spectrum at 41 days post maximum; detailed analysis of the additional data will be presented in future work.

\subsection{Spectroscopy}
\begin{figure*}
    \centering
    \includegraphics[width=\textwidth]{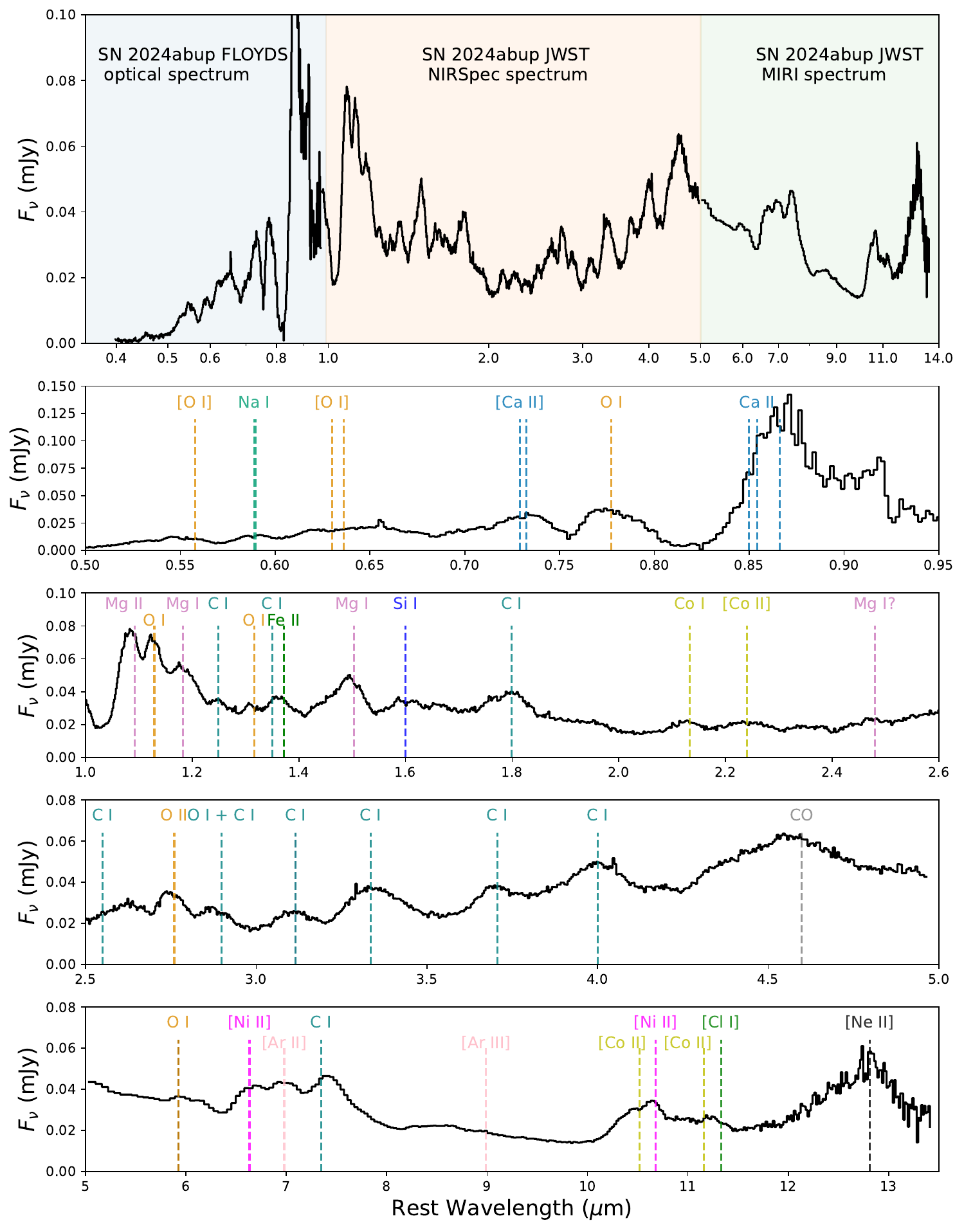}

    \caption{SN~2024abup spectrum spanning optical (FLOYDS) to MIR (JWST) $+$ 41 days after the V band maximum. The top panel covers 0.4 to 14 $\mu$ m and subsequent panels zoom in on the optical spectrum, the third covers 1 to 2.6 $\mu$ m, the fourth is for 2.5 to 5 $\mu$ m, and the last one covers 5 to 14 $\mu$m. All the major lines identified using the \texttt{SUMO} model (see Section \ref{subsec:spectral_modeling} and Figure \ref{fig:ion_model}) are marked with dashed vertical lines. }
    \label{fig:spec_jwst_opt-mir}
\end{figure*}
We obtained optical, NIR, and MIR spectra of SN~2024abup with JWST and the Las Cumbres Observatory (LCO). An optical spectrum near the peak (60649 MJD) in V band and one contemporaneous to JWST observation at $+$41\,days post V band maximum was obtained with the FLOYDS spectrographs \citep{Brown_2013} on the LCO's 2m Faulkes Telescopes North and South (FTN/FTS) as part of the Global Supernova Project. This spectrum was reduced following standard procedures using the FLOYDS pipeline \citep{Valenti_2014}.

We also obtained \textit{JWST} NIRSpec $+$ MIRI spectroscopy of SN\,2024abup at $+$41\,days post maximum through Director's Discretionary program JWST-DD-6803\footnote{All \textit{JWST} data is publicly available on MAST at DOI: \href{https://archive.stsci.edu/doi/resolve/resolve.html?doi=10.17909/0shq-ex65}{10.17909/0shq-ex65}} \citep{Shrestha2024} on 2025-01-15 UT. Combining the {\it JWST} data with our contemporaneous FLOYDS spectrum observed on 2025-01-09 UT, \autoref{fig:spec_jwst_opt-mir} shows the resulting panchromatic spectrum spanning 0.4$-$13.5\,\micron. The NIRSpec observations used the Fixed Slit (FS) medium resolution ($R\sim1000$) G140M, G235M, and G395M gratings with the F100LP, F170LP, and F290LP filters, respectively, for continuous coverage from 1$-$5\,\micron. The MIRI observations used the Low Resolution Spectrograph (LRS; R$\sim100$), covering $\sim$ 5$-$14\,\micron. The total NIRSpec science exposure time, split evenly between the three gratings, was 657\,s (11\,min) each, and the total MIRI/LRS science exposure time was 1112\,s (18.5\,min).

SN\,2024abup is located in a thick dust lane of its host galaxy, and we find that the \texttt{stage3} MIRI/LRS spectral reductions available on MAST, produced through the automatic JWST pipeline, do not properly account for nearby host galaxy flux. The standard background subtraction is done by subtracting the two along-slit-nod dithers and then co-adding along the source trace; however, in this case, directly co-adding the nod-subtracted images results in oversubtraction in regions on either side of the source. Inspecting the \texttt{stage2} two-dimensional (2D) individual nod images---after the nods have been subtracted, but prior to co-adding---we find a nearby, faint, extended trace that we identify as the host galaxy. Extracting along this faint trace reveals that the host galaxy spectrum is dominated by emission lines from polycyclic aromatic hydrocarbons (PAHs) (see \autoref{sec:dust}), consistent with expectations for an underlying dust lane. 

Starting from the \texttt{stage2} products for each individual nod, we manually reextract the one-dimensional (1D) source spectrum using the JWST pipeline \texttt{extract1d} command with custom background and aperture regions chosen to avoid the host galaxy and more cleanly extract the supernova only, following methods from the public MIRI/LRS spectral extraction notebook by S.~Kendrew and I.~Wang.\footnote{\url{https://github.com/spacetelescope/jdat_notebooks/blob/main/notebooks/MIRI/MIRI_LRS_spectral_extraction/miri_lrs_advanced_extraction.ipynb}} Both individually extracted nod spectra agree within a few percent of each other, and we average them to produce the final spectrum. In the NIR, the SN trace strongly dominates over the background regions, and the spacing of the three dithered spectra makes it difficult to define a separate, uncontaminated background region, so we use the automatic JWST pipeline \texttt{stage3} product for the NIRSpec observations.

\subsubsection{Photometry}
High-cadence multi-wavelength observations were taken as part of the Global Supernova project with the Las Cumbres Observatory telescope network of 0.4~m and 1.0~m telescopes \citep{Brown_2013} in the $B$, $V$, $g$, $r$, and $i$ bands. The PyRAF-based photometric reduction pipeline was used to reduce the images as described by \citet{Valenti_2016}. We calibrate apparent magnitudes using the APASS catalog for $g$, $r$, and $i$ and using Landolt standard fields observed with the same telescope on the same night for $B$ and $V$.

We also obtained ultraviolet images with the Ultraviolet/Optical Telescope \citep[UVOT;][]{Roming05} on the Neil Gehrels \textit{Swift} Observatory \citep{Gehrels_2004} until the SN was no longer detectable. These images were reduced using the High-Energy Astrophysics software (HEASoft\footnote{\url{https://heasarc.gsfc.nasa.gov/docs/software/heasoft/}}). We centered at the position of the SN with an aperture size of 3$\arcsec$, and the background is measured from a region without contamination from other stars with an aperture size of 5$\arcsec$. Zero points for photometry were chosen from \cite{Breeveld_2010} with time-dependent sensitivity corrections updated in 2020. In the current work, we use this multiwavelength data to calculate the explosion properties, and we present the data in \autoref{fig:lc}. 

\subsection{Radio observations}
We acquired radio observations of SN~2024abup on 2025 Feb 14 (84 days after explosion) and 2025 Jun 21 (211 days after the explosion) with the Karl G. Jansky Very Large Array (VLA) under the program 24B-533 (PI: Christy). The data were taken during the A$\rightarrow$D and C configurations, respectively, using the L, S, C, X receiver bands. We used 3C147 as the bandpass and flux-density calibrator, and J0204-1701 as the complex gain calibrator for all frequencies. The data were processed in the Common Astronomy Software Application (CASA; \citealp{Casa_07,CASA_team}) following standard reduction procedures. We imaged the data with the CASA task \verb'tclean', and derived flux densities and their uncertainties using \verb'imgfit', by fitting an elliptical Gaussian fixed to the size of the synthesized beam at the phase center of each image.

\begin{table}
 \caption{Properties of SN~2024abup} \label{tab:results}
 \begin{tabular}{ll}
    \hline
    Parameter & Value \\
    \hline
    R.A. (J2000) & 01:49:11.31 \\
    Dec. (J2000) & $-$10:25:27.4 \\
    Last Nondetection (MJD) & 60635.16 \\ 
    First Detection (MJD) & 60636.35 \\
    Explosion Epoch (MJD)\tablenotemark{a} & 60635.75 \\
    Time of $V_{\mathrm{max}}$ (MJD) & 60649.40 \\
    Peak Absolute Magnitude ($V_{\mathrm{max}}$) & $-18.75$ mag\\
    Redshift ($z$)\tablenotemark{b} & 0.00582  \\
    Distance\tablenotemark{c} & 23.3  $\pm$ 1.6 Mpc\\
    Distance modulus ($\mu$)\tablenotemark{c} & 31.84 $\pm$ 0.15 mag\\
    $E(B-V)_\mathrm{MW}$ \tablenotemark{b}& $0.0296 \pm 0.0344$ mag\\
    $E(B-V)_\mathrm{host}\tablenotemark{c}$ & $0.79 \pm 0.06$ mag\\
   % $E(B-V)_\mathrm{tot}$\tablenotemark{c} & $0.82 \pm 0.06$ mag\\
    Expansion velocity\tablenotemark{d} & $18200 \pm 3200$ km/s \\
    $M_{Ni}$ & $0.24 \pm 0.01 M_\odot$\\
    $M_{ej}$ & $3.53 \pm 0.15 M_\odot$ \\
    $E_K$ & $8.9 \pm 3.1 \times 10^{51}$erg\\
   % Rise time ($V$) & 14 days \\
    \hline
 \end{tabular}
 \tablenotetext{a}{mid point of last nondetection and first detection}
 \tablenotetext{b}{from \citet{Schlafly_2011} from https://irsa.ipac.caltech.edu/applications/DUST/}
 \tablenotetext{c}{from the \ion{Na}{1}~D lines of the host galaxy and \citet{Gordon_2024}}
  \tablenotetext{d}{at V band maximum}

\end{table}

\section{Host galaxy \& extinction} \label{sec:host}
% \begin{figure}
%     \centering
%     \includegraphics[width=\columnwidth]{figure/HostProperties.pdf}
%     \caption{Properties of the host galaxy of SN~2024abup with respect to other SN Ic and SN Ic-bl host galaxies. Here we present the star formation rate (SFR) with respect to the mass of the host galaxies. The values SN Ic and SN Ic-bl population is from \citet{Modjaz_2016}. We find the SFR for SN~2024abup host is similar to general population, however, the mass of the host galaxy is slightly higher than SN Ic-bl's host and similar to SN Ic's host.}
%     \label{fig:hp}
% \end{figure}
We examine the properties of the host galaxy of SN\,2024abup. \citet{Kalinova_2021_host} estimated the stellar mass ($M_*$), and star formation rate (SFR) of 238 galaxies---including NGC\,0681, the host galaxy of SN\,2024abup---assuming a Chabrier initial mass function (IMF) \citep{Chabrier_2003_imf}. We adopt their values for NGC\,0681 of $\log(\frac{\mathrm{SFR}}{\mathrm{M_\odot yr^{-1}}})=-0.245$  and $\log(\frac{\mathrm{M_*}}{M_\odot}) = 10.16$. Comparing to the SN Ic-bl host galaxy sample of \citet{Modjaz_2020, Japelj_2018}, which spans a median mass of $\log(\frac{\mathrm{M_*}}{M_\odot}) = 8.9$ and a maximum of $\log(\frac{\mathrm{M_*}}{M_\odot}) = 9.46$, we find that NGC\,0681 is relatively massive. The SFR of NGC\,0681 is consistent with the general SN Ic-bl host population \citep{Modjaz_2020}, whose median and maximum values are $\log(\frac{\mathrm{SFR}}{\mathrm{M_\odot yr^{-1}}})=-$0.41  and $\log(\frac{\mathrm{SFR}}{\mathrm{M_\odot yr^{-1}}})=$0.42, respectively. We use the host redshift from the literature $z = 0.00582$ reported by \citet{Gordon_2024} and velocity of  $1704 \pm 9~\mathrm{km/s}$, and we adopt it for all the analyses presented in this paper.

%A wide range of redshifts for the host are presented in the literature\footnote{\href{https://ned.ipac.caltech.edu/cgi\-bin/datasearchsearch\_type=z\_id\&objid=4936\&objname=NGC\%200681\&img\_stamp=YES\&hconst=73.0\&omegam=0.27\&omegav=0.73\&corr\_z=1\&of=table}{Redshifts of SN~2024abup in NED}}. We used Na I D lines in a high-resolution spectrum of SN~2024abup to estimate the redshift of the host galaxy to be $z = 0.005821$ as shown in \autoref{fig:redshift}. This value agrees with a previously reported measurement by \citet{Driel_2016}, and we adopt it for all the analyses presented in this paper.

\subsection{Line-of-sight extinction}
We assume the interstellar dust extinction due to Milky Way dust to be $E(B-V)_{\rm MW}=$0.0297 $\pm$ 0.0019 mag from \citet{Schlafly_2011}. SN~2024abup is located on a dust lane in NGC\,0681, complicating the calculation of host galaxy dust extinction. Traditionally, the equivalent width of Na I D absorption features is used to calculate the dust extinction value; however, in this case, the absorption feature is saturated, so we cannot implement this method. Hence, we follow the prescription presented by \cite{Stritzinger_2018}, \cite{Taddia_2019}, and \cite{Drout_2011} utilizing the SN color. In short, this method compares the $B-V$ color evolution from 0 to $+$20\,days after the $B$-band peak of stripped-envelope SN with insignificant host extinction. The difference in weighted average in this time range between the two SNe is used to estimate the host extinction value. For SN~2024abup, we compared the $B-V$ evolution with SN~2007ru, an SN Ic-bl with no host galaxy component detected \citep{Sahu_2009}. From this method, we find the host dust extinction contribution to be $E(B-V)_{\rm host}=$0.79 $\pm$ 0.06\,mag. We compared the $B-V$ evolution of SN~2024abup with several other SN Ic-bl to check the validity of this method and found that they agree well. We therefore use these extinction values for our analysis. We note that, as the focus of this paper is on the infrared wavelength range, these high extinction values do not impact the analysis significantly.

\section{Explosion properties}\label{sec:photometry}

\begin{figure}
    \centering
    \includegraphics[width=\columnwidth]{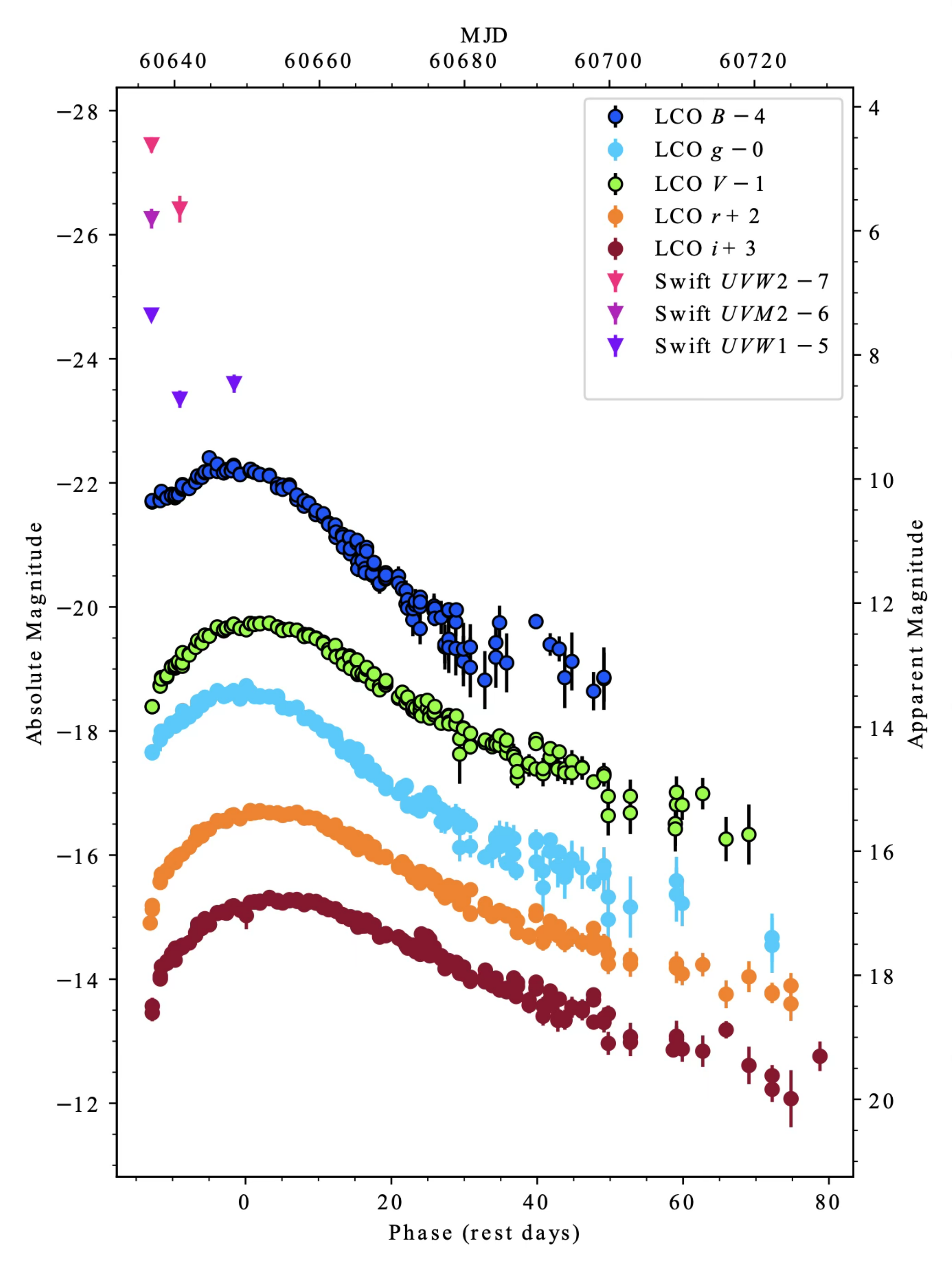}
    \caption{Multi-wavelength extinction corrected photometry of SN\,2024abup spanning UV to optical observed using Swfit and LCO. The light-curve is well sampled to 80 days after the peak in the V band.  }
    \label{fig:lc}
\end{figure}

\begin{figure}
    \centering
    \includegraphics[width=\columnwidth]{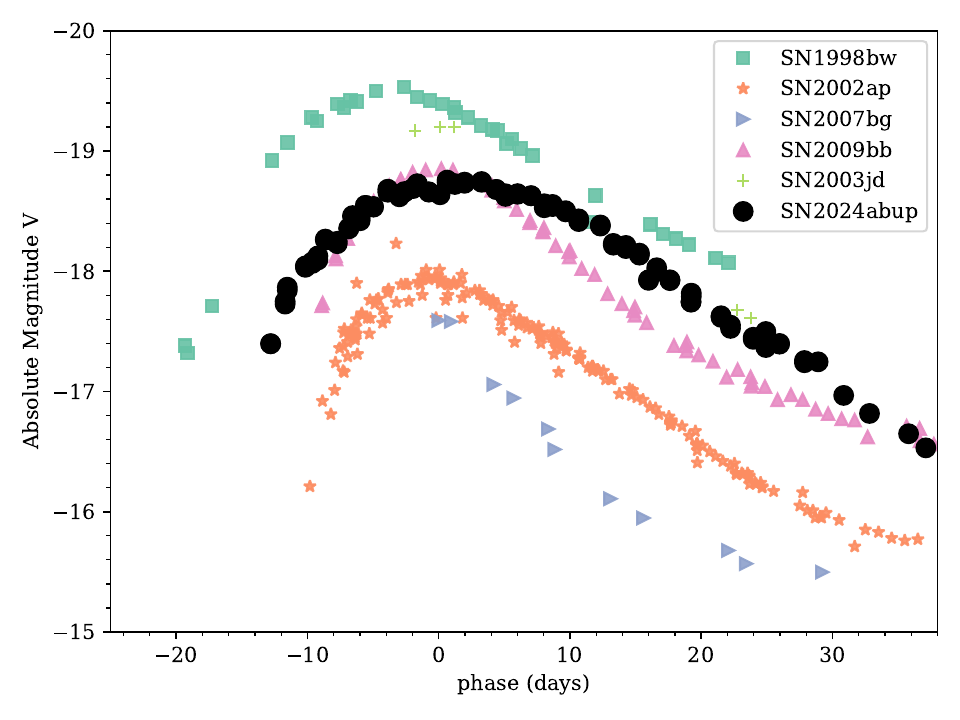}
    \caption{ Comparison of extinction corrected absolute V band magnitudes for a few different SN Ic-bl with (SN~1998bw) and without an associated GRB. The phases are given with respect to the date of $V_{\mathrm{max}}$. SN~2024abup (black circle) falls within a reasonable range with SN~1998bw, which is associated with a low-luminosity GRB, being the brightest.}
    \label{fig:V-lightcurve}
\end{figure}

From the near-peak optical spectrum of SN\,2024abup (\autoref{fig:spec_max}), we estimate the expansion velocity using the \ion{Fe}{2} absorption velocity as a proxy, as described by \citet{Modjaz_2016}. We find $v=$18200 $\pm$ 3200\,$\mathrm{km/s}$, which falls in a normal range for SN Ic-bl with no associated GRB \citep{Modjaz_2016}. 

We use this expansion velocity and the light curve (\autoref{fig:lc}) to calculate other properties of SN~2024abup, such as nickel mass ($M_{\rm Ni}$), ejecta mass ($M_{\rm ej}$), and kinetic energy ($E_K$) using analytic models from \citet{Arnett_1982} and \citet{Valenti_2008}. This approach relies on some simplified assumptions, such as spherical symmetry, homologous expansion of the ejecta, a single opacity value for different compositions, and full mixing. 
We used the light-curve fitting package \citet{hosseinzadeh_light_2023} to generate a bolometric light curve, which is used to fit an Arnett model \citep{Arnett_1982} using a Markov chain Monte Carlo (MCMC) method. We fit the light-curve only up to the photospheric phase ($<60$\,days) to a one-component model as described in \citet{Valenti_2008} where opacity $\kappa$ is a free parameter. From this method, we obtain $M_{\rm Ni} = 0.2 \pm 0.1 M_\odot$, $M_{\rm ej} = 3.6 \pm 0.2 M_\odot$, $\kappa = 0.02 \pm 0.01$ and $E_{K} = 8.9 \pm 3.1 \times 10^{51}$\,erg as presented in \autoref{tab:results}. The corner plot for the fit is shown in \autoref{fig:corner}. Comparison of explosion properties of SN\,2024abup with other SN\,Ic-bl presented by \citet{Taddia_2019} shows that this transient falls in a normal range across all the properties for the SNe Ic-BL population. Additionally, the evolution of the light curve of SN\,2024abup is similar to other SN Ic-bl population as show in \autoref{fig:V-lightcurve}.

\section{Spectral analysis} \label{sec:spectra}

%In this paper, we present the optical to MIR spectrum of SN~2024abup in \autoref{fig:spec_jwst_opt-mir}. 
We mark the dominant contributing ions/molecules to prominent spectral features in the JWST spectrum, including previously unidentified lines at $\lambda>2.5$\,$\mu$m as shown in \autoref{fig:spec_jwst_opt-mir}. Most of these line identifications are based on a \texttt{SUMO} NLTE spectral model (see Section \ref{subsec:spectral_modeling} and Figure \ref{fig:ion_model}). As in the optical spectrum, and characteristic of the Ic-bl class as a whole, SN\,2024abup also exhibits broad and blended features at these longer wavelengths that are difficult to separate cleanly.

A large fraction of previous observations of SNe Ic-bl have been limited to optical and NIR wavelengths ($<$2.5 $\mu m$) at comparable epochs. \citet{Siebert_2024} and \citet{Rastinejad_2025_25kg} have recently presented JWST spectra of two SNe Ic-bl spanning 1 to 5 $\mathrm{\mu m}$; however, the spectrum of SN 2025kg was obtained near peak brightness, and that of SN 2023adta suffers from low signal-to-noise.  Here we present a spectrum of SN 2024abup spanning 0.4--14 micron (\autoref{fig:spec_jwst_opt-mir}), and we employ the spectral modeling described below to aid in identifying the MIR features.

%SN~2025kg JWST spectrum is during the peak in magnitude, and SN~2023adta JWST spectrum is low signal-to-noise. Hence, these spectra are not ideal to perform a one-to-one comparison with SN~2024abup data. The spectrum covering from $0.4 \mu m$ to $14 \mu m$ is presented in \autoref{fig:spec_jwst_opt-mir}. %One defining characteristic of SN Ic-bl is the presence of broad spectral features.  
%We use spectral modeling, described below, to aid in identifying the new MIR lines.

\subsection{Spectral Modeling with \texttt{SUMO}}
\label{subsec:spectral_modeling}

We compare the JWST NIRSpec and MIRI spectra to a model created using the \texttt{SUMO} spectral synthesis code \citep{Jerkstrand_2011_SUMOa, Jerkstrand_2012_SUMOb} in order to identify the contribution of different ions in the production of spectral features and to constrain potential $r$-process signatures. \texttt{SUMO} is a Monte Carlo radiative transfer code that takes as input an ejecta model of the exploded star and then solves the non-local thermodynamic equilibrium (NLTE) equations for the temperature, excitation, and ionization structure in the SN ejecta. We summarize our spectral modeling setup below, and give a detailed description in \autoref{appen:sumo}.

To model the $+$41-day spectrum of SN\,2024abup, we use a massive star model with $M_{\text{He,i}}=$\,12.0\,$M_{\odot}$ ($M_{\mathrm{ej}}$) evolved until core collapse by \citet{Woosley_2019_ejectamodels}, and exploded by \citet{Ertl_2020_ejectamodels}. After prompt removal of the hydrogen envelope (mimicking Roche-lobe overflow to a binary companion but see \citep{Laplace_2021}), this star lost its entire helium envelope effectively due to stellar winds. This is one of the reasons for choosing this particular model; lower-mass models retained some of their helium envelope until explosion, resembling Type Ib SNe rather than Type Ic SNe. Another reason is the model's large expansion velocities ($v \sim$ 8500 km s$^{-1}$), leading to a better match to a Type Ic/Ic-bl than its lower-mass counterparts. The main ejecta properties of this model are summarized in Table \ref{tab:summary_table_SUMO_model}. As this ejecta model contains no $r$-process elements, it enables a search for $r$-process signatures in SN\,2024abup; when a feature present in SN 2024abup is not present in the model spectrum, and this feature matches an $r$-process transition, it would then be a good candidate for an $r$-process signature. 

\begin{figure*}
    \centering
    \includegraphics[width=\textwidth]{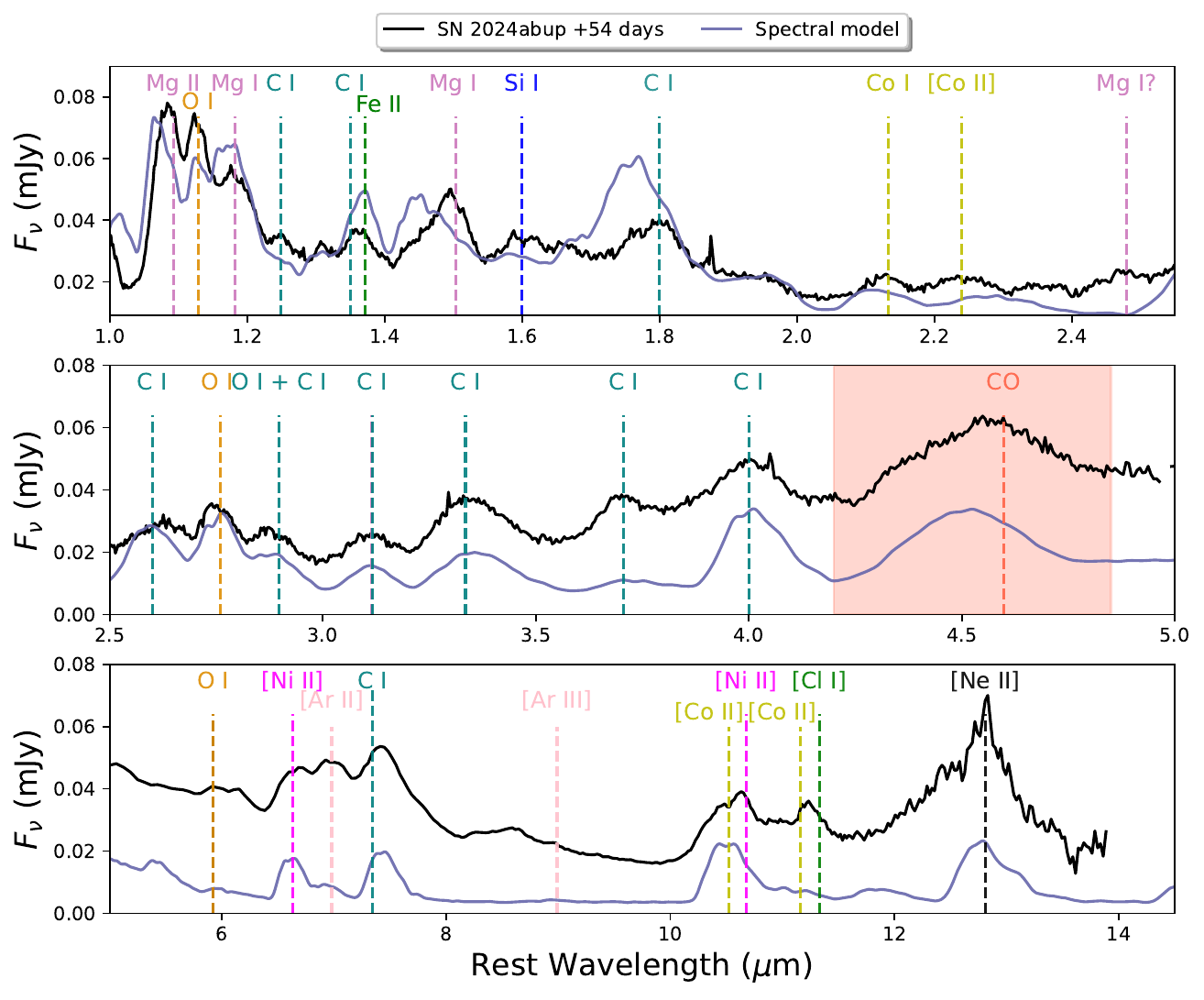}

    \caption{SN~2024abup compared to model from \texttt{SUMO}. The model is able to reproduce several observed spectral features. However, there are differences in the strength, which is expected as this model is not fine-tuned for this particular case. Overall, the observed continuum is higher than the model for wavelengths greater than $\sim$3 $\mu m $, which could be due to the presence of dust in the SN or the host galaxy or both, which is not included in the model. The shaded region in red shows the CO feature predicted by the model and the observed spectral feature in the same wavelength regime. }
    \label{fig:ion_model}
\end{figure*}

In \autoref{fig:ion_model}, we show the \texttt{SUMO} model spectrum compared to the observations, with the dominant ion contributions in the model marked. While the strengths of the model lines do not match the observed spectrum for every feature, the model successfully reproduces most of the spectral features seen in SN\,2024abup, providing reliable line identifications. We note that no extensive grid of model comparisons or parameter fine-tuning was performed to try to match the SN 2024abup spectrum perfectly; our interest here is primarily one of line identification. A more extensive modeling effort is deferred to a future article (Barmentloo et al., in prep.).  

 %Thus, it is difficult to separate different lines clearly.

%To identify the lines as shown in \autoref{fig:spec_jwst_opt-mir}, we employ a few different techniques. First, we use previous observations of SNe Ic-BL in optical and NIR wavelengths and their line identifications in the literature to find the observed lines in SN~2024abup. For the MIR, there are not any published JWST spectra of SN Ic; however, there are multiple publications on SNe Ia \citep{Kwok_2023,Kwok_2024,Kwok_2025}. Hence, we use the lines identified for SN Ia \citep{Kwok_2023,Kwok_2024,Kwok_2025} to identify the lines that are common for both. Finally, there are features in $2.25$ to $5$ $\mu m$ wavelength range which are not observed in the SN Ia JWST spectrum. Hence, to identify these lines, we use the atomic line list \footnote{https://linelist.pa.uky.edu/atomic/} and search for the strongest lines predicted for the given wavelength regime.  

\subsection{$r$-process elements}
%SN Ic-bl have been proposed as a primary alternative to BNS mergers for $r$-process element production \citep{Siegel_2019,Barnes_2022,Ricigliano_2025}. 
To identify possible $r$-process lines in the SN\,2024abup spectrum, we use a line list generated by \citet{Ricigliano_2025}. All the lines from \citet{Ricigliano_2025} are plotted along with the SN~2024abup spectrum \autoref{fig:all_rprocess}. We note that \citet{Ricigliano_2025} computed these lines for the nebular phase of SN Ic-bl; however, the JWST spectrum of SN~2024abup is not fully nebular because we still observe absorption features. 

\begin{figure*}
    \centering

    \includegraphics[width=\textwidth]{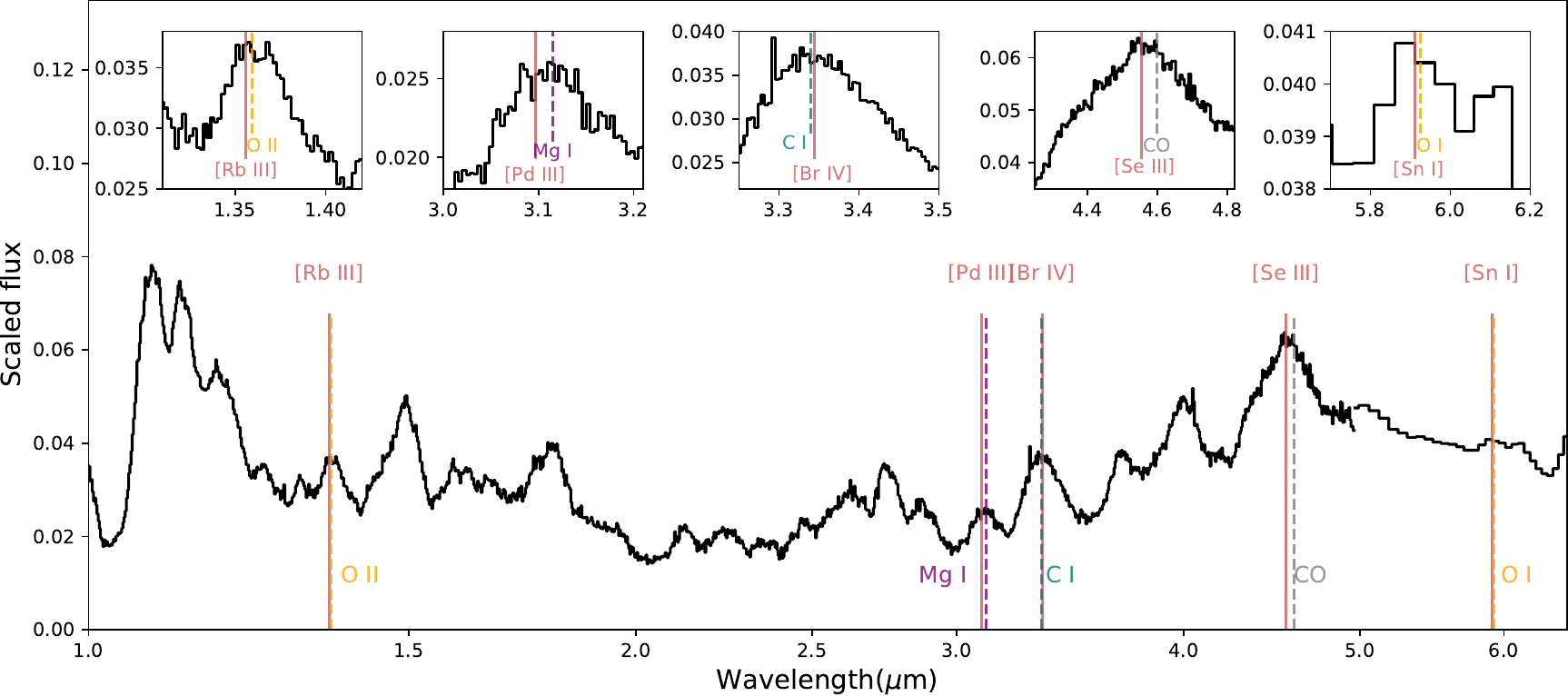}

    \caption{JWST spectrum of SN\,2024abup at $+$41 days after V band maximum, overlaid with $r$-process element lines from \citet{Ricigliano_2025} (solid red) and light element lines (colored dashed). The spectral features are more consistent with light elements, though line broadening may produce blending that complicates this identification. Insets show zoomed views of each individual feature.}
    \label{fig:only_rprocess}
\end{figure*}

We find several potential matches for lines predicted by \citet{Ricigliano_2025} for SN Ic-bl in the case with original and boosted ionization rate of heavy elements that align with spectral features in the spectrum, shown in \autoref{fig:only_rprocess}. We find that [\ion{Rb}{3}] ($\lambda 1.35 \mathrm{\mu m}$), [\ion{Pd}{3}] ($\lambda 3.09 \mathrm{\mu m}$), [\ion{Br}{4}] ($\lambda 3.34 \mathrm{\mu m}$), [\ion{Se}{3}] ($\lambda 4.55 \mathrm{\mu m}$) and [\ion{Sn}{1}] ($\lambda 5.91 \mathrm{\mu m}$) align with prominent features we see in the spectra. Out of these lines, only [\ion{Sn}{1}] could be produced without artificially boosting the ionization of these elements. For all of these features, we find that alternatively they could be attributed to non-$r$-process elements as we illustrate in  \autoref{fig:only_rprocess}. \texttt{SUMO} can reproduce these features without $r$-process ions in the code. Though, the strength and the shape of these features do not match accurately, these lines are more likely to be produced by non-$r$-process elements. Nevertheless, we cannot rule out the possibility that contributions from $r$-process elements are blended with those of other ions present in the \texttt{SUMO} model. 

We further investigate possible $r$-process element signatures in SN\,2024abup in the following sections. In \autoref{sec:ion} we calculate the $r$-process ionic mass production in the limit that the features in SN\,2024abup, which aligned with predictions, are produced entirely by heavy elements. Next, in \autoref{sec:comp_trans}, we compare the NIR spectrum of SN\,2024abup with observations of transients that have been proposed to possess $r$-process signatures.  %, and finally in \autoref{sec:comp_23dbc}, we compare SN~2024abup with SN~2023dbc (type Ic).

% In \autoref{fig:rprocess_velcomp}, we compare these lines in velocity space centering both on probable $r$-process element (left) and non-$r$-process element (right). All the lines appear to be redshifted except for CO, this redshift in lines has been previously seen for various lines around similar epoch for SN~1998bw \citep{Patat_2001}. Additionally, the spectral features are broad ($\sim$ 10,000 to 20,000 $\mathrm{Km/s}$), which indicates that multiple lines could be contributing to one feature. Hence, to definitively separate $r$-process element signatures, we need multiple spectra at different epochs going into the nebular phase. Furthermore, computational models designed for SN~2024abup will be useful in separating these lines which will be carried out in future work. 

\subsubsection{Ionic mass limits}\label{sec:ion}

At the relatively early epoch of $+$41d, the IR lines arising from low-level intra-multiplet transitions can be expected to be formed in LTE (see e.g. Barmentloo \& Jerkstrand 2026, submitted). If the line is optically thin, the luminosity is then 
\begin{equation}
L = M_{\rm ion} \left( \mu m_p\right)^{-1}
 A h \nu \frac{g_u}{Z(T)} e^{-E_u/kT}.
\end{equation}
Here, $\mu$ is the atomic weight of the ion, $A$ is the Einstein coefficient for spontaneous emission, $g_{u}$ is the statistical weight of the line's upper level, $E_{u}$ its energy and $Z(T)$ is the partition function \citep{Jerkstrand2017}. The exponential factor will likely be close to unity at these epochs in the MIR ($\gtrsim 3\ \mu m$) as the ejecta are still quite hot ($kT \gtrsim hc/(3\ \mu m)$ for $T>5000$ K). Then, rearranging in terms of typical values,
\begin{multline}
L \approx 10^{39} \left(\frac{M_{\rm ion}}{10^{-4}\ M_\odot}\right) \left(\frac{\mu}{80}\right)^{-1}\frac{g_u} {Z(T)} \\
\times \left(\frac{\lambda}{3\ \mu m}\right)^{-1}\times A ~\mbox{erg s}^{-1}.
\end{multline}
The characteristic line luminosity of the features measured in the observed spectrum is a few times $10^{39}$ erg s$^{-1}$, which is consistent with optically thin LTE emission from $\mathcal{O}(10^{-4}$)\,$M_\odot$ of the corresponding ion if the Einstein coefficient for spontaneous emission (the A-value) is of order unity. In Table \ref{table:LTEmasses} we give the specific corresponding ionic masses if the luminosity in a feature is assumed to come from this element, and the line is formed under LTE, optically thin conditions. Here, the specific line luminosity measurements were done by integrating flux over the wavelength for the lines with respect to the continuum and converting the flux to luminosity using the distance of SN~2024abup.
%Some examples of A-values are 7.2\,s$^{-1}$ for [\ion{Rb}{3}]\,1.37\,$\mu$m and 0.32\,s$^{-1}$ for [\ion{Br}{4}]\,3.81\,$\mu$m.

For a line to be optically thin, the Sobolev optical depth \citep{Sobolev_1957_Canon}, given by (assuming most ions are in the ground state)
\begin{multline}
\tau_S = 0.4 \times A \times \left(\frac{\lambda}{3\ \mu m}\right)^{3} 
\times \left(\frac{M_{\rm ion}}{10^{-4}\ M_\odot}\right) \left(\frac{\mu}{80}\right)^{-1} \\
\times \left(\frac{v_{\rm exp}}{8000\  \mbox{km s}^{-1}}\right)^{-3} \left(\frac{f}{0.1}\right)^{-1} \frac{g_u}{g_l}\left(\frac{t}{50\ \mbox{d}}\right)^{-2},
\label{eq:tau}
\end{multline}
should be $<$ 1. Here, $f$ is the filling factor (fraction of volume occupied by material containing the element). Thus, for IR lines with $A \approx 1$ s$^{-1}$, 41\,d after the V band maximum can be a typical time at which transition to optical thinness occurs ($\tau_S$ becomes $<1$). If it has not occurred, we would expect the line to form more of a P-Cygni like feature than a pure emission feature, and higher mass solutions can be allowed compared to the optically thin case. 

%In the optically thick case, even these forbidden lines could produce P-Cygni like features (normally arising in allowed transitions). In the optically thin case, the feature would be composed of emission without an absorption component. The relatively poorly constrained underlying continuum---from some combination of residual background contamination, a pseudo-continuum from SN line flux, and dust---is an additional challenge that makes it difficult to assess this and to isolate line flux.

%From these considerations, it would seem the observed JWST spectra can probe $r$-process ionic masses down to about $10^{-4}$ $M_\odot$. A hypothesis that a feature is due to a forbidden line with $A \approx 1$ s$^{-1}$ and $M \gtrsim 10^{-4}$ $M_\odot$ is viable. Higher masses than the limiting value can be allowed if absorption is seen in the line; thus, the limiting value is a lower mass limit. Analogously, if limits on non-observed features can be put to $<10^{39}$\,erg\,s$^{-1}$, then the limiting value becomes an upper limit for the ion mass (optically thin case). 

The solar abundance mass ratio of e.g., Se/Fe is around $10^{-4}$. If we take 0.1\,$M_\odot$ as the characteristic iron production, an average SN would then need to make $\sim 10^{-5}$ $M_\odot$ of Se to generate the solar abundance (the contribution by Type Ia SNe to the iron production, and the s-process to the Se production, are both roughly 50\% so they roughly cancel out). If only SN\,Ic-bl produce the Se, then Se/Fe ratio increases to $\sim 10^{-3}$\,$M_\odot$ as these are about a factor 100 rarer. These estimates, combined with the ionic mass limits derived in Table \ref{table:LTEmasses}, show that the JWST observations are sensitive enough to make meaningful tests of $r$-process production. A model or assumption is still needed about the ionization state of the r-process material. Neutral, singly and doubly ionized states are most common in SNe at these phases, so the [\ion{Sn}{1}], [\ion{Pd}{3}], and [\ion{Se}{3}] limits are likely more constraining than the [\ion{Br}{4}] one.  

Ultimately, we cannot put clear upper limits on Pd, Br, Se content in the ejecta of SN 2024abup, because the epoch of our JWST spectrum is such that the ejecta may still be optically thick in the candidate lines. Observations at somewhat later epochs would alleviate this issue. For Sn, however, if a mass more than $5 \times 10^{-5}$ $M_\odot$ were present, and to a significant extent in the neutral state, the table shows that a much stronger feature would have been observed at 5.91 $\mu$m. With the solar Sn/Fe mass ratio at $6 \times 10^{-6}$, this mass limit is borderline constraining for the case of only Ic-BL SNe, making Sn (expected Sn mass then $6 \times 10^{-5}$ $M_\odot$ per event).
%More analysis of the results are presented in the table \autoref{table:LTEmasses}.

\begin{table*}
\centering
\begin{tabular}{cccccccc}
\hline
Wavelength & Candidate & $L_{obs}$ (erg s$^{-1}$) & A (s$^{-1}$) & $g_u/Z(T)$ & Ion mass (LTE,thin) ($M_\odot$) & $\tau_s$ & Comment  \\
\hline
%[\ion{Si}{1}] 1.64 & $2.7 \times 10^{-3}$ & 5/9 & 0.015\\
\hline
%1.86 $\mu m$ & [\ion{Kr}{2}] & ? & 2.78 & 2/6  & $2 \times 10^{-5}$ (for $L=1E39$) &  \\
%\hline
%2.71 $\mu m$ & [\ion{Br}{1}] & ? &  0.90 & 2/6 & $1.0 \times 10^{-4}$ (for $L=1E39$)  &  \\
%\hline 
3.10 $\mu m$ & [\ion{Pd}{3}] & $1 \times 10^{39}$ & 0.86 & 0.52 & $\leq 4.5 \times 10^{-4}$ & $\leq 0.42$ & C I contam.\\
%A=0.76 from FAC model, rescale (3.23/3.10)^3.
\hline
%
%3.19 $\mu m$ & [\ion{Br}{2}] & $\ll 10^{39}$ & 0.67 & 3/5 & $\ll 2.3 \times 10^{-4}$\\
%\hline
%
3.34 $\mu m$ & [\ion{Br}{4}] & $4 \times 10^{39}$ & 0.33 & 1.45 & $\leq 4.7 \times 10^{-3}$ & $\leq 2.9$ & C I contam.\\
\hline
4.55 $\mu m$ & [\ion{Se}{3}]  & $5 \times 10^{39}$ & 0.14 & 1.10 & $\leq 1.5 \times 10^{-2}$ & $\leq 9.7$ & CO contam.\\
\hline
%
%5.03 $\mu m$ & [\ion{Se}{1}] & ? & 0.17 & 3/9 & $1.0 \times 10^{-3}$ (for $L=1E39$) \\
%\hline 
%
5.91 $\mu m$ & [\ion{Sn}{1}] & $1 \times 10^{37}$ & 0.084 & 0.58 & $ \leq 4.4 \times 10^{-5}$ & $\leq 0.018$ & No contam.?\\
\hline
%
%6.99 $\mu m$ & [\ion{Ar}{2}] & X & $5.3 \times 10^{-2}$ & 1/2 &$4.6 \times 10^{-3}$\\ 
%\hline
\end{tabular}
\caption{Constraints on ionic masses under the assumption of LTE, optically thin formation, for $T = 5000$ K. The $\tau_s$ column shows the optical depths for the limiting masses using the Eq. \ref{eq:tau} parameters. We limit the list to $>$3 $\mu$m lines as it's mainly these that give temperature-independent results.}
\label{table:LTEmasses}
\end{table*}

\subsubsection{Comparison with other transients with $r$-process lines}\label{sec:comp_trans}

\begin{figure*}
    \centering
    \includegraphics[width=\textwidth]{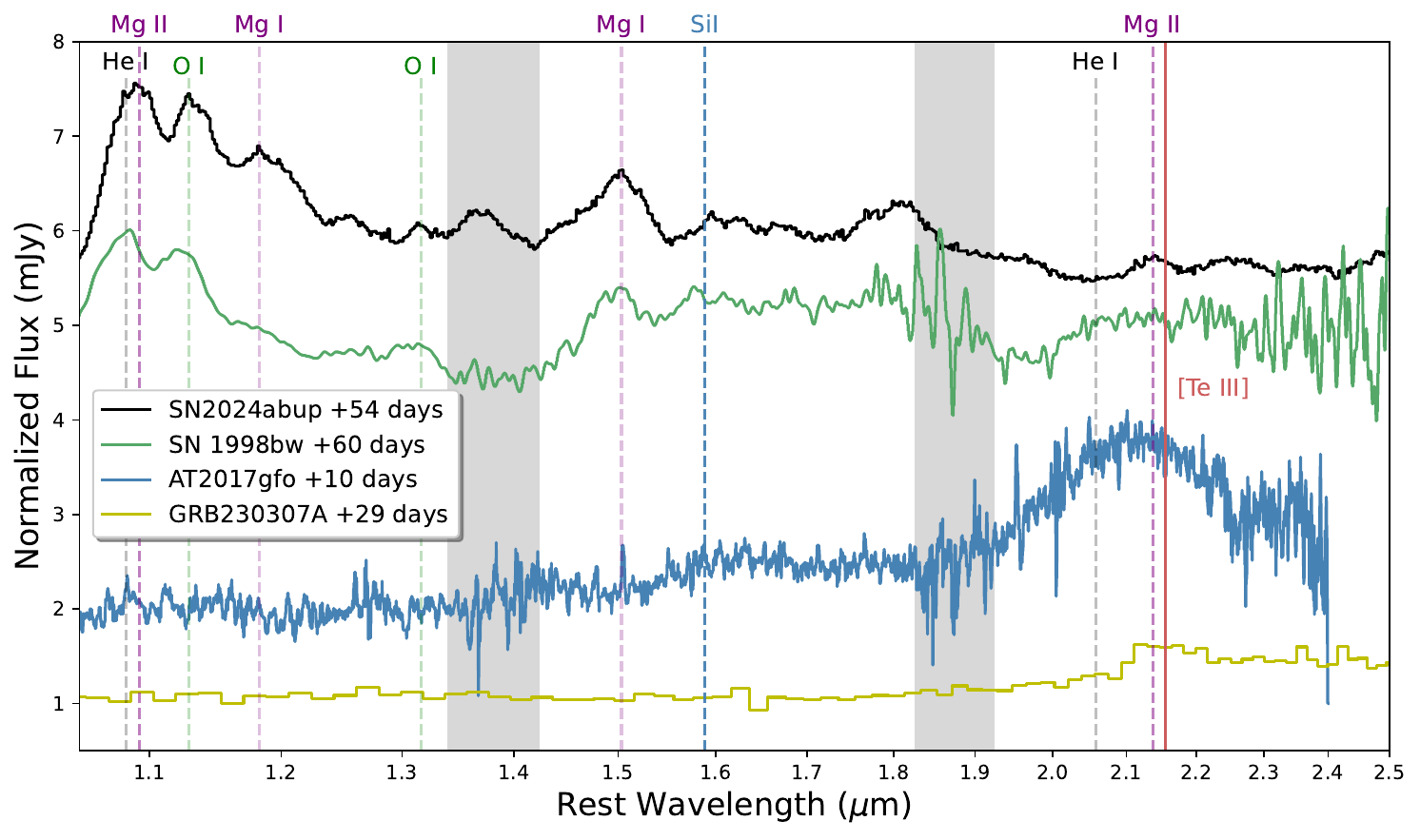}

    \caption{Comparison of the NIR spectrum of SN~2024abup at $+$41\,days with SN~1998bw \citep{Patat_2001}, AT~2017gfo \citep{Pian_2017}, and GRB~230307A \citep{Levan_2024_GRB230307A}. Similar lines are seen for both SN~1998bw and SN~2024abup. The [\ion{Te}{3}] line is identified in AT2017gfo and GRB230307A as a signature of $r$-process element production. There is a bump in SN~2024abup near the [\ion{Te}{3}] line, but it is better explained by \ion{Mg}{2} in this case.}
    \label{fig:spec_icomp}
\end{figure*}
Observational signatures of $r$-process elements were detected in NIR spectra of AT~2017gfo \citep{Kasliwal2022} and GRB~230307A \citep{Levan_2024_GRB230307A, Gillanders_2025} (but see \citealt{Arunachalam_2025}). In search of similar features in SN\,2024abup, we compare the NIR spectrum to three different types of transients (\autoref{fig:spec_icomp}): 1) SN\,1998bw \citep{Patat_2001}, the well-studied, canonical SN Ic-bl associated with a low-luminosity GRB; 2) AT\,2017gfo \citep{Pian_2017}, a kilonova associated with a binary neutron star merger; and 3) GRB\,230307A \citep{Levan_2024_GRB230307A}, an LGRB with kilonova-like emission. The two SN Ic-bls have similar prominent lines present. In contrast, AT\,2017gfo and GRB\,20230307A do not contain many spectral features reminiscent of SN\,Ic-bl, and instead exhibit a feature at $\sim$2.1\,$\mathrm{\mu m}$ that has been interpreted as [\ion{Te}{3}] in both \citep{Kasliwal_2017, Levan_2024_GRB230307A, Gillanders_2025}. This [\ion{Te}{3}] feature is significantly broader and more prominent in AT\,2017gfo than in GRB\,230307A, which might be due in part to the earlier phase. 

There is also a bump in the spectrum of  SN\,2024abup and a broader feature in SN\,1998bw (though with lower signal-to-noise) near the [\ion{Te}{3}] line. We investigate if another line from lighter elements could be responsible for the bump in SN\,2024abup and find that \ion{Mg}{2} is a likely alternative (\autoref{fig:spec_icomp}). The bump is more centered on the \ion{Mg}{2} line, and blue-shifted relative to the [\ion{Te}{3}] line. Most of the other lines in SN\,2024abup are redshifted, and other \ion{Mg}{1} and \ion{Mg}{2} lines are present elsewhere in the spectrum, so we attribute this feature to be \ion{Mg}{2}, rather than [\ion{Te}{3}]. However, we cannot rule out a weak contribution from [\ion{Te}{3}]. Future fully nebular phase IR spectroscopy of SN Ic-bl would strengthen constraints on the production of $r$-process elements in collapsars. We note that this feature could potentially account for the $\sim$2.1 $\mu m$ signal observed in GRB~230307A; however, this interpretation is unlikely, given the overall interpretation of the source being a neutron star merger (in which alpha elements are not generally synthesized).

\subsubsection{Comparison with SN~2023dbc} \label{sec:comp_23dbc}
\begin{figure*}
    \centering
    \includegraphics[width=\textwidth]{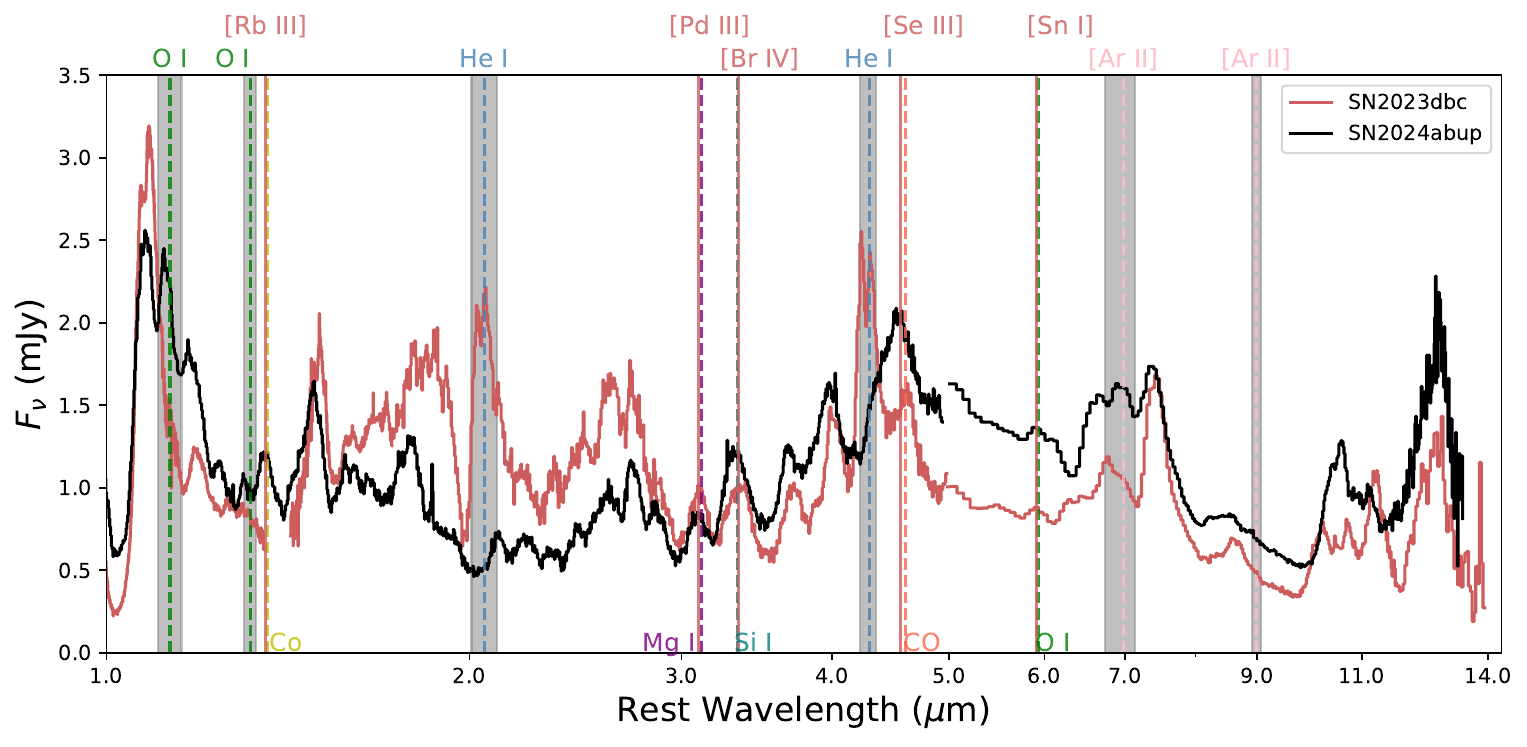}

    \caption{Comparison of the JWST NIR $+$ MIR spectra of SN~ (+41 days) and SN\,Ic~2023dbc (+36 days) \citep{Shahbandeh_2023}. Major lines that are different between the two are shown in shaded gray regions. Lines from light and heavy species are indicated by dashed and solid lines, respectively. One of the major differences is the presence of prominent \ion{He}{1} features in SN~2023dbc, which are absent in the case of SN~2024abup. Additionally, we detect \ion{O}{1} lines in SN~2024abup, which are not present in SN~2023dbc. The CO feature in SN~2024abup is more prominent than in SN~2023dbc.}
    \label{fig:spec_comp}
\end{figure*}

Since the launch of JWST, there is a growing sample of observations of SNe at IR wavelengths. However, there are very few JWST observations of stripped envelope SN (SESN). So far, the Type Ic SN\,2023dbc \citep{Shahbandeh_2023} is the only SESN with 1--14\,$\mathrm{\mu m}$ coverage at a similar phase as SN\,2024abup. We note that SN\,2023dbc has also been classified as Type Ib by \citet{Yamanaka_2026}. In either case, it is an SESN and is expected to have similar lines as SN Ic-bl; however, these features are broader and more blueshifted in SN Ic-bl due to the higher velocity of the ejecta, as shown by a large sample study by \citet{Modjaz_2016}. Hence, this makes comparison with SN\,2023dbc an ideal case to search for $r$-process signatures. SN\,2023dbc is located in the very nearby galaxy M108 ($\sim$10 Mpc). We compare the JWST spectra of SN\,2023dbc \citep{Shahbandeh_2023} taken at +36 days since the r band maximum (we use the r band maximum to be at MJD 60030.6 from \citealt{Yamanaka_2026}) with SN\,2024abup at +41 days (\autoref{fig:spec_comp}). The comparison of these two spectra is valuable to study the properties in the MIR wavelength regime. Additionally, we can simultaneously investigate the presence of any $r$-process element contribution in the SN~2023dbc spectrum that coincides with predicted features, as shown in \autoref{fig:only_rprocess}. Notably, for normal SNe Ic, \citet{Barnes_2022} predict that hints for the presence of hypothetical $r$-process material would appear only for the compositional mixing in the ejecta being particularly efficient.

The spectra are overall similar between the two events, but with several key differences highlighted in \autoref{fig:spec_comp}. In the MIR SN\,2024abup shows stronger CO emission and a more pronounced continuum, which we associate with dust (see \autoref{sec:dust}). Another major difference is the presence of prominent \ion{He}{1} emission lines in SN\,2023dbc (shown in blue) that are not present in SN\,2024abup, indicating a higher degree of envelope stripping in the progenitor of SN\,2024abup. In contrast, the [\ion{Ar}{2}] line (and to a lesser extent the weak [\ion{Ar}{3}] line) is more pronounced in SN\,2024abup. The presence of Ar lines could also indicate explosive O burning, which is expected in massive star explosions \citep{Woosley_2002}.

Comparison of spectra in \autoref{fig:spec_comp} reveals that the broad features in SN~2023dbc align with those in SN~2024abup at wavelengths consistent with theoretically predicted $r$-process elements. The presence of these features in SN~2023dbc casts doubt on the formation of $r$-process elements in SN~2024abup. However, the features coincident with [\ion{Br}{4}] and [\ion{Se}{3}] are more pronounced in SN\,2024abup. This could potentially indicate a minor production of these heavy elements in SN~2024abup.

\subsection{CO \& Dust detection}\label{sec:dust}
CCSNe have been proposed as major producers of dust in high-redshift galaxies \citep{Nanni_2025}, due to their short lifetimes and the growing evidence for dust production in nearby CCSNe \citep{Medler_2025, Mera_2026}. The detection of CO in SN spectra is an important precursor for dust formation, because CO plays a crucial role in cooling SN ejecta to conducive temperatures and provides a seed for condensation. While CO and dust have been detected most frequently in Type II SN \citep[e.g.,][]{Sarangi_2018_dustIIn,Park25, Medler_2025,Mera_2026}, dust and molecules are also expected to form in SESN \citep[e.g.,][]{Liljegren_2020_Molecules}. Observationally, CO emission was detected in a few Type Ic SN \citep[e.g.][]{Rho_2021,Ravi_2023, Tinyanont_2026}. However, CO or dust signatures have not previously been detected in Ic-BL.   

\begin{figure*}
    
    \includegraphics[width=\textwidth]{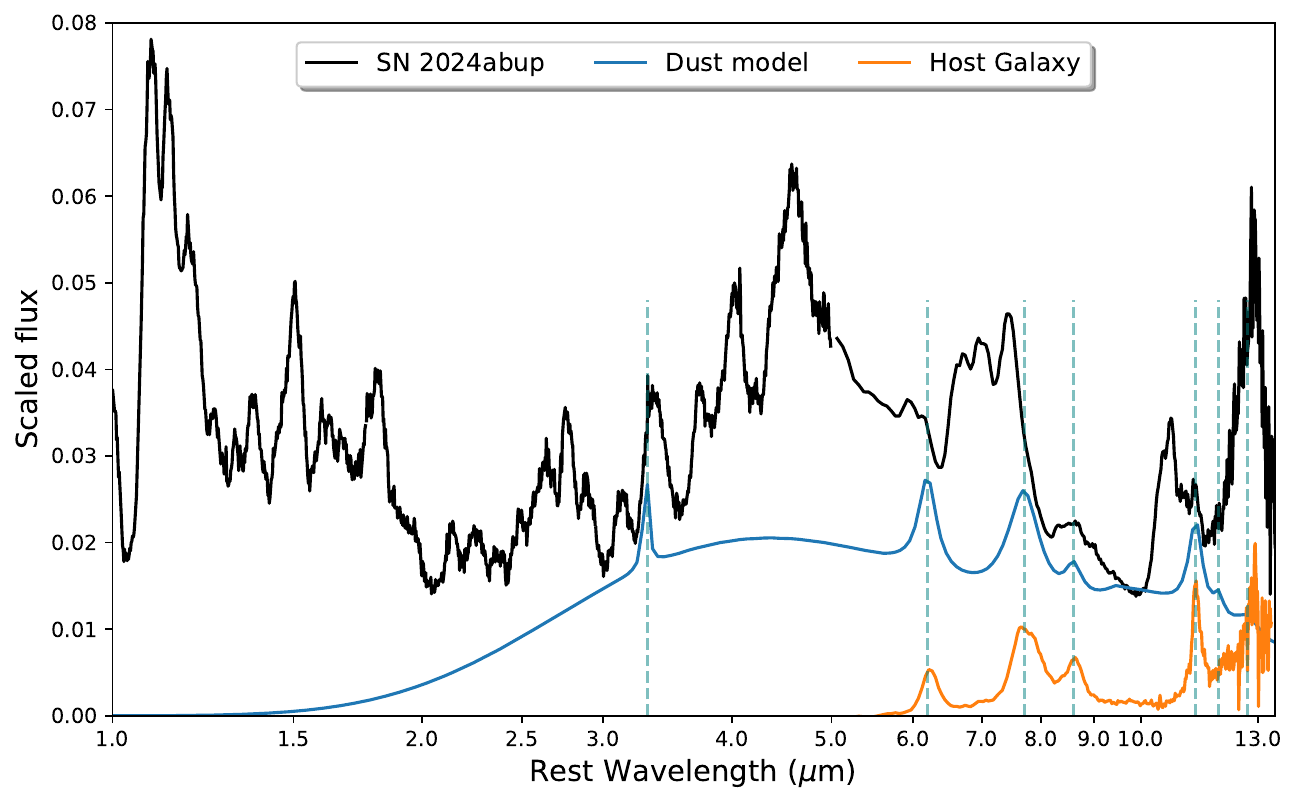}

    \caption{SN~2024abup spectrum (black) along with the model of graphite, silicate, and PAH contribution (blue) and the host galaxy (orange). The prominent PAH lines, contributed mainly by the host galaxy, are identified by dashed teal lines. There is a clear excess in the continuum of SN\,2024abup compared to the host galaxy spectrum. }
    \label{fig:spec_dust}
\end{figure*}

Given the location of SN\,2024abup in a dust lane of NGC\,0681, the dust continuum could be attributed to host galaxy dust heated by SN\,2024abup. Though we have attempted to subtract the host galaxy contribution during the data reduction, there could be some residual due to imperfect background subtraction. Alternatively (or additionally), it could be attributed to dust in the local environment produced by the previous mass loss from the progenitor or newly formed dust. The detection of CO suggests that fresh dust will form at later times; however, it is unclear how much dust may already have formed. We observe several MIR features that we attribute partially to emission from polycyclic aromatic hydrocarbons (PAHs). Comparing with the extracted host galaxy trace, we suggest that these PAH lines predominantly arise from the underlying host galaxy (\autoref{fig:spec_dust}).

The strongest CO feature observed in SN\,2024abup is the broad, fundamental ro-vibrational mode around $\mathrm{4.6\,\mu m}$. There is a prominent, broad feature in SN\,2024abup in this wavelength regime, and the \texttt{SUMO} model demonstrates a similar feature in the model, which is composed of CO emission as shown in \autoref{fig:ion_model}. 
%Following the prescription from \autoref{sec:ion}, we find that by $+$54\,days, one estimates SN\,2024abup to have produced $M \gtrsim 10^{-4}$\,$M_\odot$ of CO. However, the LTE and optically thin assumptions 
In the model, 1.9 $\times 10^{-5} M_{\odot}$ of CO is created, so that (assuming an underlying continuum in SN 2024abup around the CO feature) SN 2024abup is estimated to have produced a similar amount of CO (see Figure \ref{fig:ion_model}). \citet{Liljegren_2020_Molecules} predict that molecules like CO form in SESN between 100-600\,days post-explosion and that $10^{-4}$\,$M_\odot$ of CO is produced by $\sim$100 days. SN\,2024abup was observed at a somewhat earlier phase ($+$54\,days), so that our somewhat lower estimate can be said to be on the order of what these models predict. We note that we do not observe the CO first overtone at $\sim$2.3 $\mu m$, which is expected to be weaker in strength than the $\mathrm{4.6\,\mu m}$ feature \citep{Spyromilio_1989}. The \texttt{SUMO} model does not predict the CO feature at $\sim$2.3 $\mu m$ at this phase either. 

In addition to the CO detection, SN\,2024abup exhibits a continuum excess in the NIR and MIR (see \autoref{fig:spec_dust}) compared to the \texttt{SUMO} model as well as the continuum of SN\,2023dbc. Qualitatively, we fit the excess in continuum at wavelengths greater than 1.5 $\mathrm{\mu m}$ with carbonaceous dust (pre-existing or newly formed) heated by the SN. We use the analytical work from \citet{Henseley_2023} to model the dust continuum from a combination of graphite and silicate. For this fit, we model the dust contribution such that the continuum level matches at $\sim$10 $\mu m$. We ensured that the dust continuum did not exceed the observed spectrum at any wavelength. From this qualitative analysis, we find that silicate (graphite) dust with a mass of $\sim$0.05$M_\odot$ ($\sim$0.009 $M_\odot$) of grain size 0.1 $\mu$ at a temperature of 700\,K (300\,K) reproduces the observed continuum the best.

% \subsection{PAH} \label{sec:dust}
% We observe Polycyclic aromatic hydrocarbons (PAHs) features in the spectrum of SN~2024abup. We use an analytical model by \citet{Henseley_2023} to reproduce the PAH features seen in SN~2024abup as shown in \autoref{fig:spec_dust} in the blue line. To match the observed PAH features, we find that a mass of 0.008 $M_\odot$ at 1200K is needed.  In \autoref{fig:spec_dust}, the host galaxy spectrum is shown in an orange line. It is clear that the prominent PAH lines at wavelengths greater than 5 $\mu$m are due to the host galaxy. However, we are not able to extract the host galaxy spectrum at wavelengths where the first PAH feature is present, so we cannot definitively say if that line is from the host, the SN, or the nearby environment.

% \begin{figure*}
%     \centering

%     \includegraphics[width=\columnwidth]{figure/rprocess_velcomp.pdf} \includegraphics[width=\columnwidth]{figure/nonrprocess_velcomp.pdf}

%     \caption{}
%     \label{fig:rprocess_velcomp}
% \end{figure*}

\section{Jet constraints from Radio Observations}\label{sec:radio}
All SN associated with LGRBs have been SN\,Ic-bl; however, most SN\,Ic-bl do not have an observationally associated LGRB. Prompt gamma-ray emission from LGRBs is detected only when we are located close to the initial jets' opening angle. Thus, it could be the case that all SN Ic-bl produce LGRB, but the jet is off-axis. Radio wavelengths are favored to detect off-axis jets when the SN jet has decelerated to sub-relativistic velocity. As of now, off-axis GRBs have not been discovered in this way \citep{Corsi_2016,Corsi_2023} except, most recently, PTF10tqv, the best off-axis SN Ic-bl candidate \citep{Schroeder_2025}. 

\begin{figure}
    \centering
    \includegraphics[width=\columnwidth]{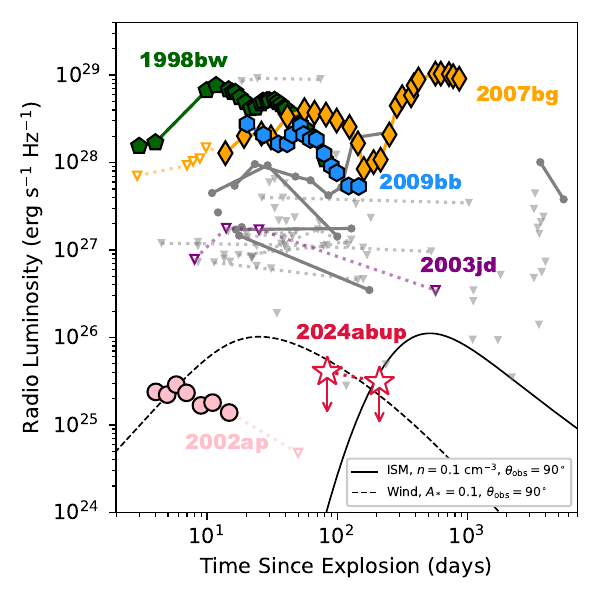}
    \caption{Radio light curve of SN\,2024abup at frequencies shown in \autoref{tab:radio}, shown alongside other well-studied SN\,Ic-bl for comparison: 1998bw \citep{Kulkarni_1998}; 2002ap \citep{Berger_2002}; 2003jd \citep{Soderberg_2006}; 2007bg \citep{Salas_2013}; 2009bb \citep{Soderberg_2010}; detections and upper limits of several other SNe are from \cite{Soderberg_2007}, \cite{Corsi_2023}, and \cite{Schroeder_2025}. The solid and dashed curves show the expected 6 GHz light curves generated using \texttt{VegasAfterglow} \citep{Vegas_afterglow} for a typical relativistic jet with energy $E_{iso}=10^{51}~\rm{erg}$, initial Lorentz factor $\Gamma_0 = 10^2$, jet opening angle $\theta_j = 0.2$, electron power-law index $p=2.3$, and microphysical parameters $\epsilon_e = 0.1$ and $\epsilon_B = 0.01$. The solid curve assumes an ISM-like CSM with density $n=0.1~\rm{cm^{-3}}$, while the dashed curve assumes a wind-like medium with $A_*=0.1$, both for a maximally off\--axis viewing angle of $\theta_{\rm obs}=90^\circ$. Our non-detections rule out these typical GRB-like jet models for any viewing angle more on-axis than $\theta_{\rm obs}=90^\circ$, implying that any relativistic jet in SN\,2024abup if present, must be either less energetic than standard GRB jets or propagating through a significantly lower-density environment.}
    \label{fig:radio-lightcurve}
\end{figure}

SN~2024abup was not associated with a LGRB, so we searched for an off-axis jet using Very Large Array (VLA) radio observations at two different epochs. We do not detect a source at the location of SN~2024abup in either epoch and report the $3\sigma$ upper limits on the flux density in \autoref{tab:radio}. We also show these limits in context with the broader SNe Ic-BL population in \autoref{fig:radio-lightcurve}. All SN Ic-bl radio detections have been attributed to the SN shock rather than an off-axis jet \citep[e.g,][]{Kulkarni_1998,Berger_2002,Soderberg_2006,Salas_2013,Soderberg_2010} except for the case of PTF10tqv \citep{Schroeder_2025}. We further compare our non-detections to \texttt{VegasAfterglow} \citep{Vegas_afterglow} models of a typical GRB-like relativistic jet viewed at a maximally off-axis angle ($\theta_{\rm obs}=90^{\circ}$). Since more on-axis viewing angles would produce brighter radio emission at all epochs, our non-detections rule out the presence of a standard jet for a wide range of viewing angles. This suggests that any jet associated with SN\,2024abup must either be intrinsically weaker than typical GRB jets or propagate through a particularly low-density circumstellar environment. The absence of detectable radio emission from both an off-axis jet and the SN shock is consistent with a progenitor that lacked significant pre-explosion mass loss.

\section{Discussions \& Conclusions} \label{sec:conclusions}
In this work, we present near and mid-IR JWST observations of the very nearby Type Ic-bl, SN~2024abup. We obtained a JWST spectrum spanning 1--14\,$\mu m$ at $+$41 days after the maximum. The spectrum contains many broad, previously unobserved features that we identify using ion contributions from a \texttt{SUMO}  spectral synthesis model \citep{Jerkstrand_2011_SUMOa}. The IR spectrum of  SN\,2024abup is well modeled with a massive star model with $M_{\text{He,i}}=$\,12.0\,$M_{\odot}$ ($M_{\mathrm{ej}}$) without any $r$-process elements in the ejecta. We find that most of the observed features are also present in the model without inclusion of $r$-process elements \citep{Ertl_2020_ejectamodels}, although the model does not yet match the detailed shapes and strengths of individual lines.

We find the photometric evolution of SN~2024abup to be similar to that of other SN Ic-bl. %The peak in the $V$ band is reached $\sim$ 14 days after the explosion with a peak magnitude of $M_V$=$-$18.75, similar to other SN Ic-bl without an associated LGRB. We calculate the explosion properties of SN~2024abup such as $M_{\rm ej}$, $E_{\rm ej}$, $M_{\rm Ni}$ and \ion{Fe}{2} velocity. 
The values for various explosion properties we obtain for SN~2024abup fall in a normal range compared to a large population of SN Ic-bl. We find the host galaxy of SN~2024abup to be on the more massive side than the general SN Ic-bl population; however, the SFR of the galaxy is normal.

Additionally, we search for $r$-process lines in the SN~2024abup spectrum based on theoretical models by \citet{Ricigliano_2025}. We find some broad features overlapping $r$-process lines that are predicted by the model, as shown in \autoref{fig:only_rprocess}. We further constrain the $r$-process ionic mass in SN~2024abup by considering that the observed features are consistent with the lines predicted by \citet{Ricigliano_2025}. We find the JWST spectrum can probe this mass down to $\sim$$10^{-4} M_{\odot}$. However, the \texttt{SUMO} model is able to reproduce most of these lines without $r$-process elements. Since the exact strength and shape of features produced by the model do not match the observed spectrum, there is a possibility of blending of lines with $r$-process ions.  This work shows that we need more observations of SN Ic-bl during the nebular phase and more models, including $r$-process ions, to separate the blending of lines, to answer definitively whether SN Ic-bl produces $r$-process elements or not. %Additionally, a stronger signature of $r$-process elements is expected during the nebular phase \citep{Ricigliano_2025}.

To put the observed behavior of SN~2024abup in context with other transients, we compare NIR spectra of SN~2024abup with three other transients: SN~1998bw, AT~2017gfo, and GRB~230307A. %We find similar broad features in both SN~2024abup and SN~1998bw; however, AT~2017gfo and GRB~230307A only have one broad feature at $\sim$ 2.1 $\mu m$. 
The broad feature at 2.1 $\mathrm{\mu m}$ has been attributed to the $r$-process line of [\ion{Te}{3}] in the case of AT~2017gfo and GRB~230307A. We see a feature at the same wavelength for SN~2024abup; however, the width of the feature is much smaller than in the case of AT~2017gfo and could be due to \ion{Mg}{2}. This is a more favorable case for SN~2024abup because Mg lines are prominent at other wavelengths as well. We do note that in SN~2024abup, there is a possibility of blending of lines produced by both [\ion{Te}{3}] and \ion{Mg}{2}. Nonetheless, we do not see a clear sign of $r$-process elements in the NIR spectrum of SN~2024abup.

We observe a broad feature at $\sim$4.5 $\mathrm{\mu m}$ which is produced by \ion{CO}{1} as shown by a \texttt{SUMO} model. This is the first-ever detection of CO in any SN Ic-bl and one of the earliest for any CCSN. Since CO is a coolant that can effectively decrease the temperature of SN ejecta to be ideal for dust formation, the detection of this ion signifies that dust formation can commence. We estimate a CO mass of $\sim 2 \times 10^{-5}\, \mathrm{M_{\odot}}$ from the observed spectrum. Additionally, we observe some excess in the continuum in the MIR. This excess could be explained as dust heated by the SN, and is not likely due to host galaxy contamination. We find a combination of silicate and graphite dust of mass 0.05 and 0.009 $\mathrm{M_\odot}$ heated to temperatures of 300 K and 700 K, respectively, reproduces the observed excess in the continuum. %This dust could be pre-existing dust or newly formed dust. 
We find a hint of dust formation in SN Ic-bl for the first time, and it could play an important role in producing dust in the early universe. 

We also present a comparison of the JWST spectra of SN 2023dbc (Type Ic/Ib) and SN 2024abup obtained at similar epochs, finding many similarities between the two SNe. However, there are strong \ion{He}{1} lines in SN~2023dbc, which are absent in SN~2024abup. This could indicate that the progenitor of SN~2024abup is more stripped of its He envelope than the progenitor of SN~2024abup. Additionally, we detect [\ion{Ar}{2}] and [\ion{Ar}{3}] lines in SN~2024abup, which are much weaker in SN~2023dbc. This could indicate that SN~2024abup is at a higher ionization state compared to SN~2023dbc. The CO feature seen in SN~2024abup is much more prominent than the feature seen in SN~2023dbc. Hence, there could be a small amount of CO in SN~2023dbc. These observed differences point to different levels of stripping and possibly dust production in the SN Ic-bl and SN Ic classes, which future observations should verify. 

 %Most of the SNe associated with LGRBs have been classified as SN Ic-bl. However, there is a large fraction of SNe Ic-BL ($\gtrsim$81\%) without an associated LGRB detection \citep{Corsi_2023}. It has been proposed that all SNe Ic-BL launch jets, but they are not pointed at our line of sight; thus, we do not detect the prompt emission. 
 We search for an off-axis jet using radio observations with the VLA. We do not detect radio emission coincident with SN~2024abup and obtain a deep limit in our VLA radio observations. %Many SN Ic-bl with and without LGRBs have been observed in the radio, as shown in \autoref{fig:radio-lightcurve}. However, all the cases where radio emission was detected were better explained by a shock due to interaction with a dense medium instead of an off-axis jet. 
Here, non-detection in the radio for SN~2024abup could mean that the jet is not launched or pointed at a viewing angle not visible even in radio frequencies. Additionally, it could mean that there is no dense medium near the explosion.

This work shows the power of one JWST SN Ic-bl spectrum in furthering our understanding of chemical enrichment and dust formation from this class of transients. We do not detect any clear sign of $r$-process element production in SN~2024abup. Future observations in the IR during nebular epochs could provide a clearer answer to the question if collapsars could produce $r$-process elements. However, it is clear that SN Ic-bl could produce dust, and more work needs to be done to constrain the amount of dust produced by these transients.

%\section{Software and third party data repository citations} \label{sec:cite}

%\url{https://github.com/AASJournals/Tutorials/tree/master/Repositories}.

%% Also note that the akcnowlodgment environment does not support long amounts of text. If you have a lot of people and institutions to acknowledge, do not use this command. Instead, create a new \section{Acknowledgments}.
\begin{acknowledgments}
The work presented here is based on observations made with the NASA/ESA/CSA JWST as part of program \#06803. We thank William Januszewski for the help with observation scheduling. The data were obtained from the Mikulski Archive for Space Telescopes at the Space Telescope Science Institute (STScI), which is operated by the Association of Universities for Research in Astronomy (AURA), Inc., under National Aeronautics and Space Administration (NASA) contract NAS 5-03127 for JWST. Support for this program at the University of Arizona was provided by NASA through grants JWST-GO-06803.003.

MS acknowledges funding from the Australian Research Council (ARC) Centre of Excellence CE230100016.
Time-domain research by the University of Arizona
team and D.J.S. is supported by National Science Foundation
(NSF) grants 2308181, 2407566, and 2432036. The research by Y.D., S.V., N.M., and E.H. is supported by NSF grant AST-2008108. 

This work makes use of data from the Las Cumbres Observatory global telescope network.  The LCO team is supported by NSF grants AST-1911225 and AST-1911151.

K.A.B. is supported by an LSSTC Catalyst Fellowship; this publication was thus made possible through the support of Grant 62192 from the John Templeton Foundation to LSSTC. The opinions expressed in this publication are those of the authors and do not necessarily reflect the views of LSSTC or the John Templeton Foundation.  

M.M. acknowledges support in part from ADAP program grant No. 80NSSC22K0486, from the NSF grant AST-2206657 and from the National Science Foundation under Cooperative Agreement 2421782 and the Simons Foundation grant MPS-AI-00010515 awarded to the NSF-Simons AI Institute for Cosmic Origins (CosmicAI), \url{https://www.cosmicai.org/}.

A.J. acknowledges funding by the Swedish Research Council (Grant 2018-03799). The \texttt{SUMO} computations in this work
were enabled by resources provided by the Swedish National
Infrastructure for Computing (SNIC), the National Academic Infrastructure for Supercomputing in Sweden (NAISS), and at the Parallelldatorcentrum (PDC) Center for High Performance Computing, Royal Institute of Technology (KTH), partially funded by the Swedish Research Council through grant agreements nos 2022-
06725 and 2018-05973.

This research has made use of the NASA Astrophysics Data System (ADS) Bibliographic Services, and the NASA/IPAC Infrared Science Archive (IRSA), which is funded by the National Aeronautics and Space Administration and operated by the California Institute of Technology.   This work made use of data supplied by the UK Swift Science Data Centre at the University of Leicester. BM acknowledges support from the ARC through Discovery Projects DP240101786 and DP260104967. S.W.J. gratefully acknowledges support from a Guggenheim Fellowship.

\end{acknowledgments}


\vspace{5mm}
\facilities{JWST(NIRSpec+MIRI), LCOGT, Swift (UVOT), VLA}


\software{Astropy \citep{astropy:2013,astropy:2018, astropy:2022}, Photutils \citep{Bradley_2019}, Panacea, BANZAI \citep{Banzai}, Light Curve Fitting \citep{lightcurvefitting}, Matplotlib \citep{mpl}, Numpy \citep{numpy}, Scipy \citep{scipy}, IRAF \citep{iraf1,iraf2}, \texttt{lcogtsnpipe} \citep{Valenti_2016}
          }



\appendix
\counterwithin{figure}{section}
\section{Corner plot}
\begin{figure*}
    \centering
    \includegraphics[width=\textwidth]{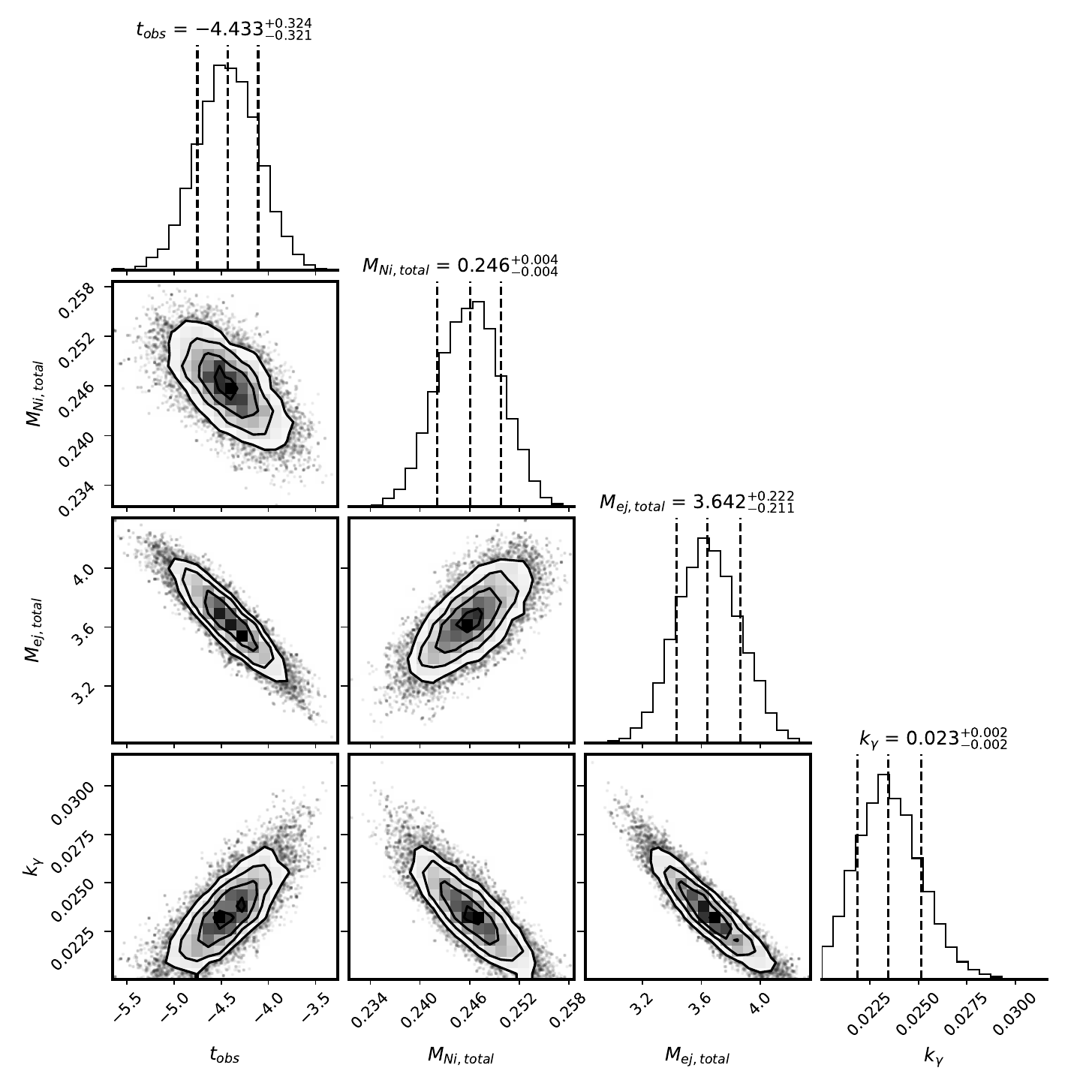}  
    \caption{The corner plot for the fit using analytic models from \citet{Arnett_1982} and \citet{Valenti_2008}.}
    \label{fig:corner}
\end{figure*}

\section{$r$-process lines}
\begin{figure*}
    \centering
    \includegraphics[width=\textwidth]{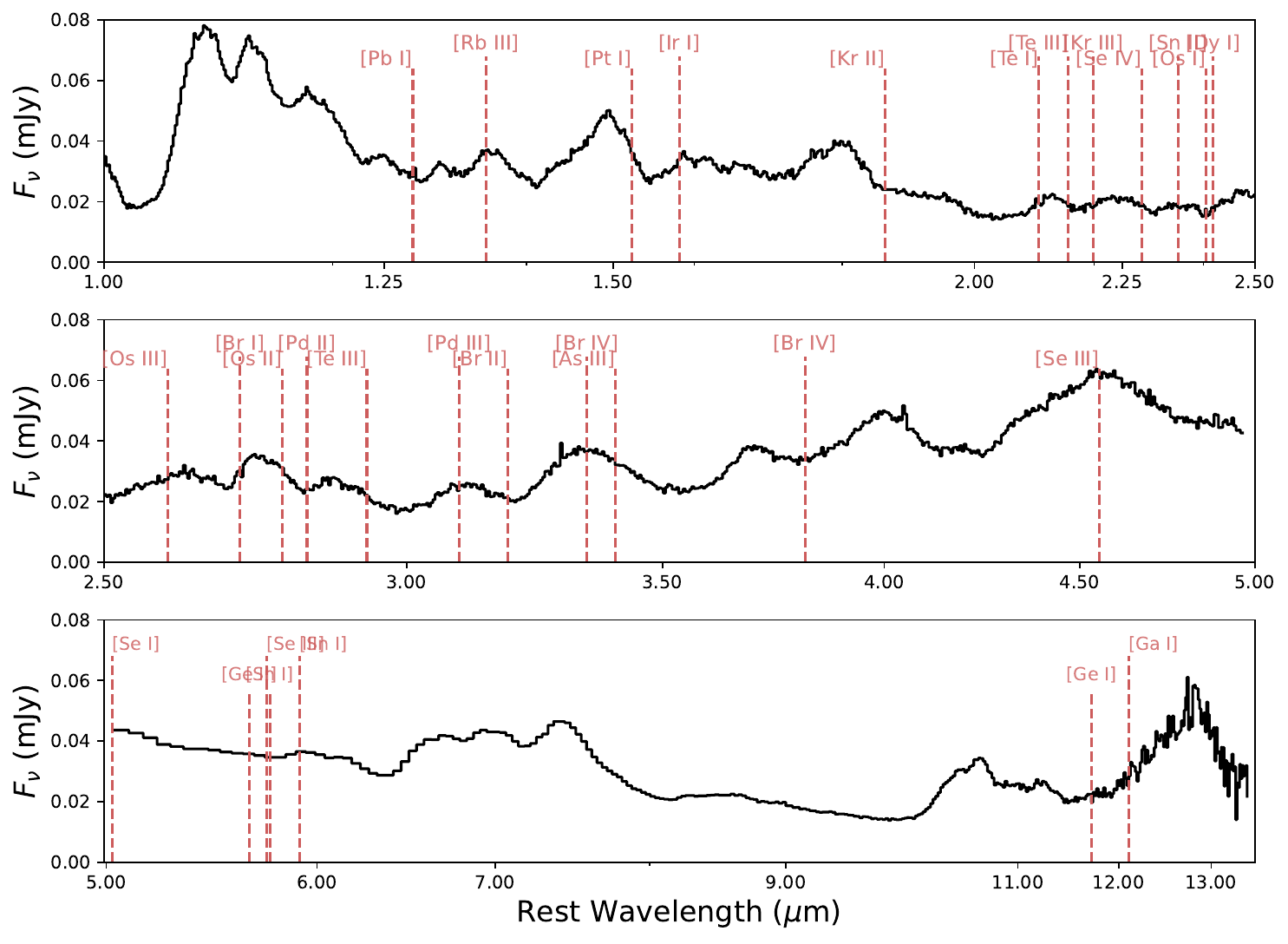}  
    \caption{The SN~2024abup JWST spectrum overlayed with all the prominent $r$-process lines identified and presented in Table 3 in \cite{Ricigliano_2025} (red dashed lines).}
    \label{fig:all_rprocess}
\end{figure*}

\autoref{fig:all_rprocess} shows the JWST spectrum of SN~2024abup covering the wavelength range from 1 to 6.5 $\mathrm{\mu m}$ with all the prominent $r$-process lines predicted by \citet{Ricigliano_2025} as shown in their table 3. All of the lines are forbidden lines from heavy elements.

\section{Radio observations}
In \autoref{tab:radio}, we present the summary of VLA observations performed for SN~2024abup in two different epochs. We present the upper limit for our non-detection in different frequency ranges. This deep limit, along with other well-studied SNe Ic-BL in the radio, are presented in \autoref{fig:radio-lightcurve}. We see that if the radio emission from SN~2024abup was similar to other SN Ic-bl, then we should have detected it.
\begin{table*}
\centering
 \caption{Summary of VLA Observations. Non-detections are reported as
3$\sigma$ upper limits.}
 \begin{tabular}{ c c c c c c}
    \hline
    Date (UTC) & MJD &  Configuration & Phase (d) & $\nu$ (GHz)  &  $F_\nu$ (mJy) \\
    \hline 
    2025-02-14 & 60720 & A$\rightarrow$D & 84 & 1.52 & $<$0.16 \\ 
    2025-02-14 & 60720 & A$\rightarrow$D & 84 & 3.0 & $<$0.10 \\
    2025-02-14 & 60720 & A$\rightarrow$D & 84 & 6.0 & $<$0.06 \\
    2025-02-14 & 60720 & A$\rightarrow$D & 84 & 10.0 & $<$0.06 \\ [5pt]
    2025-06-21 & 60848 & C & 211 & 1.52 & $<$0.22 \\
    2025-06-21 & 60848 & C & 211 & 3.0 & $<$0.11 \\
    2025-06-21 & 60848 & C & 211 & 6.0 & $<$0.05 \\
    2025-06-21 & 60848 & C & 211 & 10.0 & $<$0.05 \\[5pt]
    \hline
 \end{tabular}
 
 \label{tab:radio}
\end{table*}

\section{\texttt{SUMO} model}\label{appen:sumo}

In this appendix, we will provide a more detailed description of the \texttt{SUMO} spectral modeling effort presented in Section \ref{subsec:spectral_modeling}. As mentioned there, we selected the $M_{\text{He, i}}$ = 12.0 $M_{\odot}$ model from \citet{Woosley_2019_ejectamodels} and \citet{Ertl_2020_ejectamodels} as our base ejecta model. This model was cut up into different compositional zones following the methodology described in \citet{Barmentloo_2024_NIIpaper}. These zones are dubbed the Fe/He, Si/S, O/Si/S, O/Ne/Mg, O/C, and C/O zones. The O/C + C/O split (not present in \citealt{Barmentloo_2024_NIIpaper}) was not only a better empirical description of the compositional profiles (see \autoref{fig:ion_model}), it was also found to be essential for proper (CO) molecule formation in \texttt{SUMO}, as a single O/C zone would be too polluted by He$^{+}$, causing too efficient destruction of CO molecules \citep{Liu_1992_Molecules, Liljegren_2020_Molecules}. To obtain a model completely free of a helium envelope, the C/O zone was cut off as soon as the mass abundance of helium surpassed 0.2 (removing the $\sim$ 2.5\% fastest moving ejecta). Additionally, the $^{56}$Ni mass of the model was increased to 0.0791 $M_{\odot}$ (following \citealt{Dessart_2021_SESNspectra}; see their table A.2) by simply increasing the mass of $^{56}$Ni in the Fe/He zone by the required amount, adjusting all other mass abundances so that all other elemental masses are conserved.
% \begin{figure*}
%     \centering
%     \includegraphics[width=\textwidth]{figure/SN2024abup_comp_model.pdf}

%     \caption{SN~2024abup compared to model.}
%     \label{fig:spec_jwst_comp_model}
% \end{figure*}

\begin{table}[]
\centering
    \begin{tabular}{ll}
        
        \hline Parameter & Value \\
        \hline $E_{\text{kin}}$ & 0.81 B \\
        $M_{\text{preSN}}$ & 7.24 $M_{\odot}$\\
        $M_{\text{ej}}$  & 5.18 $M_{\odot}$\\
        $M_{\text{$^{56}$Ni}}$ & 0.0791 $M_{\odot}$ \\
        $V_{\text{core}}$ & 8 547 km s$^{-1}$ \\
        \hline
    \end{tabular}
    
    \caption{Summary table of the properties of the ejecta model used to simulate the \texttt{SUMO} model spectrum presented in \autoref{fig:ion_model}. This ejecta model is an adoptation of the $M_{\text{He, i}}$ = 12.0 $M_{\odot}$ model from \citet{Woosley_2019_ejectamodels} and \citet{Ertl_2020_ejectamodels} (see text)}
    \label{tab:summary_table_SUMO_model}
\end{table}

With the compositional zones defined, we now macroscopically mix the ejecta, randomly distributing 1000 `clumps' per compositional zone within a sphere at radii ranging from $R = 0$ to $R = V_{\text{core}}t$.\footnote{Here, $V_{\text{core}}$ is the outer velocity of the C/O zone (see Table \ref{tab:summary_table_SUMO_model}).} Following earlier works with \texttt{SUMO} \citep[e.g.,][]{Jerkstrand_2015_IIb}, we assume the Fe/He and Si/S zones to expand due to radioactive heating \citep{Herant_1991_Nibubble}. As in \citet{Jerkstrand_2015_IIb}, we assume an expansion such that $\rho_{\text{FeHE}}$ = $\rho_{\text{SiS}}$/10 = $\rho_{\text{other zones}}$/30.

\bibliography{icbl}{}

@ARTICLE{Park25,
       author = {{Park}, Seong Hyun and {Rho}, Jeonghee and {Yoon}, Sung-Chul and {Pearson}, Jeniveve and {Shrestha}, Manisha and {Tinyanont}, Samaporn and {Geballe}, T.~R. and {Foley}, Ryan J. and {Ravi}, Aravind P. and {Andrews}, Jennifer and {Sand}, David J. and {Azalee Bostroem}, K. and {Ashall}, Chris and {Hoeflich}, Peter and {Valenti}, Stefano and {Dong}, Yize and {Retamal}, Nicolas Meza and {Hoang}, Emily and {Mehta}, Darshana and {Andrew Howell}, D. and {Farah}, Joseph R. and {Terreran}, Giacomo and {Padilla Gonzalez}, Estefania and {Andrews}, Moira and {Newsome}, Megan and {Shahbandeh}, Melissa and {Smith}, Nathan and {Hwan Kang}, Jae and {Suntzeff}, Nick and {Baron}, Eddie and {Medler}, Kyle and {Mera Evans}, Tyco and {DerKacy}, James M. and {Larison}, Conor and {Galbany}, Llu{\'\i}s and {Jacobson-Gal{\'a}n}, Wynn},
        title = "{Near-infrared spectroscopy and detection of carbon monoxide in the Type II supernova SN 2023ixf}",
      journal = {\aap},
         year = 2025,
        month = nov,
       volume = {703},
          eid = {A227},
        pages = {A227},
          doi = {10.1051/0004-6361/202555244},
archivePrefix = {arXiv},
       eprint = {2507.11877},
 primaryClass = {astro-ph.HE},
       adsurl = {https://ui.adsabs.harvard.edu/abs/2025A&A...703A.227P}
}

@MISC{Shrestha2024,
       author = {{Shrestha}, Manisha and {Alexander}, Kate Denham and {Andrews}, Jennifer and {Bostroem}, Kyra Azalee and {Christy}, Collin and {Dong}, Yize and {Fields}, Carl Edward and {Hoang}, Emily and {Hosseinzadeh}, Griffin and {Janzen}, Daryl and {Jha}, Saurabh W. and {Kwok}, Lindsey and {Lundquist}, Michael Jon and {Mehta}, Darshana and {Meza Retamal}, Nicolas Eduardo and {Pazhayath Ravi}, Aravind and {Pearson}, Jeniveve and {Renzo}, Mathieu and {Sand}, David J. and {Subrayan}, Bhagya and {Valenti}, Stefano},
        title = "{NIR+MIR Spectroscopy of the Nearby Broad Line Type Ic SN 2024abup: r-process, Dust and Explosion Physics}",
 howpublished = {JWST Proposal. Cycle 3, ID. \#6803},
         year = 2024,
        month = dec,
        pages = {6803},
       adsurl = {https://ui.adsabs.harvard.edu/abs/2024jwst.prop.6803S}
}

@ARTICLE{Rho_2021,
       author = {{Rho}, J. and {Evans}, A. and {Geballe}, T.~R. and {Banerjee}, D.~P.~K. and {Hoeflich}, P. and {Shahbandeh}, M. and {Valenti}, S. and {Yoon}, S.-C. and {Jin}, H. and {Williamson}, M. and {Modjaz}, M. and {Hiramatsu}, D. and {Howell}, D.~A. and {Pellegrino}, C. and {Vink{\'o}}, J. and {Cartier}, R. and {Burke}, J. and {McCully}, C. and {An}, H. and {Cha}, H. and {Pritchard}, T. and {Wang}, X. and {Andrews}, J. and {Galbany}, L. and {Van Dyk}, S. and {Graham}, M.~L. and {Blinnikov}, S. and {Joshi}, V. and {P{\'a}l}, A. and {Kriskovics}, L. and {Ordasi}, A. and {Szakats}, R. and {Vida}, K. and {Chen}, Z. and {Li}, X. and {Zhang}, J. and {Yan}, S.},
        title = "{Near-infrared and Optical Observations of Type Ic SN 2020oi and Broad-lined Type Ic SN 2020bvc: Carbon Monoxide, Dust, and High-velocity Supernova Ejecta}",
      journal = {\apj},
         year = 2021,
        month = feb,
       volume = {908},
       number = {2},
          eid = {232},
        pages = {232},
          doi = {10.3847/1538-4357/abd850},
archivePrefix = {arXiv},
       eprint = {2010.00662},
 primaryClass = {astro-ph.SR},
       adsurl = {https://ui.adsabs.harvard.edu/abs/2021ApJ...908..232R}
}

@ARTICLE{Mera_2026,
       author = {{Mera}, T. and {Ashall}, C. and {Hoeflich}, P. and {Medler}, K. and {Shahbandeh}, M. and {Burns}, C.~R. and {Baron}, E. and {DerKacy}, J.~M. and {Morrell}, N. and {Lu}, J. and {Hinkle}, J.~T. and {Mazzali}, P.~A. and {Fereidouni}, E. and {Pfeffer}, C.~M. and {Shiber}, S. and {Temim}, T. and {Galbany}, L. and {Coulter}, D.~A. and {Ferrari}, L. and {Hoogendam}, W.~B. and {Hsiao}, E.~Y. and {Phillips}, M.~M. and {Shappee}, B.~J.},
        title = "{JWST Observations of SN 2024ggi. II. NIRSpec Spectroscopy and CO Modeling at +285─385 Days past the Explosion}",
      journal = {\apj},
         year = 2026,
        month = feb,
       volume = {997},
       number = {2},
          eid = {330},
        pages = {330},
          doi = {10.3847/1538-4357/ae317e},
archivePrefix = {arXiv},
       eprint = {2510.09600},
 primaryClass = {astro-ph.SR},
       adsurl = {https://ui.adsabs.harvard.edu/abs/2026ApJ...997..330M}
}

@INPROCEEDINGS{Spyromilio_1989,
       author = {{Spyromilio}, J. and {Meikle}, W.~P.~S. and {Learner}, R.~C.~M. and {Allen}, D.~A.},
        title = "{Carbon monoxide in Supernova 1987a}",
    booktitle = {Infrared Spectroscopy in Astronomy},
         year = 1989,
       editor = {{B{\"o}hm-Vitense}, Erika},
        month = sep,
        pages = {381},
       adsurl = {https://ui.adsabs.harvard.edu/abs/1989ESASP.290..381S}
}

@ARTICLE{Yamanaka_2026,
       author = {{Yamanaka}, Masayuki and {Nagayama}, Takahiro and {Kumano}, Akari and {Sahu}, Devendra Kumar and {Singh}, Avinash and {Das}, Hrishav and {Anupama}, G.~C.},
        title = "{SN 2023dbc in M108: Optical and Near-Infrared Observations of a Highly-Obscured, Moderately Energetic Stripped-Envelope Supernova}",
      journal = {arXiv e-prints},
         year = 2026,
        month = may,
          eid = {arXiv:2605.16916},
        pages = {arXiv:2605.16916},
archivePrefix = {arXiv},
       eprint = {2605.16916},
 primaryClass = {astro-ph.HE},
       adsurl = {https://ui.adsabs.harvard.edu/abs/2026arXiv260516916Y}
}

@ARTICLE{Miller_2020,
       author = {{Miller}, Jonah M. and {Sprouse}, Trevor M. and {Fryer}, Christopher L. and {Ryan}, Benjamin R. and {Dolence}, Joshua C. and {Mumpower}, Matthew R. and {Surman}, Rebecca},
        title = "{Full Transport General Relativistic Radiation Magnetohydrodynamics for Nucleosynthesis in Collapsars}",
      journal = {\apj},
         year = 2020,
        month = oct,
       volume = {902},
       number = {1},
          eid = {66},
        pages = {66},
          doi = {10.3847/1538-4357/abb4e3},
archivePrefix = {arXiv},
       eprint = {1912.03378},
 primaryClass = {astro-ph.HE},
       adsurl = {https://ui.adsabs.harvard.edu/abs/2020ApJ...902...66M}
}

@ARTICLE{Medler_2025,
       author = {{Medler}, K. and {Ashall}, C. and {Hoeflich}, P. and {Baron}, E. and {DerKacy}, J.~M. and {Shahbandeh}, M. and {Mera}, T. and {Pfeffer}, C.~M. and {Hoogendam}, W.~B. and {Jones}, D.~O. and {Shiber}, S. and {Fereidouni}, E. and {Fox}, O.~D. and {Jencson}, J. and {Galbany}, L. and {Hinkle}, J.~T. and {Tucker}, M.~A. and {Shappee}, B.~J. and {Huber}, M.~E. and {Auchettl}, K. and {Angus}, C.~R. and {Desai}, D.~D. and {Do}, A. and {Payne}, A.~V. and {Shi}, J. and {Kong}, M.~Y. and {Romagnoli}, S. and {Syncatto}, A. and {Burns}, C.~R. and {Clayton}, G. and {Dulude}, M. and {Engesser}, M. and {Filippenko}, A.~V. and {Gomez}, S. and {Hsiao}, E.~Y. and {de Jaeger}, T. and {Johansson}, J. and {Krisciunas}, K. and {Kumar}, S. and {Lu}, J. and {Matsuura}, M. and {Mazzali}, P.~A. and {Milisavljevic}, D. and {Morrell}, N. and {O'Steen}, R. and {Park}, S. and {Phillips}, M.~M. and {Ravi}, A.~P. and {Rest}, A. and {Rho}, J. and {Suntzeff}, N.~B. and {Sarangi}, A. and {Smith}, N. and {Stritzinger}, M.~D. and {Strolger}, L. and {Szalai}, T. and {Temim}, T. and {Tinyanont}, S. and {Van Dyk}, S.~D. and {Wang}, L. and {Wang}, Q. and {Wesson}, R. and {Yang}, Y. and {Zs{\'\i}ros}, S.},
        title = "{JWST Observations of SN 2023ixf. II. The Panchromatic Evolution between 250 and 720 Days after the Explosion}",
      journal = {\apj},
         year = 2025,
        month = nov,
       volume = {993},
       number = {2},
          eid = {191},
        pages = {191},
          doi = {10.3847/1538-4357/ae0736},
archivePrefix = {arXiv},
       eprint = {2507.19727},
 primaryClass = {astro-ph.SR},
       adsurl = {https://ui.adsabs.harvard.edu/abs/2025ApJ...993..191M}
}

@ARTICLE{Nanni_2025,
       author = {{Nanni}, Ambra and {Romano}, Michael and {Donevski}, Darko and {Witstok}, Joris and {Shivaei}, Irene and {Fioc}, Michel and {Sawant}, Prasad},
        title = "{Origins of Carbon Dust in a JWST-observed Primeval Galaxy at z {\ensuremath{\sim}} 6.7}",
      journal = {\apjl},
         year = 2025,
        month = jul,
       volume = {988},
       number = {1},
          eid = {L5},
        pages = {L5},
          doi = {10.3847/2041-8213/ade2e5},
archivePrefix = {arXiv},
       eprint = {2505.10701},
 primaryClass = {astro-ph.GA},
       adsurl = {https://ui.adsabs.harvard.edu/abs/2025ApJ...988L...5N}
}

@ARTICLE{Sarangi_2018_dustIIn,
       author = {{Sarangi}, Arkaprabha and {Dwek}, Eli and {Arendt}, Richard G.},
        title = "{Delayed Shock-induced Dust Formation in the Dense Circumstellar Shell Surrounding the Type IIn Supernova SN 2010jl}",
      journal = {\apj},
         year = 2018,
        month = may,
       volume = {859},
       number = {1},
          eid = {66},
        pages = {66},
          doi = {10.3847/1538-4357/aabfc3},
archivePrefix = {arXiv},
       eprint = {1804.06878},
 primaryClass = {astro-ph.SR},
       adsurl = {https://ui.adsabs.harvard.edu/abs/2018ApJ...859...66S}
}

@ARTICLE{Ventura_2014,
       author = {{Ventura}, P. and {Dell'Agli}, F. and {Schneider}, R. and {Di Criscienzo}, M. and {Rossi}, C. and {La Franca}, F. and {Gallerani}, S. and {Valiante}, R.},
        title = "{Dust from asymptotic giant branch stars: relevant factors and modelling uncertainties}",
      journal = {\mnras},
         year = 2014,
        month = mar,
       volume = {439},
       number = {1},
        pages = {977-989},
          doi = {10.1093/mnras/stu028},
archivePrefix = {arXiv},
       eprint = {1401.1332},
 primaryClass = {astro-ph.SR},
       adsurl = {https://ui.adsabs.harvard.edu/abs/2014MNRAS.439..977V}
}

@ARTICLE{Dellagli_2018,
       author = {{Dell'Agli}, F. and {Di Criscienzo}, M. and {Ventura}, P. and {Limongi}, M. and {Garc{\'\i}a-Hern{\'a}ndez}, D.~A. and {Marini}, E. and {Rossi}, C.},
        title = "{Evolved stars in the Local Group galaxies - II. AGB, RSG stars, and dust production in IC10}",
      journal = {\mnras},
         year = 2018,
        month = oct,
       volume = {479},
       number = {4},
        pages = {5035-5048},
          doi = {10.1093/mnras/sty1614},
archivePrefix = {arXiv},
       eprint = {1806.04160},
 primaryClass = {astro-ph.SR},
       adsurl = {https://ui.adsabs.harvard.edu/abs/2018MNRAS.479.5035D}
}

@INPROCEEDINGS{Gehrz_1989,
       author = {{Gehrz}, R.},
        title = "{Sources of Stardust in the Galaxy}",
    booktitle = {Interstellar Dust},
         year = 1989,
       editor = {{Allamandola}, Louis J. and {Tielens}, A.~G.~G.~M.},
       series = {IAU Symposium},
       volume = {135},
        month = jan,
        pages = {445},
       adsurl = {https://ui.adsabs.harvard.edu/abs/1989IAUS..135..445G}
}

@ARTICLE{Siebert_2024,
       author = {{Siebert}, M.~R. and {DeCoursey}, C. and {Coulter}, D.~A. and {Engesser}, M. and {Pierel}, J.~D.~R. and {Rest}, A. and {Egami}, E. and {Shahbandeh}, M. and {Chen}, W. and {Fox}, O.~D. and {Zenati}, Y. and {Moriya}, T.~J. and {Bunker}, A.~J. and {Cargile}, P.~A. and {Curti}, M. and {Eisenstein}, D.~J. and {Gezari}, S. and {Gomez}, S. and {Guolo}, M. and {Johnson}, B.~D. and {Joshi}, B.~A. and {Karmen}, M. and {Maiolino}, R. and {Quimby}, R.~M. and {Robertson}, B. and {Strolger}, L.~G. and {Sun}, F. and {Wang}, Q. and {Wevers}, T.},
        title = "{Discovery of a Relativistic Stripped-envelope Type Ic-BL Supernova at z = 2.83 with JWST}",
      journal = {\apjl},
         year = 2024,
        month = sep,
       volume = {972},
       number = {1},
          eid = {L13},
        pages = {L13},
          doi = {10.3847/2041-8213/ad6c32},
archivePrefix = {arXiv},
       eprint = {2406.05076},
 primaryClass = {astro-ph.HE},
       adsurl = {https://ui.adsabs.harvard.edu/abs/2024ApJ...972L..13S}
}

@ARTICLE{Woosley_2002,
       author = {{Woosley}, S.~E. and {Heger}, A. and {Weaver}, T.~A.},
        title = "{The evolution and explosion of massive stars}",
      journal = {Reviews of Modern Physics},
         year = 2002,
        month = nov,
       volume = {74},
       number = {4},
        pages = {1015-1071},
          doi = {10.1103/RevModPhys.74.1015},
       adsurl = {https://ui.adsabs.harvard.edu/abs/2002RvMP...74.1015W}
}

@ARTICLE{Corsi_2016,
       author = {{Corsi}, A. and {Gal-Yam}, A. and {Kulkarni}, S.~R. and {Frail}, D.~A. and {Mazzali}, P.~A. and {Cenko}, S.~B. and {Kasliwal}, M.~M. and {Cao}, Y. and {Horesh}, A. and {Palliyaguru}, N. and {Perley}, D.~A. and {Laher}, R.~R. and {Taddia}, F. and {Leloudas}, G. and {Maguire}, K. and {Nugent}, P.~E. and {Sollerman}, J. and {Sullivan}, M.},
        title = "{Radio Observations of a Sample of Broad-line Type IC Supernovae Discovered by PTF/IPTF: A Search for Relativistic Explosions}",
      journal = {\apj},
         year = 2016,
        month = oct,
       volume = {830},
       number = {1},
          eid = {42},
        pages = {42},
          doi = {10.3847/0004-637X/830/1/42},
archivePrefix = {arXiv},
       eprint = {1512.01303},
 primaryClass = {astro-ph.HE},
       adsurl = {https://ui.adsabs.harvard.edu/abs/2016ApJ...830...42C}
}

@ARTICLE{Arunachalam_2025,
       author = {{Arunachalam}, Prasiddha and {Macias}, Phillip and {Foley}, Ryan. J.},
        title = "{GRB 230307A Formed No Dust or Was Not a Binary Neutron Star Merger}",
      journal = {arXiv e-prints},
         year = 2025,
        month = oct,
          eid = {arXiv:2510.16121},
        pages = {arXiv:2510.16121},
          doi = {10.48550/arXiv.2510.16121},
archivePrefix = {arXiv},
       eprint = {2510.16121},
 primaryClass = {astro-ph.HE},
       adsurl = {https://ui.adsabs.harvard.edu/abs/2025arXiv251016121A}
}

@ARTICLE{Rastinejad_2025_25kg,
       author = {{Rastinejad}, Jillian C. and {Levan}, Andrew J. and {Jonker}, Peter G. and {Kilpatrick}, Charles D. and {Fryer}, Christopher L. and {Sarin}, Nikhil and {Gompertz}, Benjamin P. and {Liu}, Chang and {Eyles-Ferris}, Rob A.~J. and {Fong}, Wen-fai and {Burns}, Eric and {Gillanders}, James H. and {Mandel}, Ilya and {Malesani}, Daniele Bj{\o}rn and {O'Brien}, Paul T. and {Tanvir}, Nial R. and {Ackley}, Kendall and {Aryan}, Amar and {Bauer}, Franz E. and {Bloemen}, Steven and {de Boer}, Thomas and {Bom}, Cl{\'e}cio R. and {Chac{\'o}n}, Jennifer A. and {Chambers}, Ken and {Chen}, Ting-Wan and {Chrimes}, Ashley A. and {van Dalen}, Joyce N.~D. and {D'Elia}, Valerio and {De Pasquale}, Massimiliano and {Fulton}, Michael D. and {Groot}, Paul J. and {Gupta}, Rahul and {Hartmann}, Dieter H. and {van Hoof}, Agnes P.~C. and {Huber}, Mark E. and {Izzo}, Luca and {Jacobson-Galan}, Wynn and {Jakobsson}, P{\'a}ll and {Kong}, Albert and {Laskar}, Tanmoy and {Lowe}, Thomas B. and {Magnier}, Eugene A. and {Maiorano}, Elisabetta and {Martin-Carrillo}, Antonio and {Mas-Ribas}, Lluis and {Mata S{\'a}nchez}, Daniel and {Nicholl}, Matt and {Nixon}, Christopher J. and {Oates}, Samantha R. and {Paek}, Gregory and {Palmerio}, Jesse and {Paris}, Diego and {Pieterse}, Dani{\"e}lle L.~A. and {Pugliese}, Giovanna and {Quirola Vasquez}, Jonathan A. and {van Roestel}, Jan and {Rossi}, Andrea and {Rouco Escorial}, Alicia and {Salvaterra}, Ruben and {Schneider}, Benjamin and {Smartt}, Stephen J. and {Smith}, Ken and {Smith}, Ian A. and {Srivastav}, Shubham and {Torres}, Manuel A.~P. and {Ventura}, Chiara and {Vreeswijk}, Paul and {Wainscoat}, Richard and {Yang}, Yi-Jung and {Yang}, Sheng},
        title = "{EP 250108a/SN 2025kg: Observations of the Most Nearby Broad-line Type Ic Supernova Following an Einstein Probe Fast X-Ray Transient}",
      journal = {\apjl},
         year = 2025,
        month = jul,
       volume = {988},
       number = {1},
          eid = {L13},
        pages = {L13},
          doi = {10.3847/2041-8213/ade7f9},
archivePrefix = {arXiv},
       eprint = {2504.08889},
 primaryClass = {astro-ph.HE},
       adsurl = {https://ui.adsabs.harvard.edu/abs/2025ApJ...988L..13R}
}

@ARTICLE{Balcon_2024_classification,
       author = {{Balcon}, C.},
        title = "{Transient Classification Report for 2024-11-26}",
      journal = {Transient Name Server Classification Report},
         year = 2024,
        month = nov,
       volume = {2024-4655},
        pages = {1},
       adsurl = {https://ui.adsabs.harvard.edu/abs/2024TNSCR4655....1B}
}

@ARTICLE{Lidman_2024_classification_confirm,
       author = {{Lidman}, C. and {Rauf}, L. and {Auchettl}, K. and {Romagnoli}, S. and {Shi}, J. and {Schmidt}, B.~P. and {Armstrong}, P. and {Martin}, B. and {Dove}, K. and {Tucker}, B.~E. and {Timmermans}, T. and {Fulton}, M. and {Srivastav}, S. and {Smartt}, S.~J.},
        title = "{Transient Classification Report for 2024-11-27}",
      journal = {Transient Name Server Classification Report},
         year = 2024,
        month = nov,
       volume = {2024-4668},
        pages = {1},
       adsurl = {https://ui.adsabs.harvard.edu/abs/2024TNSCR4668....1L}
}

@ARTICLE{Roming05,
       author = {{Roming}, Peter W.~A. and {Kennedy}, Thomas E. and {Mason}, Keith O. and {Nousek}, John A. and {Ahr}, Lindy and {Bingham}, Richard E. and {Broos}, Patrick S. and {Carter}, Mary J. and {Hancock}, Barry K. and {Huckle}, Howard E. and {Hunsberger}, S.~D. and {Kawakami}, Hajime and {Killough}, Ronnie and {Koch}, T. Scott and {McLelland}, Michael K. and {Smith}, Kelly and {Smith}, Philip J. and {Soto}, Juan Carlos and {Boyd}, Patricia T. and {Breeveld}, Alice A. and {Holland}, Stephen T. and {Ivanushkina}, Mariya and {Pryzby}, Michael S. and {Still}, Martin D. and {Stock}, Joseph},
        title = "{The Swift Ultra-Violet/Optical Telescope}",
      journal = {\ssr},
         year = 2005,
        month = oct,
       volume = {120},
       number = {3-4},
        pages = {95-142},
          doi = {10.1007/s11214-005-5095-4},
archivePrefix = {arXiv},
       eprint = {astro-ph/0507413},
 primaryClass = {astro-ph},
       adsurl = {https://ui.adsabs.harvard.edu/abs/2005SSRv..120...95R}
}

@ARTICLE{Tonry_2018,
       author = {{Tonry}, J.~L. and {Denneau}, L. and {Heinze}, A.~N. and {Stalder}, B. and {Smith}, K.~W. and {Smartt}, S.~J. and {Stubbs}, C.~W. and {Weiland}, H.~J. and {Rest}, A.},
        title = "{ATLAS: A High-cadence All-sky Survey System}",
      journal = {\pasp},
         year = 2018,
        month = jun,
       volume = {130},
       number = {988},
        pages = {064505},
          doi = {10.1088/1538-3873/aabadf},
archivePrefix = {arXiv},
       eprint = {1802.00879},
 primaryClass = {astro-ph.IM},
       adsurl = {https://ui.adsabs.harvard.edu/abs/2018PASP..130f4505T}
}

@ARTICLE{Tonry_2011,
       author = {{Tonry}, John L.},
        title = "{An Early Warning System for Asteroid Impact}",
      journal = {\pasp},
         year = 2011,
        month = jan,
       volume = {123},
       number = {899},
        pages = {58},
          doi = {10.1086/657997},
archivePrefix = {arXiv},
       eprint = {1011.1028},
 primaryClass = {astro-ph.IM},
       adsurl = {https://ui.adsabs.harvard.edu/abs/2011PASP..123...58T}
}

@ARTICLE{Smith_2020,
       author = {{Smith}, K.~W. and {Smartt}, S.~J. and {Young}, D.~R. and {Tonry}, J.~L. and {Denneau}, L. and {Flewelling}, H. and {Heinze}, A.~N. and {Weiland}, H.~J. and {Stalder}, B. and {Rest}, A. and {Stubbs}, C.~W. and {Anderson}, J.~P. and {Chen}, T. -W. and {Clark}, P. and {Do}, A. and {F{\"o}rster}, F. and {Fulton}, M. and {Gillanders}, J. and {McBrien}, O.~R. and {O'Neill}, D. and {Srivastav}, S. and {Wright}, D.~E.},
        title = "{Design and Operation of the ATLAS Transient Science Server}",
      journal = {\pasp},
         year = 2020,
        month = aug,
       volume = {132},
       number = {1014},
          eid = {085002},
        pages = {085002},
          doi = {10.1088/1538-3873/ab936e},
archivePrefix = {arXiv},
       eprint = {2003.09052},
 primaryClass = {astro-ph.IM},
       adsurl = {https://ui.adsabs.harvard.edu/abs/2020PASP..132h5002S}
}

@ARTICLE{Li_2011,
       author = {{Li}, Weidong and {Leaman}, Jesse and {Chornock}, Ryan and {Filippenko}, Alexei V. and {Poznanski}, Dovi and {Ganeshalingam}, Mohan and {Wang}, Xiaofeng and {Modjaz}, Maryam and {Jha}, Saurabh and {Foley}, Ryan J. and {Smith}, Nathan},
        title = "{Nearby supernova rates from the Lick Observatory Supernova Search - II. The observed luminosity functions and fractions of supernovae in a complete sample}",
      journal = {\mnras},
         year = 2011,
        month = apr,
       volume = {412},
       number = {3},
        pages = {1441-1472},
          doi = {10.1111/j.1365-2966.2011.18160.x},
archivePrefix = {arXiv},
       eprint = {1006.4612},
 primaryClass = {astro-ph.SR},
       adsurl = {https://ui.adsabs.harvard.edu/abs/2011MNRAS.412.1441L}
}

@ARTICLE{Brown_2013,
       author = {{Brown}, T.~M. and {Baliber}, N. and {Bianco}, F.~B. and {Bowman}, M. and {Burleson}, B. and {Conway}, P. and {Crellin}, M. and {Depagne}, {\'E}. and {De Vera}, J. and {Dilday}, B. and {Dragomir}, D. and {Dubberley}, M. and {Eastman}, J.~D. and {Elphick}, M. and {Falarski}, M. and {Foale}, S. and {Ford}, M. and {Fulton}, B.~J. and {Garza}, J. and {Gomez}, E.~L. and {Graham}, M. and {Greene}, R. and {Haldeman}, B. and {Hawkins}, E. and {Haworth}, B. and {Haynes}, R. and {Hidas}, M. and {Hjelstrom}, A.~E. and {Howell}, D.~A. and {Hygelund}, J. and {Lister}, T.~A. and {Lobdill}, R. and {Martinez}, J. and {Mullins}, D.~S. and {Norbury}, M. and {Parrent}, J. and {Paulson}, R. and {Petry}, D.~L. and {Pickles}, A. and {Posner}, V. and {Rosing}, W.~E. and {Ross}, R. and {Sand}, D.~J. and {Saunders}, E.~S. and {Shobbrook}, J. and {Shporer}, A. and {Street}, R.~A. and {Thomas}, D. and {Tsapras}, Y. and {Tufts}, J.~R. and {Valenti}, S. and {Vander Horst}, K. and {Walker}, Z. and {White}, G. and {Willis}, M.},
        title = "{Las Cumbres Observatory Global Telescope Network}",
      journal = {\pasp},
         year = 2013,
        month = sep,
       volume = {125},
       number = {931},
        pages = {1031},
          doi = {10.1086/673168},
archivePrefix = {arXiv},
       eprint = {1305.2437},
 primaryClass = {astro-ph.IM},
       adsurl = {https://ui.adsabs.harvard.edu/abs/2013PASP..125.1031B}
}

@ARTICLE{Cherchneff_2026,
       author = {{Cherchneff}, I. and {Talbi}, D. and {Cernicharo}, J.},
        title = "{Revisiting the formation of molecules and dust in core collapse supernovae}",
      journal = {\aap},
         year = 2026,
        month = mar,
       volume = {708},
          eid = {A76},
        pages = {A76},
          doi = {10.1051/0004-6361/202557490},
archivePrefix = {arXiv},
       eprint = {2510.01079},
 primaryClass = {astro-ph.GA},
       adsurl = {https://ui.adsabs.harvard.edu/abs/2026A&A...708A..76C}
}

@ARTICLE{Dwek_2011,
       author = {{Dwek}, Eli and {Cherchneff}, Isabelle},
        title = "{The Origin of Dust in the Early Universe: Probing the Star Formation History of Galaxies by Their Dust Content}",
      journal = {\apj},
         year = 2011,
        month = feb,
       volume = {727},
       number = {2},
          eid = {63},
        pages = {63},
          doi = {10.1088/0004-637X/727/2/63},
archivePrefix = {arXiv},
       eprint = {1011.1303},
 primaryClass = {astro-ph.CO},
       adsurl = {https://ui.adsabs.harvard.edu/abs/2011ApJ...727...63D}
}

@ARTICLE{Schneider_2024,
       author = {{Schneider}, Raffaella and {Maiolino}, Roberto},
        title = "{The formation and cosmic evolution of dust in the early Universe: I. Dust sources}",
      journal = {\aapr},
         year = 2024,
        month = apr,
       volume = {32},
       number = {1},
          eid = {2},
        pages = {2},
          doi = {10.1007/s00159-024-00151-2},
archivePrefix = {arXiv},
       eprint = {2310.00053},
 primaryClass = {astro-ph.GA},
       adsurl = {https://ui.adsabs.harvard.edu/abs/2024A&ARv..32....2S}
}

@ARTICLE{Ravi_2023,
       author = {{Ravi}, Aravind P. and {Rho}, Jeonghee and {Park}, Sangwook and {Park}, Seong Hyun and {Yoon}, Sung-Chul and {Geballe}, T.~R. and {Vink{\'o}}, Jozsef and {Tinyanont}, Samaporn and {Bostroem}, K. Azalee and {Burke}, Jamison and {Hiramatsu}, Daichi and {Howell}, D. Andrew and {McCully}, Curtis and {Newsome}, Megan and {Padilla Gonzalez}, Estefania and {Pellegrino}, Craig and {Cartier}, Regis and {Pritchard}, Tyler and {Andersen}, Morten and {Blinnikov}, Sergey and {Dong}, Yize and {Blanchard}, Peter and {Kilpatrick}, Charles D. and {Hoeflich}, Peter and {Valenti}, Stefano and {Filippenko}, Alexei V. and {Suntzeff}, Nicholas B. and {Seok}, Ji Yeon and {K{\"o}nyves-T{\'o}th}, R. and {Foley}, Ryan J. and {Siebert}, Matthew R. and {Jones}, David O.},
        title = "{Near-infrared and Optical Observations of Type Ic SN 2021krf: Luminous Late-time Emission and Dust Formation}",
      journal = {\apj},
         year = 2023,
        month = jun,
       volume = {950},
       number = {1},
          eid = {14},
        pages = {14},
          doi = {10.3847/1538-4357/accddc},
archivePrefix = {arXiv},
       eprint = {2211.00205},
 primaryClass = {astro-ph.HE},
       adsurl = {https://ui.adsabs.harvard.edu/abs/2023ApJ...950...14R}
}

@ARTICLE{Issa_2025,
       author = {{Issa}, Danat and {Gottlieb}, Ore and {Metzger}, Brian D. and {Jacquemin-Ide}, Jonatan and {Liska}, Matthew and {Foucart}, Francois and {Halevi}, Goni and {Tchekhovskoy}, Alexander},
        title = "{Magnetically Driven Neutron-rich Ejecta Unleashed: Global 3D Neutrino─General Relativistic Magnetohydrodynamic Simulations of Collapsars Probe the Conditions for r-process Nucleosynthesis}",
      journal = {\apjl},
         year = 2025,
        month = jun,
       volume = {985},
       number = {2},
          eid = {L26},
        pages = {L26},
          doi = {10.3847/2041-8213/adc694},
archivePrefix = {arXiv},
       eprint = {2410.02852},
 primaryClass = {astro-ph.HE},
       adsurl = {https://ui.adsabs.harvard.edu/abs/2025ApJ...985L..26I}
}

@ARTICLE{Hotokezaka_2022,
       author = {{Hotokezaka}, Kenta and {Tanaka}, Masaomi and {Kato}, Daiji and {Gaigalas}, Gediminas},
        title = "{Tungsten versus Selenium as a potential source of kilonova nebular emission observed by Spitzer}",
      journal = {\mnras},
         year = 2022,
        month = sep,
       volume = {515},
       number = {1},
        pages = {L89-L93},
          doi = {10.1093/mnrasl/slac071},
archivePrefix = {arXiv},
       eprint = {2204.00737},
 primaryClass = {astro-ph.HE},
       adsurl = {https://ui.adsabs.harvard.edu/abs/2022MNRAS.515L..89H}
}

@ARTICLE{Corsi_2023,
       author = {{Corsi}, Alessandra and {Ho}, Anna Y.~Q. and {Cenko}, S. Bradley and {Kulkarni}, Shrinivas R. and {Anand}, Shreya and {Yang}, Sheng and {Sollerman}, Jesper and {Srinivasaragavan}, Gokul P. and {Omand}, Conor M.~B. and {Balasubramanian}, Arvind and {Frail}, Dale A. and {Fremling}, Christoffer and {Perley}, Daniel A. and {Yao}, Yuhan and {Dahiwale}, Aishwarya S. and {De}, Kishalay and {Dugas}, Alison and {Hankins}, Matthew and {Jencson}, Jacob and {Kasliwal}, Mansi M. and {Tzanidakis}, Anastasios and {Bellm}, Eric C. and {Laher}, Russ R. and {Masci}, Frank J. and {Purdum}, Josiah N. and {Regnault}, Nicolas},
        title = "{A Search for Relativistic Ejecta in a Sample of ZTF Broad-lined Type Ic Supernovae}",
      journal = {\apj},
         year = 2023,
        month = aug,
       volume = {953},
       number = {2},
          eid = {179},
        pages = {179},
          doi = {10.3847/1538-4357/acd3f2},
archivePrefix = {arXiv},
       eprint = {2210.09536},
 primaryClass = {astro-ph.HE},
       adsurl = {https://ui.adsabs.harvard.edu/abs/2023ApJ...953..179C}
}

@ARTICLE{Laporte_2017,
       author = {{Laporte}, N. and {Ellis}, R.~S. and {Boone}, F. and {Bauer}, F.~E. and {Qu{\'e}nard}, D. and {Roberts-Borsani}, G.~W. and {Pell{\'o}}, R. and {P{\'e}rez-Fournon}, I. and {Streblyanska}, A.},
        title = "{Dust in the Reionization Era: ALMA Observations of a z = 8.38 Gravitationally Lensed Galaxy}",
      journal = {\apjl},
         year = 2017,
        month = mar,
       volume = {837},
       number = {2},
          eid = {L21},
        pages = {L21},
          doi = {10.3847/2041-8213/aa62aa},
archivePrefix = {arXiv},
       eprint = {1703.02039},
 primaryClass = {astro-ph.GA},
       adsurl = {https://ui.adsabs.harvard.edu/abs/2017ApJ...837L..21L}
}

@ARTICLE{Tinyanont_2026,
       author = {{Tinyanont}, Samaporn and {Wangnok}, Kittipong and {Andrews}, Jennifer E. and {Foley}, Ryan J. and {Kaewmookda}, Methawee and {Jencson}, Jacob E. and {Rest}, Armin and {Auchettl}, Katie and {Bostroem}, K.~A. and {Coulter}, David A. and {Chainakun}, Poemwai and {Chornock}, Ryan and {Davis}, Kyle W. and {Fox}, Ori D. and {Galbany}, Llu{\'\i}s and {Geballe}, Thomas R. and {Hsu}, Brian and {Jacobson-Gal{\'a}n}, Wynn and {Jha}, Saurabh W. and {Kaur}, Ravjit and {Kasliwal}, Mansi M. and {Lau}, Ryan M. and {LeBaron}, Natalie and {Margutti}, Raffaella and {Park}, Seong Hyun and {Pearson}, Jeniveve and {Piro}, Anthony L. and {Ransome}, Conor L. and {Ravi}, Aravind P. and {Rho}, Jeonghee and {Rojas-Bravo}, C{\'e}sar and {Rose}, Sam and {Sand}, David J. and {Smith}, Nathan and {Shrestha}, Manisha and {Subrayan}, Bhagya M. and {Valenti}, Stefano},
        title = "{An infrared echo from a circumstellar disk in the hydrogen- and helium-poor SN 2024aecx}",
      journal = {arXiv e-prints},
         year = 2026,
        month = feb,
          eid = {arXiv:2602.02691},
        pages = {arXiv:2602.02691},
          doi = {10.48550/arXiv.2602.02691},
archivePrefix = {arXiv},
       eprint = {2602.02691},
 primaryClass = {astro-ph.HE},
       adsurl = {https://ui.adsabs.harvard.edu/abs/2026arXiv260202691T}
}

@ARTICLE{Shivvers_2017,
       author = {{Shivvers}, Isaac and {Modjaz}, Maryam and {Zheng}, WeiKang and {Liu}, Yuqian and {Filippenko}, Alexei V. and {Silverman}, Jeffrey M. and {Matheson}, Thomas and {Pastorello}, Andrea and {Graur}, Or and {Foley}, Ryan J. and {Chornock}, Ryan and {Smith}, Nathan and {Leaman}, Jesse and {Benetti}, Stefano},
        title = "{Revisiting the Lick Observatory Supernova Search Volume-limited Sample: Updated Classifications and Revised Stripped-envelope Supernova Fractions}",
      journal = {\pasp},
         year = 2017,
        month = may,
       volume = {129},
       number = {975},
        pages = {054201},
          doi = {10.1088/1538-3873/aa54a6},
archivePrefix = {arXiv},
       eprint = {1609.02922},
 primaryClass = {astro-ph.HE},
       adsurl = {https://ui.adsabs.harvard.edu/abs/2017PASP..129e4201S}
}

@misc{hosseinzadeh_light_2023,
  title = {Light {{Curve Fitting}} v0.9.0},
  author = {Hosseinzadeh, Griffin and Bostroem, K. Azalee and Gomez, Sebastian},
  year = {2023},
  month = jun,
  doi = {10.5281/zenodo.8049154},
  url = {https://doi.org/10.5281/zenodo.8049154},
  adsurl = {https://ui.adsabs.harvard.edu/abs/2023zndo...8049154H},
  howpublished = {Zenodo}
}

@article{Valenti_2014,
	title = {The first month of evolution of the slow-rising {Type} {IIP} {SN} 2013ej in {M74}},
	volume = {438},
	doi = {10.1093/mnrasl/slt171},
	urldate = {2017-08-16},
	journal = {MNRAS},
	author = {Valenti, S and Sand, D and Pastorello, A and Graham, M L and Howell, D A and Parrent, J T and Tomasella, L and Ochner, P and Fraser, M and Benetti, S and Yuan, F and Smartt, S J and Maund, J R and Arcavi, I and Gal-Yam, A and Inserra, C and Young, D},
	year = {2014},
	pages = {101--105},
}

@article{astropy:2018,
Adsurl = {https://ui.adsabs.harvard.edu/#abs/2018AJ....156..123T},
Author = {{Price-Whelan}, A.~M. and {Sip{\H{o}}cz}, B.~M. and {G{\"u}nther}, H.~M. and {Lim}, P.~L. and {Crawford}, S.~M. and {Conseil}, S. and {Shupe}, D.~L. and {Craig}, M.~W. and {Dencheva}, N. and {Ginsburg}, A. and {VanderPlas}, J.~T. and {Bradley}, L.~D. and {P{\'e}rez-Su{\'a}rez}, D. and {de Val-Borro}, M. and {Paper Contributors}, (Primary and {Aldcroft}, T.~L. and {Cruz}, K.~L. and {Robitaille}, T.~P. and {Tollerud}, E.~J. and {Coordination Committee}, (Astropy and {Ardelean}, C. and {Babej}, T. and {Bach}, Y.~P. and {Bachetti}, M. and {Bakanov}, A.~V. and {Bamford}, S.~P. and {Barentsen}, G. and {Barmby}, P. and {Baumbach}, A. and {Berry}, K.~L. and {Biscani}, F. and {Boquien}, M. and {Bostroem}, K.~A. and {Bouma}, L.~G. and {Brammer}, G.~B. and {Bray}, E.~M. and {Breytenbach}, H. and {Buddelmeijer}, H. and {Burke}, D.~J. and {Calderone}, G. and {Cano Rodr{\'\i}guez}, J.~L. and {Cara}, M. and {Cardoso}, J.~V.~M. and {Cheedella}, S. and {Copin}, Y. and {Corrales}, L. and {Crichton}, D. and {D{\textquoteright}Avella}, D. and {Deil}, C. and {Depagne}, {\'E}. and {Dietrich}, J.~P. and {Donath}, A. and {Droettboom}, M. and {Earl}, N. and {Erben}, T. and {Fabbro}, S. and {Ferreira}, L.~A. and {Finethy}, T. and {Fox}, R.~T. and {Garrison}, L.~H. and {Gibbons}, S.~L.~J. and {Goldstein}, D.~A. and {Gommers}, R. and {Greco}, J.~P. and {Greenfield}, P. and {Groener}, A.~M. and {Grollier}, F. and {Hagen}, A. and {Hirst}, P. and {Homeier}, D. and {Horton}, A.~J. and {Hosseinzadeh}, G. and {Hu}, L. and {Hunkeler}, J.~S. and {Ivezi{\'c}}, {\v{Z}}. and {Jain}, A. and {Jenness}, T. and {Kanarek}, G. and {Kendrew}, S. and {Kern}, N.~S. and {Kerzendorf}, W.~E. and {Khvalko}, A. and {King}, J. and {Kirkby}, D. and {Kulkarni}, A.~M. and {Kumar}, A. and {Lee}, A. and {Lenz}, D. and {Littlefair}, S.~P. and {Ma}, Z. and {Macleod}, D.~M. and {Mastropietro}, M. and {McCully}, C. and {Montagnac}, S. and {Morris}, B.~M. and {Mueller}, M. and {Mumford}, S.~J. and {Muna}, D. and {Murphy}, N.~A. and {Nelson}, S. and {Nguyen}, G.~H. and {Ninan}, J.~P. and {N{\"o}the}, M. and {Ogaz}, S. and {Oh}, S. and {Parejko}, J.~K. and {Parley}, N. and {Pascual}, S. and {Patil}, R. and {Patil}, A.~A. and {Plunkett}, A.~L. and {Prochaska}, J.~X. and {Rastogi}, T. and {Reddy Janga}, V. and {Sabater}, J. and {Sakurikar}, P. and {Seifert}, M. and {Sherbert}, L.~E. and {Sherwood-Taylor}, H. and {Shih}, A.~Y. and {Sick}, J. and {Silbiger}, M.~T. and {Singanamalla}, S. and {Singer}, L.~P. and {Sladen}, P.~H. and {Sooley}, K.~A. and {Sornarajah}, S. and {Streicher}, O. and {Teuben}, P. and {Thomas}, S.~W. and {Tremblay}, G.~R. and {Turner}, J.~E.~H. and {Terr{\'o}n}, V. and {van Kerkwijk}, M.~H. and {de la Vega}, A. and {Watkins}, L.~L. and {Weaver}, B.~A. and {Whitmore}, J.~B. and {Woillez}, J. and {Zabalza}, V. and {Contributors}, (Astropy},
Doi = {10.3847/1538-3881/aabc4f},
Eid = {123},
Journal = {\aj},
Month = Sep,
Pages = {123},
Primaryclass = {astro-ph.IM},
Title = {{The Astropy Project: Building an Open-science Project and Status of the v2.0 Core Package}},
Volume = {156},
Year = 2018}

@ARTICLE{Fulton_2023,
       author = {{Fulton}, M.~D. and {Smartt}, S.~J. and {Rhodes}, L. and {Huber}, M.~E. and {Villar}, A.~V. and {Moore}, T. and {Srivastav}, S. and {Schultz}, A.~S.~B. and {Chambers}, K.~C. and {Izzo}, L. and {Hjorth}, J. and {Chen}, T. -W. and {Nicholl}, M. and {Foley}, R.~J. and {Rest}, A. and {Smith}, K.~W. and {Young}, D.~R. and {Sim}, S.~A. and {Bright}, J. and {Zenati}, Y. and {de Boer}, T. and {Bulger}, J. and {Fairlamb}, J. and {Gao}, H. and {Lin}, C. -C. and {Lowe}, T. and {Magnier}, E.~A. and {Smith}, I.~A. and {Wainscoat}, R. and {Coulter}, D.~A. and {Jones}, D.~O. and {Kilpatrick}, C.~D. and {McGill}, P. and {Ramirez-Ruiz}, E. and {Lee}, K. -S. and {Narayan}, G. and {Ramakrishnan}, V. and {Ridden-Harper}, R. and {Singh}, A. and {Wang}, Q. and {Kong}, A.~K.~H. and {Ngeow}, C. -C. and {Pan}, Y. -C. and {Yang}, S. and {Davis}, K.~W. and {Piro}, A.~L. and {Rojas-Bravo}, C. and {Sommer}, J. and {Yadavalli}, S.~K.},
        title = "{The optical light curve of GRB 221009A: the afterglow and detection of the emerging supernova SN 2022xiw}",
      journal = {arXiv e-prints},
         year = 2023,
        month = jan,
          eid = {arXiv:2301.11170},
        pages = {arXiv:2301.11170},
archivePrefix = {arXiv},
       eprint = {2301.11170},
 primaryClass = {astro-ph.HE},
       adsurl = {https://ui.adsabs.harvard.edu/abs/2023arXiv230111170F}
}

@misc{Bradley_2019,
   author = {Larry Bradley and Brigitta Sip{\H o}cz and Thomas Robitaille and
             Erik Tollerud and Z\`e Vin{\'{\i}}cius and Christoph Deil and
             Kyle Barbary and Hans Moritz G{\"u}nther and Mihai Cara and
             Ivo Busko and Simon Conseil and Michael Droettboom and
             Azalee Bostroem and E. M. Bray and Lars Andersen Bratholm and
             Tom Wilson and Matt Craig and Geert Barentsen and
             Sergio Pascual and Axel Donath and Johnny Greco and
             Gabriel Perren and P. L. Lim and Wolfgang Kerzendorf},
    title = {astropy/photutils: v0.6},
    month = jan,
     year = 2019,
      doi = {10.5281/zenodo.2533376},
      url = {https://doi.org/10.5281/zenodo.2533376}
}

@preamble{ " \newcommand{\noop}[1]{} " }

@article{astropy:2013,
Adsurl = {http://adsabs.harvard.edu/abs/2013A%26A...558A..33A},
Archiveprefix = {arXiv},
Author = {{Astropy Collaboration} and {Robitaille}, T.~P. and {Tollerud}, E.~J. and {Greenfield}, P. and {Droettboom}, M. and {Bray}, E. and {Aldcroft}, T. and {Davis}, M. and {Ginsburg}, A. and {Price-Whelan}, A.~M. and {Kerzendorf}, W.~E. and {Conley}, A. and {Crighton}, N. and {Barbary}, K. and {Muna}, D. and {Ferguson}, H. and {Grollier}, F. and {Parikh}, M.~M. and {Nair}, P.~H. and {Unther}, H.~M. and {Deil}, C. and {Woillez}, J. and {Conseil}, S. and {Kramer}, R. and {Turner}, J.~E.~H. and {Singer}, L. and {Fox}, R. and {Weaver}, B.~A. and {Zabalza}, V. and {Edwards}, Z.~I. and {Azalee Bostroem}, K. and {Burke}, D.~J. and {Casey}, A.~R. and {Crawford}, S.~M. and {Dencheva}, N. and {Ely}, J. and {Jenness}, T. and {Labrie}, K. and {Lim}, P.~L. and {Pierfederici}, F. and {Pontzen}, A. and {Ptak}, A. and {Refsdal}, B. and {Servillat}, M. and {Streicher}, O.},
Doi = {10.1051/0004-6361/201322068},
Eid = {A33},
Eprint = {1307.6212},
Journal = {\aap},
Month = oct,
Pages = {A33},
Primaryclass = {astro-ph.IM},
Title = {{Astropy: A community Python package for astronomy}},
Volume = 558,
Year = 2013}

@INPROCEEDINGS{iraf1,
       author = {{Tody}, Doug},
        title = "{The IRAF Data Reduction and Analysis System}",
    booktitle = {Instrumentation in astronomy VI},
         year = 1986,
       editor = {{Crawford}, David L.},
       series = {Society of Photo-Optical Instrumentation Engineers (SPIE) Conference Series},
       volume = {627},
        month = jan,
        pages = {733},
          doi = {10.1117/12.968154},
       adsurl = {https://ui.adsabs.harvard.edu/abs/1986SPIE..627..733T}
}

@INPROCEEDINGS{iraf2,
       author = {{Tody}, Doug},
        title = "{IRAF in the Nineties}",
    booktitle = {Astronomical Data Analysis Software and Systems II},
         year = 1993,
       editor = {{Hanisch}, R.~J. and {Brissenden}, R.~J.~V. and {Barnes}, J.},
       series = {Astronomical Society of the Pacific Conference Series},
       volume = {52},
        month = jan,
        pages = {173},
       adsurl = {https://ui.adsabs.harvard.edu/abs/1993ASPC...52..173T}
}

@ARTICLE{Kalinova_2021_host,
       author = {{Kalinova}, V. and {Colombo}, D. and {S{\'a}nchez}, S.~F. and {Kodaira}, K. and {Garc{\'\i}a-Benito}, R. and {Gonz{\'a}lez Delgado}, R. and {Rosolowsky}, E. and {Lacerda}, E.~A.~D.},
        title = "{Star formation quenching stages of active and non-active galaxies}",
      journal = {\aap},
         year = 2021,
        month = apr,
       volume = {648},
          eid = {A64},
        pages = {A64},
          doi = {10.1051/0004-6361/202039896},
archivePrefix = {arXiv},
       eprint = {2101.10019},
 primaryClass = {astro-ph.GA},
       adsurl = {https://ui.adsabs.harvard.edu/abs/2021A&A...648A..64K}
}

@ARTICLE{Srinivasaragavan_2024,
       author = {{Srinivasaragavan}, Gokul P. and {Yang}, Sheng and {Anand}, Shreya and {Sollerman}, Jesper and {Ho}, Anna Y.~Q. and {Corsi}, Alessandra and {Cenko}, S. Bradley and {Perley}, Daniel and {Schulze}, Steve and {Sanchez-Fleming}, Marquice and {Pope}, Jack and {Sarin}, Nikhil and {Omand}, Conor and {Das}, Kaustav K. and {Fremling}, Christoffer and {Andreoni}, Igor and {Bruch}, Rachel and {Burdge}, Kevin B. and {De}, Kishalay and {Gal-Yam}, Avishay and {Gangopadhyay}, Anjasha and {Graham}, Matthew J. and {Jencson}, Jacob E. and {Karambelkar}, Viraj and {Kasliwal}, Mansi M. and {Kulkarni}, S.~R. and {Martikainen}, Julia and {Sharma}, Yashvi S. and {Tzanidakis}, Anastasios and {Yan}, Lin and {Yao}, Yuhan and {Bellm}, Eric C. and {Groom}, Steven L. and {Masci}, Frank J. and {Nir}, Guy and {Purdum}, Josiah and {Smith}, Roger and {Sravan}, Niharika},
        title = "{Optical and Radio Analysis of Systematically Classified Broad-lined Type Ic Supernovae from the Zwicky Transient Facility}",
      journal = {\apj},
         year = 2024,
        month = nov,
       volume = {976},
       number = {1},
          eid = {71},
        pages = {71},
          doi = {10.3847/1538-4357/ad7fde},
archivePrefix = {arXiv},
       eprint = {2408.14586},
 primaryClass = {astro-ph.HE},
       adsurl = {https://ui.adsabs.harvard.edu/abs/2024ApJ...976...71S}
}

@ARTICLE{Ricigliano_2025,
       author = {{Ricigliano}, Giacomo and {Hotokezaka}, Kenta and {Arcones}, Almudena},
        title = "{Modeling the emission lines from r-process elements in Supernova nebulae}",
      journal = {\mnras},
         year = 2025,
        month = sep,
          doi = {10.1093/mnras/staf1577},
       adsurl = {https://ui.adsabs.harvard.edu/abs/2025MNRAS.tmp.1499R}
}

@ARTICLE{Henseley_2023,
       author = {{Hensley}, Brandon S. and {Draine}, B.~T.},
        title = "{The Astrodust+PAH Model: A Unified Description of the Extinction, Emission, and Polarization from Dust in the Diffuse Interstellar Medium}",
      journal = {\apj},
         year = 2023,
        month = may,
       volume = {948},
       number = {1},
          eid = {55},
        pages = {55},
          doi = {10.3847/1538-4357/acc4c2},
archivePrefix = {arXiv},
       eprint = {2208.12365},
 primaryClass = {astro-ph.GA},
       adsurl = {https://ui.adsabs.harvard.edu/abs/2023ApJ...948...55H}
}

@ARTICLE{Arnett_1982,
       author = {{Arnett}, W.~D.},
        title = "{Type I supernovae. I - Analytic solutions for the early part of the light curve}",
      journal = {\apj},
         year = 1982,
        month = feb,
       volume = {253},
        pages = {785-797},
          doi = {10.1086/159681},
       adsurl = {https://ui.adsabs.harvard.edu/abs/1982ApJ...253..785A}
}

@ARTICLE{Sahu_2009,
       author = {{Sahu}, D.~K. and {Tanaka}, Masaomi and {Anupama}, G.~C. and {Gurugubelli}, Uday K. and {Nomoto}, Ken'ichi},
        title = "{The Broad-Line Type Ic Supernova SN 2007ru: Adding to the Diversity of Type Ic Supernovae}",
      journal = {\apj},
         year = 2009,
        month = may,
       volume = {697},
       number = {1},
        pages = {676-683},
          doi = {10.1088/0004-637X/697/1/676},
archivePrefix = {arXiv},
       eprint = {0805.3201},
 primaryClass = {astro-ph},
       adsurl = {https://ui.adsabs.harvard.edu/abs/2009ApJ...697..676S}
}

@ARTICLE{Stritzinger_2018,
       author = {{Stritzinger}, M.~D. and {Taddia}, F. and {Burns}, C.~R. and {Phillips}, M.~M. and {Bersten}, M. and {Contreras}, C. and {Folatelli}, G. and {Holmbo}, S. and {Hsiao}, E.~Y. and {Hoeflich}, P. and {Leloudas}, G. and {Morrell}, N. and {Sollerman}, J. and {Suntzeff}, N.~B.},
        title = "{The Carnegie Supernova Project I. Methods to estimate host-galaxy reddening of stripped-envelope supernovae}",
      journal = {\aap},
         year = 2018,
        month = feb,
       volume = {609},
          eid = {A135},
        pages = {A135},
          doi = {10.1051/0004-6361/201730843},
archivePrefix = {arXiv},
       eprint = {1707.07615},
 primaryClass = {astro-ph.HE},
       adsurl = {https://ui.adsabs.harvard.edu/abs/2018A&A...609A.135S}
}

@ARTICLE{Taddia_2019,
       author = {{Taddia}, F. and {Sollerman}, J. and {Fremling}, C. and {Barbarino}, C. and {Karamehmetoglu}, E. and {Arcavi}, I. and {Cenko}, S.~B. and {Filippenko}, A.~V. and {Gal-Yam}, A. and {Hiramatsu}, D. and {Hosseinzadeh}, G. and {Howell}, D.~A. and {Kulkarni}, S.~R. and {Laher}, R. and {Lunnan}, R. and {Masci}, F. and {Nugent}, P.~E. and {Nyholm}, A. and {Perley}, D.~A. and {Quimby}, R. and {Silverman}, J.~M.},
        title = "{Analysis of broad-lined Type Ic supernovae from the (intermediate) Palomar Transient Factory}",
      journal = {\aap},
         year = 2019,
        month = jan,
       volume = {621},
          eid = {A71},
        pages = {A71},
          doi = {10.1051/0004-6361/201834429},
archivePrefix = {arXiv},
       eprint = {1811.09544},
 primaryClass = {astro-ph.HE},
       adsurl = {https://ui.adsabs.harvard.edu/abs/2019A&A...621A..71T}
}

@ARTICLE{Chabrier_2003_imf,
       author = {{Chabrier}, Gilles},
        title = "{Galactic Stellar and Substellar Initial Mass Function}",
      journal = {\pasp},
         year = 2003,
        month = jul,
       volume = {115},
       number = {809},
        pages = {763-795},
          doi = {10.1086/376392},
archivePrefix = {arXiv},
       eprint = {astro-ph/0304382},
 primaryClass = {astro-ph},
       adsurl = {https://ui.adsabs.harvard.edu/abs/2003PASP..115..763C}
}

@MISC{lightcurvefitting,
       author = {{Hosseinzadeh}, Griffin and {Gomez}, Sebastian},
        title = "{Light Curve Fitting}",
 howpublished = {Zenodo},
         year = 2020,
        month = dec,
          eid = {10.5281/zenodo.4312178},
          doi = {10.5281/zenodo.4312178},
      version = {v0.2.0},
    publisher = {Zenodo},
       adsurl = {https://ui.adsabs.harvard.edu/abs/2020zndo...4312178H}
}

@ARTICLE{mpl,
       author = {{Hunter}, John D.},
        title = "{Matplotlib: A 2D Graphics Environment}",
      journal = {Computing in Science and Engineering},
         year = 2007,
        month = may,
       volume = {9},
       number = {3},
        pages = {90-95},
          doi = {10.1109/MCSE.2007.55},
       adsurl = {https://ui.adsabs.harvard.edu/abs/2007CSE.....9...90H}
}

@ARTICLE{Modjaz_2019,
       author = {{Modjaz}, Maryam and {Guti{\'e}rrez}, Claudia P. and {Arcavi}, Iair},
        title = "{New regimes in the observation of core-collapse supernovae}",
      journal = {Nature Astronomy},
         year = 2019,
        month = aug,
       volume = {3},
        pages = {717-724},
          doi = {10.1038/s41550-019-0856-2},
archivePrefix = {arXiv},
       eprint = {1908.02476},
 primaryClass = {astro-ph.HE},
       adsurl = {https://ui.adsabs.harvard.edu/abs/2019NatAs...3..717M}
}

@INCOLLECTION{Jerkstrand2017,
       author = {{Jerkstrand}, Anders},
        title = "{Spectra of Supernovae in the Nebular Phase}",
    booktitle = {Handbook of Supernovae},
         year = 2017,
       editor = {{Alsabti}, Athem W. and {Murdin}, Paul},
        pages = {795},
          doi = {10.1007/978-3-319-21846-5_29},
       adsurl = {https://ui.adsabs.harvard.edu/abs/2017hsn..book..795J}
}

@ARTICLE{Japelj_2018,
       author = {{Japelj}, J. and {Vergani}, S.~D. and {Salvaterra}, R. and {Renzo}, M. and {Zapartas}, E. and {de Mink}, S.~E. and {Kaper}, L. and {Zibetti}, S.},
        title = "{Host galaxies of SNe Ic-BL with and without long gamma-ray bursts}",
      journal = {\aap},
         year = 2018,
        month = sep,
       volume = {617},
          eid = {A105},
        pages = {A105},
          doi = {10.1051/0004-6361/201833209},
archivePrefix = {arXiv},
       eprint = {1806.10613},
 primaryClass = {astro-ph.HE},
       adsurl = {https://ui.adsabs.harvard.edu/abs/2018A&A...617A.105J}
}

@ARTICLE{GW_2025_mergerrate,
       author = {{The LIGO Scientific Collaboration} and {the Virgo Collaboration} and {the KAGRA Collaboration} and {Abac}, A.~G. and {Abouelfettouh}, I. and {Acernese}, F. and {Ackley}, K. and {Adamcewicz}, C. and {Adhicary}, S. and {Adhikari}, D. and {Adhikari}, N. and {Adhikari}, R.~X. and {Adkins}, V.~K. and {Afroz}, S. and {Agarwal}, D. and {Agathos}, M. and {Aghaei Abchouyeh}, M. and {Aguiar}, O.~D. and {Ahmadzadeh}, S. and {Aiello}, L. and {Ain}, A. and {Ajith}, P. and {Akutsu}, T. and {Albanesi}, S. and {Alfaidi}, R.~A. and {Al-Jodah}, A. and {All{\'e}n{\'e}}, C. and {Allocca}, A. and {Al-Shammari}, S. and {Altin}, P.~A. and {Alvarez-Lopez}, S. and {Amarasinghe}, O. and {Amato}, A. and {Amra}, C. and {Ananyeva}, A. and {Anderson}, S.~B. and {Anderson}, W.~G. and {Andia}, M. and {Ando}, M. and {Andrade}, T. and {Andr{\'e}s-Carcasona}, M. and {Andri{\'c}}, T. and {Anglin}, J. and {Ansoldi}, S. and {Antelis}, J.~M. and {Antier}, S. and {Aoumi}, M. and {Appavuravther}, E.~Z. and {Appert}, S. and {Apple}, S.~K. and {Arai}, K. and {Araya}, A. and {Araya}, M.~C. and {Arca Sedda}, M. and {Areeda}, J.~S. and {Argianas}, L. and {Aritomi}, N. and {Armato}, F. and {Armstrong}, S. and {Arnaud}, N. and {Arogeti}, M. and {Aronson}, S.~M. and {Arun}, K.~G. and {Ashton}, G. and {Aso}, Y. and {Assiduo}, M. and {Assis de Souza Melo}, S. and {Aston}, S.~M. and {Astone}, P. and {Attadio}, F. and {Aubin}, F. and {AultONeal}, K. and {Avallone}, G. and {Babak}, S. and {Badaracco}, F. and {Badger}, C. and {Bae}, S. and {Bagnasco}, S. and {Bagui}, E. and {Baiotti}, L. and {Bajpai}, R. and {Baka}, T. and {Baker}, T. and {Ball}, M. and {Ballardin}, G. and {Ballmer}, S.~W. and {Banagiri}, S. and {Banerjee}, B. and {Bankar}, D. and {Baptiste}, T.~M. and {Baral}, P. and {Barayoga}, J.~C. and {Barish}, B.~C. and {Barker}, D. and {Barman}, N. and {Barneo}, P. and {Barone}, F. and {Barr}, B. and {Barsotti}, L. and {Barsuglia}, M. and {Barta}, D. and {Bartoletti}, A.~M. and {Barton}, M.~A. and {Bartos}, I. and {Basak}, S. and {Basalaev}, A. and {Bassiri}, R. and {Basti}, A. and {Bates}, D.~E. and {Bawaj}, M. and {Baxi}, P. and {Bayley}, J.~C. and {Baylor}, A.~C. and {Baynard}, II, P.~A. and {Bazzan}, M. and {Bedakihale}, V.~M. and {Beirnaert}, F. and {Bejger}, M. and {Belardinelli}, D. and {Bell}, A.~S. and {Bellie}, D.~S. and {Bellizzi}, L. and {Beltran-Martinez}, D. and {Benoit}, W. and {Bentara}, I. and {Bentley}, J.~D. and {Ben Yaala}, M. and {Bera}, S. and {Bergamin}, F. and {Berger}, B.~K. and {Bernuzzi}, S. and {Beroiz}, M. and {Berry}, C.~P.~L. and {Bersanetti}, D. and {Bertolini}, A. and {Betzwieser}, J. and {Beveridge}, D. and {Bevilacqua}, G. and {Bevins}, N. and {Bhandare}, R. and {Bhatt}, R. and {Bhattacharjee}, D. and {Bhaumik}, S. and {Bhowmick}, S. and {Biancalana}, V. and {Bianchi}, A. and {Bilenko}, I.~A. and {Billingsley}, G. and {Binetti}, A. and {Bini}, S. and {Binu}, C. and {Birnholtz}, O. and {Biscoveanu}, S. and {Bisht}, A. and {Bitossi}, M. and {Bizouard}, M. -A. and {Blaber}, S. and {Blackburn}, J.~K. and {Blagg}, L.~A. and {Blair}, C.~D. and {Blair}, D.~G. and {Bobba}, F. and {Bode}, N. and {Boileau}, G. and {Boldrini}, M. and {Bolingbroke}, G.~N. and {Bolliand}, A. and {Bonavena}, L.~D. and {Bondarescu}, R. and {Bondu}, F. and {Bonilla}, E. and {Bonilla}, M.~S. and {Bonino}, A. and {Bonnand}, R. and {Booker}, P. and {Borchers}, A. and {Borhanian}, S. and {Boschi}, V. and {Bose}, S. and {Bossilkov}, V. and {Boudon}, A. and {Bozzi}, A. and {Bradaschia}, C. and {Brady}, P.~R. and {Branch}, A. and {Branchesi}, M. and {Braun}, I. and {Briant}, T. and {Brillet}, A. and {Brinkmann}, M. and {Brockill}, P. and {Brockmueller}, E. and {Brooks}, A.~F. and {Brown}, B.~C. and {Brown}, D.~D. and {Brozzetti}, M.~L. and {Brunett}, S. and {Bruno}, G. and {Bruntz}, R. and {Bryant}, J.},
        title = "{GWTC-4.0: Population Properties of Merging Compact Binaries}",
      journal = {arXiv e-prints},
         year = 2025,
        month = aug,
          eid = {arXiv:2508.18083},
        pages = {arXiv:2508.18083},
          doi = {10.48550/arXiv.2508.18083},
archivePrefix = {arXiv},
       eprint = {2508.18083},
 primaryClass = {astro-ph.HE},
       adsurl = {https://ui.adsabs.harvard.edu/abs/2025arXiv250818083T}
}

@MISC{Shahbandeh_2023,
       author = {{Shahbandeh}, Melissa and {Ashall}, Chris and {Baade}, Dietrich and {Baron}, Eddie and {Burns}, Christopher and {Burrow}, Anthony and {DerKacy}, James M. and {Engesser}, Mike and {Engh}, Greg and {Filippenko}, Alex V. and {Foley}, Ryan and {Fox}, Ori Dosovitz and {Gezari}, Suvi and {Gomez}, Sebastian and {Hoeflich}, Peter A. and {Hsiao}, Eric and {Jencson}, Jacob and {Kumar}, Sahana and {Lu}, Jing and {Milisavljevic}, Dan and {Modjaz}, Maryam and {Phillips}, Mark M. and {Pierel}, Justin and {Rest}, Armin and {Sarangi}, Arkaprabha and {Shappee}, Benjamin John and {Stritzinger}, Maximillian and {Strolger}, Louis-Gregory and {Szalai}, Tamas and {Temim}, Tea and {Van Dyk}, Schuyler D. and {Williams}, Robert E. and {Woosley}, Stan},
        title = "{Near- and Mid-IR Observations to Probe Dust Formation in the Remarkably Nearby Stripped-Envelope Supernova 2023dbc}",
 howpublished = {JWST Proposal. Cycle 1, ID. \#4436},
         year = 2023,
        month = apr,
        pages = {4436},
       adsurl = {https://ui.adsabs.harvard.edu/abs/2023jwst.prop.4436S}
}

@ARTICLE{Gillanders_2025,
       author = {{Gillanders}, J.~H. and {Smartt}, S.~J.},
        title = "{Analysis of the JWST spectra of the kilonova AT 2023vfi accompanying GRB 230307A}",
      journal = {\mnras},
         year = 2025,
        month = apr,
       volume = {538},
       number = {3},
        pages = {1663-1689},
          doi = {10.1093/mnras/staf287},
archivePrefix = {arXiv},
       eprint = {2408.11093},
 primaryClass = {astro-ph.HE},
       adsurl = {https://ui.adsabs.harvard.edu/abs/2025MNRAS.538.1663G}
}

@ARTICLE{Levan_2024_GRB230307A,
       author = {{Levan}, Andrew J. and {Gompertz}, Benjamin P. and {Salafia}, Om Sharan and {Bulla}, Mattia and {Burns}, Eric and {Hotokezaka}, Kenta and {Izzo}, Luca and {Lamb}, Gavin P. and {Malesani}, Daniele B. and {Oates}, Samantha R. and {Ravasio}, Maria Edvige and {Rouco Escorial}, Alicia and {Schneider}, Benjamin and {Sarin}, Nikhil and {Schulze}, Steve and {Tanvir}, Nial R. and {Ackley}, Kendall and {Anderson}, Gemma and {Brammer}, Gabriel B. and {Christensen}, Lise and {Dhillon}, Vikram S. and {Evans}, Phil A. and {Fausnaugh}, Michael and {Fong}, Wen-fai and {Fruchter}, Andrew S. and {Fryer}, Chris and {Fynbo}, Johan P.~U. and {Gaspari}, Nicola and {Heintz}, Kasper E. and {Hjorth}, Jens and {Kennea}, Jamie A. and {Kennedy}, Mark R. and {Laskar}, Tanmoy and {Leloudas}, Giorgos and {Mandel}, Ilya and {Martin-Carrillo}, Antonio and {Metzger}, Brian D. and {Nicholl}, Matt and {Nugent}, Anya and {Palmerio}, Jesse T. and {Pugliese}, Giovanna and {Rastinejad}, Jillian and {Rhodes}, Lauren and {Rossi}, Andrea and {Saccardi}, Andrea and {Smartt}, Stephen J. and {Stevance}, Heloise F. and {Tohuvavohu}, Aaron and {van der Horst}, Alexander and {Vergani}, Susanna D. and {Watson}, Darach and {Barclay}, Thomas and {Bhirombhakdi}, Kornpob and {Breedt}, Elm{\'e} and {Breeveld}, Alice A. and {Brown}, Alexander J. and {Campana}, Sergio and {Chrimes}, Ashley A. and {D'Avanzo}, Paolo and {D'Elia}, Valerio and {De Pasquale}, Massimiliano and {Dyer}, Martin J. and {Galloway}, Duncan K. and {Garbutt}, James A. and {Green}, Matthew J. and {Hartmann}, Dieter H. and {Jakobsson}, P{\'a}ll and {Kerry}, Paul and {Kouveliotou}, Chryssa and {Langeroodi}, Danial and {Le Floc'h}, Emeric and {Leung}, James K. and {Littlefair}, Stuart P. and {Munday}, James and {O'Brien}, Paul and {Parsons}, Steven G. and {Pelisoli}, Ingrid and {Sahman}, David I. and {Salvaterra}, Ruben and {Sbarufatti}, Boris and {Steeghs}, Danny and {Tagliaferri}, Gianpiero and {Th{\"o}ne}, Christina C. and {de Ugarte Postigo}, Antonio and {Kann}, David Alexander},
        title = "{Heavy-element production in a compact object merger observed by JWST}",
      journal = {\nat},
         year = 2024,
        month = feb,
       volume = {626},
       number = {8000},
        pages = {737-741},
          doi = {10.1038/s41586-023-06759-1},
archivePrefix = {arXiv},
       eprint = {2307.02098},
 primaryClass = {astro-ph.HE},
       adsurl = {https://ui.adsabs.harvard.edu/abs/2024Natur.626..737L}
}

@ARTICLE{Milisavljevic_2015,
       author = {{Milisavljevic}, D. and {Margutti}, R. and {Parrent}, J.~T. and {Soderberg}, A.~M. and {Fesen}, R.~A. and {Mazzali}, P. and {Maeda}, K. and {Sanders}, N.~E. and {Cenko}, S.~B. and {Silverman}, J.~M. and {Filippenko}, A.~V. and {Kamble}, A. and {Chakraborti}, S. and {Drout}, M.~R. and {Kirshner}, R.~P. and {Pickering}, T.~E. and {Kawabata}, K. and {Hattori}, T. and {Hsiao}, E.~Y. and {Stritzinger}, M.~D. and {Marion}, G.~H. and {Vinko}, J. and {Wheeler}, J.~C.},
        title = "{The Broad-lined Type Ic SN 2012ap and the Nature of Relativistic Supernovae Lacking a Gamma-Ray Burst Detection}",
      journal = {\apj},
         year = 2015,
        month = jan,
       volume = {799},
       number = {1},
          eid = {51},
        pages = {51},
          doi = {10.1088/0004-637X/799/1/51},
archivePrefix = {arXiv},
       eprint = {1408.1606},
 primaryClass = {astro-ph.HE},
       adsurl = {https://ui.adsabs.harvard.edu/abs/2015ApJ...799...51M}
}

@ARTICLE{Anand2024,
       author = {{Anand}, Shreya and {Barnes}, Jennifer and {Yang}, Sheng and {Kasliwal}, Mansi M. and {Coughlin}, Michael W. and {Sollerman}, Jesper and {De}, Kishalay and {Fremling}, Christoffer and {Corsi}, Alessandra and {Ho}, Anna Y.~Q. and {Balasubramanian}, Arvind and {Omand}, Conor and {Srinivasaragavan}, Gokul P. and {Cenko}, S. Bradley and {Ahumada}, Tom{\'a}s and {Andreoni}, Igor and {Dahiwale}, Aishwarya and {Das}, Kaustav Kashyap and {Jencson}, Jacob and {Karambelkar}, Viraj and {Kumar}, Harsh and {Metzger}, Brian D. and {Perley}, Daniel and {Sarin}, Nikhil and {Schweyer}, Tassilo and {Schulze}, Steve and {Sharma}, Yashvi and {Sit}, Tawny and {Stein}, Robert and {Tartaglia}, Leonardo and {Tinyanont}, Samaporn and {Tzanidakis}, Anastasios and {van Roestel}, Jan and {Yao}, Yuhan and {Bloom}, Joshua S. and {Cook}, David O. and {Dekany}, Richard and {Graham}, Matthew J. and {Groom}, Steven L. and {Kaplan}, David L. and {Masci}, Frank J. and {Medford}, Michael S. and {Riddle}, Reed and {Zhang}, Chaoran},
        title = "{Collapsars as Sites of r-process Nucleosynthesis: Systematic Photometric Near-infrared Follow-up of Type Ic-BL Supernovae}",
      journal = {\apj},
         year = 2024,
        month = feb,
       volume = {962},
       number = {1},
          eid = {68},
        pages = {68},
          doi = {10.3847/1538-4357/ad11df},
archivePrefix = {arXiv},
       eprint = {2302.09226},
 primaryClass = {astro-ph.HE},
       adsurl = {https://ui.adsabs.harvard.edu/abs/2024ApJ...962...68A}
}

@ARTICLE{Rastinejad_2024,
       author = {{Rastinejad}, J.~C. and {Fong}, W. and {Levan}, A.~J. and {Tanvir}, N.~R. and {Kilpatrick}, C.~D. and {Fruchter}, A.~S. and {Anand}, S. and {Bhirombhakdi}, K. and {Covino}, S. and {Fynbo}, J.~P.~U. and {Halevi}, G. and {Hartmann}, D.~H. and {Heintz}, K.~E. and {Izzo}, L. and {Jakobsson}, P. and {Kangas}, T. and {Lamb}, G.~P. and {Malesani}, D.~B. and {Melandri}, A. and {Metzger}, B.~D. and {Milvang-Jensen}, B. and {Pian}, E. and {Pugliese}, G. and {Rossi}, A. and {Siegel}, D.~M. and {Singh}, P. and {Stratta}, G.},
        title = "{A Hubble Space Telescope Search for r-Process Nucleosynthesis in Gamma-Ray Burst Supernovae}",
      journal = {\apj},
         year = 2024,
        month = jun,
       volume = {968},
       number = {1},
          eid = {14},
        pages = {14},
          doi = {10.3847/1538-4357/ad409c},
archivePrefix = {arXiv},
       eprint = {2312.04630},
 primaryClass = {astro-ph.HE},
       adsurl = {https://ui.adsabs.harvard.edu/abs/2024ApJ...968...14R}
}

@ARTICLE{Gordon_2024,
       author = {{Gordon}, Alexander J. and {Ferguson}, Annette M.~N. and {Mann}, Robert G.},
        title = "{Uncovering tidal treasures: automated classification of faint tidal features in DECaLS data}",
      journal = {\mnras},
         year = 2024,
        month = oct,
       volume = {534},
       number = {2},
        pages = {1459-1480},
          doi = {10.1093/mnras/stae2169},
archivePrefix = {arXiv},
       eprint = {2404.06487},
 primaryClass = {astro-ph.GA},
       adsurl = {https://ui.adsabs.harvard.edu/abs/2024MNRAS.534.1459G}
}

@ARTICLE{scipy,
       author = {{Virtanen}, Pauli and {Gommers}, Ralf and {Oliphant}, Travis E. and {Haberland}, Matt and {Reddy}, Tyler and {Cournapeau}, David and {Burovski}, Evgeni and {Peterson}, Pearu and {Weckesser}, Warren and {Bright}, Jonathan and {van der Walt}, St{\'e}fan J. and {Brett}, Matthew and {Wilson}, Joshua and {Millman}, K. Jarrod and {Mayorov}, Nikolay and {Nelson}, Andrew R.~J. and {Jones}, Eric and {Kern}, Robert and {Larson}, Eric and {Carey}, C.~J. and {Polat}, {\.I}lhan and {Feng}, Yu and {Moore}, Eric W. and {VanderPlas}, Jake and {Laxalde}, Denis and {Perktold}, Josef and {Cimrman}, Robert and {Henriksen}, Ian and {Quintero}, E.~A. and {Harris}, Charles R. and {Archibald}, Anne M. and {Ribeiro}, Ant{\^o}nio H. and {Pedregosa}, Fabian and {van Mulbregt}, Paul and {SciPy 1. 0 Contributors}},
        title = "{SciPy 1.0: fundamental algorithms for scientific computing in Python}",
      journal = {Nature Methods},
         year = 2020,
        month = feb,
       volume = {17},
        pages = {261-272},
          doi = {10.1038/s41592-019-0686-2},
archivePrefix = {arXiv},
       eprint = {1907.10121},
 primaryClass = {cs.MS},
       adsurl = {https://ui.adsabs.harvard.edu/abs/2020NatMe..17..261V}
}

@Article{numpy,
 title         = {Array programming with {NumPy}},
 author        = {Charles R. Harris and K. Jarrod Millman and St{\'{e}}fan J.
                 van der Walt and Ralf Gommers and Pauli Virtanen and David
                 Cournapeau and Eric Wieser and Julian Taylor and Sebastian
                 Berg and Nathaniel J. Smith and Robert Kern and Matti Picus
                 and Stephan Hoyer and Marten H. van Kerkwijk and Matthew
                 Brett and Allan Haldane and Jaime Fern{\'{a}}ndez del
                 R{\'{i}}o and Mark Wiebe and Pearu Peterson and Pierre
                 G{\'{e}}rard-Marchant and Kevin Sheppard and Tyler Reddy and
                 Warren Weckesser and Hameer Abbasi and Christoph Gohlke and
                 Travis E. Oliphant},
 year          = {2020},
 month         = sep,
 journal       = {Nature},
 volume        = {585},
 number        = {7825},
 pages         = {357--362},
 doi           = {10.1038/s41586-020-2649-2},
 publisher     = {Springer Science and Business Media {LLC}},
 url           = {https://doi.org/10.1038/s41586-020-2649-2}
}

@ARTICLE{Soderberg_2006,
       author = {{Soderberg}, A.~M. and {Nakar}, E. and {Berger}, E. and {Kulkarni}, S.~R.},
        title = "{Late-Time Radio Observations of 68 Type Ibc Supernovae: Strong Constraints on Off-Axis Gamma-Ray Bursts}",
      journal = {\apj},
         year = 2006,
        month = feb,
       volume = {638},
       number = {2},
        pages = {930-937},
          doi = {10.1086/499121},
archivePrefix = {arXiv},
       eprint = {astro-ph/0507147},
 primaryClass = {astro-ph},
       adsurl = {https://ui.adsabs.harvard.edu/abs/2006ApJ...638..930S}
}

@ARTICLE{Cano_2017,
       author = {{Cano}, Zach and {Wang}, Shan-Qin and {Dai}, Zi-Gao and {Wu}, Xue-Feng},
        title = "{The Observer's Guide to the Gamma-Ray Burst Supernova Connection}",
      journal = {Advances in Astronomy},
         year = 2017,
        month = jan,
       volume = {2017},
          eid = {8929054},
        pages = {8929054},
          doi = {10.1155/2017/892905410.48550/arXiv.1604.03549},
archivePrefix = {arXiv},
       eprint = {1604.03549},
 primaryClass = {astro-ph.HE},
       adsurl = {https://ui.adsabs.harvard.edu/abs/2017AdAst2017E...5C}
}

@ARTICLE{Drout_2011,
       author = {{Drout}, Maria R. and {Soderberg}, Alicia M. and {Gal-Yam}, Avishay and {Cenko}, S. Bradley and {Fox}, Derek B. and {Leonard}, Douglas C. and {Sand}, David J. and {Moon}, Dae-Sik and {Arcavi}, Iair and {Green}, Yoav},
        title = "{The First Systematic Study of Type Ibc Supernova Multi-band Light Curves}",
      journal = {\apj},
         year = 2011,
        month = nov,
       volume = {741},
       number = {2},
          eid = {97},
        pages = {97},
          doi = {10.1088/0004-637X/741/2/9710.48550/arXiv.1011.4959},
archivePrefix = {arXiv},
       eprint = {1011.4959},
 primaryClass = {astro-ph.CO},
       adsurl = {https://ui.adsabs.harvard.edu/abs/2011ApJ...741...97D}
}

@ARTICLE{Modjaz_2020,
       author = {{Modjaz}, Maryam and {Bianco}, Federica B. and {Siwek}, Magdalena and {Huang}, Shan and {Perley}, Daniel A. and {Fierroz}, David and {Liu}, Yu-Qian and {Arcavi}, Iair and {Gal-Yam}, Avishay and {Filippenko}, Alexei V. and {Blagorodnova}, Nadia and {Cenko}, Bradley S. and {Kasliwal}, Mansi and {Kulkarni}, Shri and {Schulze}, Steve and {Taggart}, Kirsty and {Zheng}, Weikang},
        title = "{Host Galaxies of Type Ic and Broad-lined Type Ic Supernovae from the Palomar Transient Factory: Implications for Jet Production}",
      journal = {\apj},
         year = 2020,
        month = apr,
       volume = {892},
       number = {2},
          eid = {153},
        pages = {153},
          doi = {10.3847/1538-4357/ab418510.48550/arXiv.1901.00872},
archivePrefix = {arXiv},
       eprint = {1901.00872},
 primaryClass = {astro-ph.HE},
       adsurl = {https://ui.adsabs.harvard.edu/abs/2020ApJ...892..153M}
}

@ARTICLE{Modjaz_2016,
       author = {{Modjaz}, Maryam and {Liu}, Yuqian Q. and {Bianco}, Federica B. and {Graur}, Or},
        title = "{The Spectral SN-GRB Connection: Systematic Spectral Comparisons between Type Ic Supernovae and Broad-lined Type Ic Supernovae with and without Gamma-Ray Bursts}",
      journal = {\apj},
         year = 2016,
        month = dec,
       volume = {832},
       number = {2},
          eid = {108},
        pages = {108},
          doi = {10.3847/0004-637X/832/2/10810.48550/arXiv.1509.07124},
archivePrefix = {arXiv},
       eprint = {1509.07124},
 primaryClass = {astro-ph.HE},
       adsurl = {https://ui.adsabs.harvard.edu/abs/2016ApJ...832..108M}
}

@ARTICLE{Smartt_2009,
       author = {{Smartt}, Stephen J.},
        title = "{Progenitors of Core-Collapse Supernovae}",
      journal = {\araa},
         year = 2009,
        month = sep,
       volume = {47},
       number = {1},
        pages = {63-106},
          doi = {10.1146/annurev-astro-082708-10173710.48550/arXiv.0908.0700},
archivePrefix = {arXiv},
       eprint = {0908.0700},
 primaryClass = {astro-ph.SR},
       adsurl = {https://ui.adsabs.harvard.edu/abs/2009ARA&A..47...63S}
}

@ARTICLE{Gaskell_1986,
       author = {{Gaskell}, C.~M. and {Cappellaro}, E. and {Dinerstein}, H.~L. and {Garnett}, D.~R. and {Harkness}, R.~P. and {Wheeler}, J.~C.},
        title = "{Type Ib Supernovae 1983n and 1985f: Oxygen-rich Late Time Spectra}",
      journal = {\apjl},
         year = 1986,
        month = jul,
       volume = {306},
        pages = {L77},
          doi = {10.1086/184709},
       adsurl = {https://ui.adsabs.harvard.edu/abs/1986ApJ...306L..77G}
}

@ARTICLE{Podsiadlowski_1993,
       author = {{Podsiadlowski}, Ph. and {Hsu}, J.~J.~L. and {Joss}, P.~C. and {Ross}, R.~R.},
        title = "{The progenitor of supernova 1993J: a stripped supergiant in a binary system?}",
      journal = {\nat},
         year = 1993,
        month = aug,
       volume = {364},
       number = {6437},
        pages = {509-511},
          doi = {10.1038/364509a0},
       adsurl = {https://ui.adsabs.harvard.edu/abs/1993Natur.364..509P}
}

@ARTICLE{Nomoto_1995,
       author = {{Nomoto}, K.~I. and {Iwamoto}, K. and {Suzuki}, T.},
        title = "{The evolution and explosion of massive binary stars and Type Ib-Ic-IIb-IIL supernovae.}",
      journal = {\physrep},
         year = 1995,
        month = may,
       volume = {256},
       number = {1},
        pages = {173-191},
          doi = {10.1016/0370-1573(94)00107-E},
       adsurl = {https://ui.adsabs.harvard.edu/abs/1995PhR...256..173N}
}

@INPROCEEDINGS{Banzai,
       author = {{McCully}, Curtis and {Volgenau}, Nikolaus H. and {Harbeck}, Daniel-Rolf and {Lister}, Tim A. and {Saunders}, Eric S. and {Turner}, Monica L. and {Siiverd}, Robert J. and {Bowman}, Mark},
        title = "{Real-time processing of the imaging data from the network of Las Cumbres Observatory Telescopes using BANZAI}",
    booktitle = {Software and Cyberinfrastructure for Astronomy V},
         year = 2018,
       editor = {{Guzman}, Juan C. and {Ibsen}, Jorge},
       series = {Society of Photo-Optical Instrumentation Engineers (SPIE) Conference Series},
       volume = {10707},
        month = jul,
          eid = {107070K},
        pages = {107070K},
          doi = {10.1117/12.2314340},
archivePrefix = {arXiv},
       eprint = {1811.04163},
 primaryClass = {astro-ph.IM},
       adsurl = {https://ui.adsabs.harvard.edu/abs/2018SPIE10707E..0KM}
}

@ARTICLE{Patat_2001,
       author = {{Patat}, Ferdinando and {Cappellaro}, Enrico and {Danziger}, John and {Mazzali}, Paolo A. and {Sollerman}, Jesper and {Augusteijn}, Thomas and {Brewer}, James and {Doublier}, Vanessa and {Gonzalez}, Jean Fran{\c{c}}ois and {Hainaut}, Olivier and {Lidman}, Chris and {Leibundgut}, Bruno and {Nomoto}, Ken'ichi and {Nakamura}, Takayoshi and {Spyromilio}, Jason and {Rizzi}, Luca and {Turatto}, Massimo and {Walsh}, Jeremy and {Galama}, Titus J. and {van Paradijs}, Jan and {Kouveliotou}, Chryssa and {Vreeswijk}, Paul M. and {Frontera}, Filippo and {Masetti}, Nicola and {Palazzi}, Eliana and {Pian}, Elena},
        title = "{The Metamorphosis of SN 1998bw}",
      journal = {\apj},
         year = 2001,
        month = jul,
       volume = {555},
       number = {2},
        pages = {900-917},
          doi = {10.1086/321526},
archivePrefix = {arXiv},
       eprint = {astro-ph/0103111},
 primaryClass = {astro-ph},
       adsurl = {https://ui.adsabs.harvard.edu/abs/2001ApJ...555..900P}
}

@ARTICLE{Kulkarni_1998,
       author = {{Kulkarni}, S.~R. and {Frail}, D.~A. and {Wieringa}, M.~H. and {Ekers}, R.~D. and {Sadler}, E.~M. and {Wark}, R.~M. and {Higdon}, J.~L. and {Phinney}, E.~S. and {Bloom}, J.~S.},
        title = "{Radio emission from the unusual supernova 1998bw and its association with the {\ensuremath{\gamma}}-ray burst of 25 April 1998}",
      journal = {\nat},
         year = 1998,
        month = oct,
       volume = {395},
       number = {6703},
        pages = {663-669},
          doi = {10.1038/27139},
       adsurl = {https://ui.adsabs.harvard.edu/abs/1998Natur.395..663K}
}

@ARTICLE{Valenti_2017,
       author = {{Valenti}, Stefano and {Sand}, David J. and {Yang}, Sheng and {Cappellaro}, Enrico and {Tartaglia}, Leonardo and {Corsi}, Alessandra and {Jha}, Saurabh W. and {Reichart}, Daniel E. and {Haislip}, Joshua and {Kouprianov}, Vladimir},
        title = "{The Discovery of the Electromagnetic Counterpart of GW170817: Kilonova AT 2017gfo/DLT17ck}",
      journal = {\apjl},
         year = 2017,
        month = oct,
       volume = {848},
       number = {2},
          eid = {L24},
        pages = {L24},
          doi = {10.3847/2041-8213/aa8edf},
archivePrefix = {arXiv},
       eprint = {1710.05854},
 primaryClass = {astro-ph.HE},
       adsurl = {https://ui.adsabs.harvard.edu/abs/2017ApJ...848L..24V}
}

@ARTICLE{Valenti_2008,
       author = {{Valenti}, S. and {Benetti}, S. and {Cappellaro}, E. and {Patat}, F. and {Mazzali}, P. and {Turatto}, M. and {Hurley}, K. and {Maeda}, K. and {Gal-Yam}, A. and {Foley}, R.~J. and {Filippenko}, A.~V. and {Pastorello}, A. and {Challis}, P. and {Frontera}, F. and {Harutyunyan}, A. and {Iye}, M. and {Kawabata}, K. and {Kirshner}, R.~P. and {Li}, W. and {Lipkin}, Y.~M. and {Matheson}, T. and {Nomoto}, K. and {Ofek}, E.~O. and {Ohyama}, Y. and {Pian}, E. and {Poznanski}, D. and {Salvo}, M. and {Sauer}, D.~N. and {Schmidt}, B.~P. and {Soderberg}, A. and {Zampieri}, L.},
        title = "{The broad-lined Type Ic supernova 2003jd}",
      journal = {\mnras},
         year = 2008,
        month = feb,
       volume = {383},
       number = {4},
        pages = {1485-1500},
          doi = {10.1111/j.1365-2966.2007.12647.x},
archivePrefix = {arXiv},
       eprint = {0710.5173},
 primaryClass = {astro-ph},
       adsurl = {https://ui.adsabs.harvard.edu/abs/2008MNRAS.383.1485V}
}

@ARTICLE{astropy:2022,
       author = {{Astropy Collaboration} and {Price-Whelan}, Adrian M. and {Lim}, Pey Lian and {Earl}, Nicholas and {Starkman}, Nathaniel and {Bradley}, Larry and {Shupe}, David L. and {Patil}, Aarya A. and {Corrales}, Lia and {Brasseur}, C.~E. and {N{\"o}the}, Maximilian and {Donath}, Axel and {Tollerud}, Erik and {Morris}, Brett M. and {Ginsburg}, Adam and {Vaher}, Eero and {Weaver}, Benjamin A. and {Tocknell}, James and {Jamieson}, William and {van Kerkwijk}, Marten H. and {Robitaille}, Thomas P. and {Merry}, Bruce and {Bachetti}, Matteo and {G{\"u}nther}, H. Moritz and {Aldcroft}, Thomas L. and {Alvarado-Montes}, Jaime A. and {Archibald}, Anne M. and {B{\'o}di}, Attila and {Bapat}, Shreyas and {Barentsen}, Geert and {Baz{\'a}n}, Juanjo and {Biswas}, Manish and {Boquien}, M{\'e}d{\'e}ric and {Burke}, D.~J. and {Cara}, Daria and {Cara}, Mihai and {Conroy}, Kyle E. and {Conseil}, Simon and {Craig}, Matthew W. and {Cross}, Robert M. and {Cruz}, Kelle L. and {D'Eugenio}, Francesco and {Dencheva}, Nadia and {Devillepoix}, Hadrien A.~R. and {Dietrich}, J{\"o}rg P. and {Eigenbrot}, Arthur Davis and {Erben}, Thomas and {Ferreira}, Leonardo and {Foreman-Mackey}, Daniel and {Fox}, Ryan and {Freij}, Nabil and {Garg}, Suyog and {Geda}, Robel and {Glattly}, Lauren and {Gondhalekar}, Yash and {Gordon}, Karl D. and {Grant}, David and {Greenfield}, Perry and {Groener}, Austen M. and {Guest}, Steve and {Gurovich}, Sebastian and {Handberg}, Rasmus and {Hart}, Akeem and {Hatfield-Dodds}, Zac and {Homeier}, Derek and {Hosseinzadeh}, Griffin and {Jenness}, Tim and {Jones}, Craig K. and {Joseph}, Prajwel and {Kalmbach}, J. Bryce and {Karamehmetoglu}, Emir and {Ka{\l}uszy{\'n}ski}, Miko{\l}aj and {Kelley}, Michael S.~P. and {Kern}, Nicholas and {Kerzendorf}, Wolfgang E. and {Koch}, Eric W. and {Kulumani}, Shankar and {Lee}, Antony and {Ly}, Chun and {Ma}, Zhiyuan and {MacBride}, Conor and {Maljaars}, Jakob M. and {Muna}, Demitri and {Murphy}, N.~A. and {Norman}, Henrik and {O'Steen}, Richard and {Oman}, Kyle A. and {Pacifici}, Camilla and {Pascual}, Sergio and {Pascual-Granado}, J. and {Patil}, Rohit R. and {Perren}, Gabriel I. and {Pickering}, Timothy E. and {Rastogi}, Tanuj and {Roulston}, Benjamin R. and {Ryan}, Daniel F. and {Rykoff}, Eli S. and {Sabater}, Jose and {Sakurikar}, Parikshit and {Salgado}, Jes{\'u}s and {Sanghi}, Aniket and {Saunders}, Nicholas and {Savchenko}, Volodymyr and {Schwardt}, Ludwig and {Seifert-Eckert}, Michael and {Shih}, Albert Y. and {Jain}, Anany Shrey and {Shukla}, Gyanendra and {Sick}, Jonathan and {Simpson}, Chris and {Singanamalla}, Sudheesh and {Singer}, Leo P. and {Singhal}, Jaladh and {Sinha}, Manodeep and {Sip{\H{o}}cz}, Brigitta M. and {Spitler}, Lee R. and {Stansby}, David and {Streicher}, Ole and {{\v{S}}umak}, Jani and {Swinbank}, John D. and {Taranu}, Dan S. and {Tewary}, Nikita and {Tremblay}, Grant R. and {de Val-Borro}, Miguel and {Van Kooten}, Samuel J. and {Vasovi{\'c}}, Zlatan and {Verma}, Shresth and {de Miranda Cardoso}, Jos{\'e} Vin{\'\i}cius and {Williams}, Peter K.~G. and {Wilson}, Tom J. and {Winkel}, Benjamin and {Wood-Vasey}, W.~M. and {Xue}, Rui and {Yoachim}, Peter and {Zhang}, Chen and {Zonca}, Andrea and {Astropy Project Contributors}},
        title = "{The Astropy Project: Sustaining and Growing a Community-oriented Open-source Project and the Latest Major Release (v5.0) of the Core Package}",
      journal = {\apj},
         year = 2022,
        month = aug,
       volume = {935},
       number = {2},
          eid = {167},
        pages = {167},
          doi = {10.3847/1538-4357/ac7c74},
archivePrefix = {arXiv},
       eprint = {2206.14220},
 primaryClass = {astro-ph.IM},
       adsurl = {https://ui.adsabs.harvard.edu/abs/2022ApJ...935..167A}
}

@ARTICLE{Valenti_2016,
       author = {{Valenti}, S. and {Howell}, D.~A. and {Stritzinger}, M.~D. and {Graham}, M.~L. and {Hosseinzadeh}, G. and {Arcavi}, I. and {Bildsten}, L. and {Jerkstrand}, A. and {McCully}, C. and {Pastorello}, A. and {Piro}, A.~L. and {Sand}, D. and {Smartt}, S.~J. and {Terreran}, G. and {Baltay}, C. and {Benetti}, S. and {Brown}, P. and {Filippenko}, A.~V. and {Fraser}, M. and {Rabinowitz}, D. and {Sullivan}, M. and {Yuan}, F.},
        title = "{The diversity of Type II supernova versus the similarity in their progenitors}",
      journal = {\mnras},
         year = 2016,
        month = jul,
       volume = {459},
       number = {4},
        pages = {3939-3962},
          doi = {10.1093/mnras/stw870},
archivePrefix = {arXiv},
       eprint = {1603.08953},
 primaryClass = {astro-ph.SR},
       adsurl = {https://ui.adsabs.harvard.edu/abs/2016MNRAS.459.3939V}
}

@ARTICLE{Woosley_2006,
       author = {{Woosley}, S.~E. and {Bloom}, J.~S.},
        title = "{The Supernova Gamma-Ray Burst Connection}",
      journal = {\araa},
         year = 2006,
        month = sep,
       volume = {44},
       number = {1},
        pages = {507-556},
          doi = {10.1146/annurev.astro.43.072103.15055810.48550/arXiv.astro-ph/0609142},
archivePrefix = {arXiv},
       eprint = {astro-ph/0609142},
 primaryClass = {astro-ph},
       adsurl = {https://ui.adsabs.harvard.edu/abs/2006ARA&A..44..507W}
}

@ARTICLE{Gehrels_2004,
       author = {{Gehrels}, N. and {Chincarini}, G. and {Giommi}, P. and {Mason}, K.~O. and {Nousek}, J.~A. and {Wells}, A.~A. and {White}, N.~E. and {Barthelmy}, S.~D. and {Burrows}, D.~N. and {Cominsky}, L.~R. and {Hurley}, K.~C. and {Marshall}, F.~E. and {M{\'e}sz{\'a}ros}, P. and {Roming}, P.~W.~A. and {Angelini}, L. and {Barbier}, L.~M. and {Belloni}, T. and {Campana}, S. and {Caraveo}, P.~A. and {Chester}, M.~M. and {Citterio}, O. and {Cline}, T.~L. and {Cropper}, M.~S. and {Cummings}, J.~R. and {Dean}, A.~J. and {Feigelson}, E.~D. and {Fenimore}, E.~E. and {Frail}, D.~A. and {Fruchter}, A.~S. and {Garmire}, G.~P. and {Gendreau}, K. and {Ghisellini}, G. and {Greiner}, J. and {Hill}, J.~E. and {Hunsberger}, S.~D. and {Krimm}, H.~A. and {Kulkarni}, S.~R. and {Kumar}, P. and {Lebrun}, F. and {Lloyd-Ronning}, N.~M. and {Markwardt}, C.~B. and {Mattson}, B.~J. and {Mushotzky}, R.~F. and {Norris}, J.~P. and {Osborne}, J. and {Paczynski}, B. and {Palmer}, D.~M. and {Park}, H. -S. and {Parsons}, A.~M. and {Paul}, J. and {Rees}, M.~J. and {Reynolds}, C.~S. and {Rhoads}, J.~E. and {Sasseen}, T.~P. and {Schaefer}, B.~E. and {Short}, A.~T. and {Smale}, A.~P. and {Smith}, I.~A. and {Stella}, L. and {Tagliaferri}, G. and {Takahashi}, T. and {Tashiro}, M. and {Townsley}, L.~K. and {Tueller}, J. and {Turner}, M.~J.~L. and {Vietri}, M. and {Voges}, W. and {Ward}, M.~J. and {Willingale}, R. and {Zerbi}, F.~M. and {Zhang}, W.~W.},
        title = "{The Swift Gamma-Ray Burst Mission}",
      journal = {\apj},
         year = 2004,
        month = aug,
       volume = {611},
       number = {2},
        pages = {1005-1020},
          doi = {10.1086/422091},
archivePrefix = {arXiv},
       eprint = {astro-ph/0405233},
 primaryClass = {astro-ph},
       adsurl = {https://ui.adsabs.harvard.edu/abs/2004ApJ...611.1005G}
}

@ARTICLE{Breeveld_2010,
       author = {{Breeveld}, A.~A. and {Curran}, P.~A. and {Hoversten}, E.~A. and {Koch}, S. and {Landsman}, W. and {Marshall}, F.~E. and {Page}, M.~J. and {Poole}, T.~S. and {Roming}, P. and {Smith}, P.~J. and {Still}, M. and {Yershov}, V. and {Blustin}, A.~J. and {Brown}, P.~J. and {Gronwall}, C. and {Holland}, S.~T. and {Kuin}, N.~P.~M. and {McGowan}, K. and {Rosen}, S. and {Boyd}, P. and {Broos}, P. and {Carter}, M. and {Chester}, M.~M. and {Hancock}, B. and {Huckle}, H. and {Immler}, S. and {Ivanushkina}, M. and {Kennedy}, T. and {Mason}, K.~O. and {Morgan}, A.~N. and {Oates}, S. and {de Pasquale}, M. and {Schady}, P. and {Siegel}, M. and {vanden Berk}, D.},
        title = "{Further calibration of the Swift ultraviolet/optical telescope}",
      journal = {\mnras},
         year = 2010,
        month = aug,
       volume = {406},
       number = {3},
        pages = {1687-1700},
          doi = {10.1111/j.1365-2966.2010.16832.x},
archivePrefix = {arXiv},
       eprint = {1004.2448},
 primaryClass = {astro-ph.IM},
       adsurl = {https://ui.adsabs.harvard.edu/abs/2010MNRAS.406.1687B}
}

@ARTICLE{Ji_2016,
       author = {{Ji}, Alexander P. and {Frebel}, Anna and {Chiti}, Anirudh and {Simon}, Joshua D.},
        title = "{R-process enrichment from a single event in an ancient dwarf galaxy}",
      journal = {\nat},
         year = 2016,
        month = mar,
       volume = {531},
       number = {7596},
        pages = {610-613},
          doi = {10.1038/nature17425},
archivePrefix = {arXiv},
       eprint = {1512.01558},
 primaryClass = {astro-ph.GA},
       adsurl = {https://ui.adsabs.harvard.edu/abs/2016Natur.531..610J}
}

@ARTICLE{Naidu22,
       author = {{Naidu}, Rohan P. and {Ji}, Alexander P. and {Conroy}, Charlie and {Bonaca}, Ana and {Ting}, Yuan-Sen and {Zaritsky}, Dennis and {van Son}, Lieke A.~C. and {Broekgaarden}, Floor S. and {Tacchella}, Sandro and {Chandra}, Vedant and {Caldwell}, Nelson and {Cargile}, Phillip and {Speagle}, Joshua S.},
        title = "{Evidence from Disrupted Halo Dwarfs that r-process Enrichment via Neutron Star Mergers is Delayed by {\ensuremath{\gtrsim}}500 Myr}",
      journal = {\apjl},
         year = 2022,
        month = feb,
       volume = {926},
       number = {2},
          eid = {L36},
        pages = {L36},
          doi = {10.3847/2041-8213/ac5589},
archivePrefix = {arXiv},
       eprint = {2110.14652},
 primaryClass = {astro-ph.GA},
       adsurl = {https://ui.adsabs.harvard.edu/abs/2022ApJ...926L..36N}
}

@ARTICLE{Yong21,
       author = {{Yong}, D. and {Kobayashi}, C. and {Da Costa}, G.~S. and {Bessell}, M.~S. and {Chiti}, A. and {Frebel}, A. and {Lind}, K. and {Mackey}, A.~D. and {Nordlander}, T. and {Asplund}, M. and {Casey}, A.~R. and {Marino}, A.~F. and {Murphy}, S.~J. and {Schmidt}, B.~P.},
        title = "{r-Process elements from magnetorotational hypernovae}",
      journal = {\nat},
         year = 2021,
        month = jul,
       volume = {595},
       number = {7866},
        pages = {223-226},
          doi = {10.1038/s41586-021-03611-2},
archivePrefix = {arXiv},
       eprint = {2107.03010},
 primaryClass = {astro-ph.SR},
       adsurl = {https://ui.adsabs.harvard.edu/abs/2021Natur.595..223Y}
}

@ARTICLE{MacFadyen99,
       author = {{MacFadyen}, A.~I. and {Woosley}, S.~E.},
        title = "{Collapsars: Gamma-Ray Bursts and Explosions in ``Failed Supernovae''}",
      journal = {\apj},
         year = 1999,
        month = oct,
       volume = {524},
       number = {1},
        pages = {262-289},
          doi = {10.1086/307790},
archivePrefix = {arXiv},
       eprint = {astro-ph/9810274},
 primaryClass = {astro-ph},
       adsurl = {https://ui.adsabs.harvard.edu/abs/1999ApJ...524..262M}
}

@ARTICLE{Barnes_2022,
       author = {{Barnes}, Jennifer and {Metzger}, Brian D.},
        title = "{Signatures of r-process Enrichment in Supernovae from Collapsars}",
      journal = {\apjl},
         year = 2022,
        month = nov,
       volume = {939},
       number = {2},
          eid = {L29},
        pages = {L29},
          doi = {10.3847/2041-8213/ac9b41},
archivePrefix = {arXiv},
       eprint = {2205.10421},
 primaryClass = {astro-ph.HE},
       adsurl = {https://ui.adsabs.harvard.edu/abs/2022ApJ...939L..29B}
}

@ARTICLE{Siegel_2019,
       author = {{Siegel}, Daniel M. and {Barnes}, Jennifer and {Metzger}, Brian D.},
        title = "{Collapsars as a major source of r-process elements}",
      journal = {\nat},
         year = 2019,
        month = may,
       volume = {569},
       number = {7755},
        pages = {241-244},
          doi = {10.1038/s41586-019-1136-0},
archivePrefix = {arXiv},
       eprint = {1810.00098},
 primaryClass = {astro-ph.HE},
       adsurl = {https://ui.adsabs.harvard.edu/abs/2019Natur.569..241S}
}

@ARTICLE{Margutti_2021,
       author = {{Margutti}, Raffaella and {Chornock}, Ryan},
        title = "{First Multimessenger Observations of a Neutron Star Merger}",
      journal = {\araa},
         year = 2021,
        month = sep,
       volume = {59},
        pages = {155-202},
          doi = {10.1146/annurev-astro-112420-030742},
archivePrefix = {arXiv},
       eprint = {2012.04810},
 primaryClass = {astro-ph.HE},
       adsurl = {https://ui.adsabs.harvard.edu/abs/2021ARA&A..59..155M}
}

@ARTICLE{Alexander_2017,
       author = {{Alexander}, K.~D. and {Berger}, E. and {Fong}, W. and {Williams}, P.~K.~G. and {Guidorzi}, C. and {Margutti}, R. and {Metzger}, B.~D. and {Annis}, J. and {Blanchard}, P.~K. and {Brout}, D. and {Brown}, D.~A. and {Chen}, H. -Y. and {Chornock}, R. and {Cowperthwaite}, P.~S. and {Drout}, M. and {Eftekhari}, T. and {Frieman}, J. and {Holz}, D.~E. and {Nicholl}, M. and {Rest}, A. and {Sako}, M. and {Soares-Santos}, M. and {Villar}, V.~A.},
        title = "{The Electromagnetic Counterpart of the Binary Neutron Star Merger LIGO/Virgo GW170817. VI. Radio Constraints on a Relativistic Jet and Predictions for Late-time Emission from the Kilonova Ejecta}",
      journal = {\apjl},
         year = 2017,
        month = oct,
       volume = {848},
       number = {2},
          eid = {L21},
        pages = {L21},
          doi = {10.3847/2041-8213/aa905d},
archivePrefix = {arXiv},
       eprint = {1710.05457},
 primaryClass = {astro-ph.HE},
       adsurl = {https://ui.adsabs.harvard.edu/abs/2017ApJ...848L..21A}
}

@ARTICLE{Haggard_2017,
       author = {{Haggard}, Daryl and {Nynka}, Melania and {Ruan}, John J. and {Kalogera}, Vicky and {Cenko}, S. Bradley and {Evans}, Phil and {Kennea}, Jamie A.},
        title = "{A Deep Chandra X-Ray Study of Neutron Star Coalescence GW170817}",
      journal = {\apjl},
         year = 2017,
        month = oct,
       volume = {848},
       number = {2},
          eid = {L25},
        pages = {L25},
          doi = {10.3847/2041-8213/aa8ede},
archivePrefix = {arXiv},
       eprint = {1710.05852},
 primaryClass = {astro-ph.HE},
       adsurl = {https://ui.adsabs.harvard.edu/abs/2017ApJ...848L..25H}
}

@ARTICLE{Just_2015,
       author = {{Just}, O. and {Bauswein}, A. and {Ardevol Pulpillo}, R. and {Goriely}, S. and {Janka}, H. -T.},
        title = "{Comprehensive nucleosynthesis analysis for ejecta of compact binary mergers}",
      journal = {\mnras},
         year = 2015,
        month = mar,
       volume = {448},
       number = {1},
        pages = {541-567},
          doi = {10.1093/mnras/stv009},
archivePrefix = {arXiv},
       eprint = {1406.2687},
 primaryClass = {astro-ph.SR},
       adsurl = {https://ui.adsabs.harvard.edu/abs/2015MNRAS.448..541J}
}

@ARTICLE{Kasliwal_2017,
       author = {{Kasliwal}, M.~M. and {Nakar}, E. and {Singer}, L.~P. and {Kaplan}, D.~L. and {Cook}, D.~O. and {Van Sistine}, A. and {Lau}, R.~M. and {Fremling}, C. and {Gottlieb}, O. and {Jencson}, J.~E. and {Adams}, S.~M. and {Feindt}, U. and {Hotokezaka}, K. and {Ghosh}, S. and {Perley}, D.~A. and {Yu}, P. -C. and {Piran}, T. and {Allison}, J.~R. and {Anupama}, G.~C. and {Balasubramanian}, A. and {Bannister}, K.~W. and {Bally}, J. and {Barnes}, J. and {Barway}, S. and {Bellm}, E. and {Bhalerao}, V. and {Bhattacharya}, D. and {Blagorodnova}, N. and {Bloom}, J.~S. and {Brady}, P.~R. and {Cannella}, C. and {Chatterjee}, D. and {Cenko}, S.~B. and {Cobb}, B.~E. and {Copperwheat}, C. and {Corsi}, A. and {De}, K. and {Dobie}, D. and {Emery}, S.~W.~K. and {Evans}, P.~A. and {Fox}, O.~D. and {Frail}, D.~A. and {Frohmaier}, C. and {Goobar}, A. and {Hallinan}, G. and {Harrison}, F. and {Helou}, G. and {Hinderer}, T. and {Ho}, A.~Y.~Q. and {Horesh}, A. and {Ip}, W. -H. and {Itoh}, R. and {Kasen}, D. and {Kim}, H. and {Kuin}, N.~P.~M. and {Kupfer}, T. and {Lynch}, C. and {Madsen}, K. and {Mazzali}, P.~A. and {Miller}, A.~A. and {Mooley}, K. and {Murphy}, T. and {Ngeow}, C. -C. and {Nichols}, D. and {Nissanke}, S. and {Nugent}, P. and {Ofek}, E.~O. and {Qi}, H. and {Quimby}, R.~M. and {Rosswog}, S. and {Rusu}, F. and {Sadler}, E.~M. and {Schmidt}, P. and {Sollerman}, J. and {Steele}, I. and {Williamson}, A.~R. and {Xu}, Y. and {Yan}, L. and {Yatsu}, Y. and {Zhang}, C. and {Zhao}, W.},
        title = "{Illuminating gravitational waves: A concordant picture of photons from a neutron star merger}",
      journal = {Science},
         year = 2017,
        month = dec,
       volume = {358},
       number = {6370},
        pages = {1559-1565},
          doi = {10.1126/science.aap9455},
archivePrefix = {arXiv},
       eprint = {1710.05436},
 primaryClass = {astro-ph.HE},
       adsurl = {https://ui.adsabs.harvard.edu/abs/2017Sci...358.1559K}
}

@ARTICLE{Kasliwal2022,
       author = {{Kasliwal}, Mansi M. and {Kasen}, Daniel and {Lau}, Ryan M. and {Perley}, Daniel A. and {Rosswog}, Stephan and {Ofek}, Eran O. and {Hotokezaka}, Kenta and {Chary}, Ranga-Ram and {Sollerman}, Jesper and {Goobar}, Ariel and {Kaplan}, David L.},
        title = "{Spitzer mid-infrared detections of neutron star merger GW170817 suggests synthesis of the heaviest elements}",
      journal = {\mnras},
         year = 2022,
        month = feb,
       volume = {510},
       number = {1},
        pages = {L7-L12},
          doi = {10.1093/mnrasl/slz007},
archivePrefix = {arXiv},
       eprint = {1812.08708},
 primaryClass = {astro-ph.HE},
       adsurl = {https://ui.adsabs.harvard.edu/abs/2022MNRAS.510L...7K}
}

@ARTICLE{Andreoni_2017,
       author = {{Andreoni}, I. and {Ackley}, K. and {Cooke}, J. and {Acharyya}, A. and {Allison}, J.~R. and {Anderson}, G.~E. and {Ashley}, M.~C.~B. and {Baade}, D. and {Bailes}, M. and {Bannister}, K. and {Beardsley}, A. and {Bessell}, M.~S. and {Bian}, F. and {Bland}, P.~A. and {Boer}, M. and {Booler}, T. and {Brandeker}, A. and {Brown}, I.~S. and {Buckley}, D.~A.~H. and {Chang}, S. -W. and {Coward}, D.~M. and {Crawford}, S. and {Crisp}, H. and {Crosse}, B. and {Cucchiara}, A. and {Cup{\'a}k}, M. and {de Gois}, J.~S. and {Deller}, A. and {Devillepoix}, H.~A.~R. and {Dobie}, D. and {Elmer}, E. and {Emrich}, D. and {Farah}, W. and {Farrell}, T.~J. and {Franzen}, T. and {Gaensler}, B.~M. and {Galloway}, D.~K. and {Gendre}, B. and {Giblin}, T. and {Goobar}, A. and {Green}, J. and {Hancock}, P.~J. and {Hartig}, B.~A.~D. and {Howell}, E.~J. and {Horsley}, L. and {Hotan}, A. and {Howie}, R.~M. and {Hu}, L. and {Hu}, Y. and {James}, C.~W. and {Johnston}, S. and {Johnston-Hollitt}, M. and {Kaplan}, D.~L. and {Kasliwal}, M. and {Keane}, E.~F. and {Kenney}, D. and {Klotz}, A. and {Lau}, R. and {Laugier}, R. and {Lenc}, E. and {Li}, X. and {Liang}, E. and {Lidman}, C. and {Luvaul}, L.~C. and {Lynch}, C. and {Ma}, B. and {Macpherson}, D. and {Mao}, J. and {McClelland}, D.~E. and {McCully}, C. and {M{\"o}ller}, A. and {Morales}, M.~F. and {Morris}, D. and {Murphy}, T. and {Noysena}, K. and {Onken}, C.~A. and {Orange}, N.~B. and {Os{\l}owski}, S. and {Pallot}, D. and {Paxman}, J. and {Potter}, S.~B. and {Pritchard}, T. and {Raja}, W. and {Ridden-Harper}, R. and {Romero-Colmenero}, E. and {Sadler}, E.~M. and {Sansom}, E.~K. and {Scalzo}, R.~A. and {Schmidt}, B.~P. and {Scott}, S.~M. and {Seghouani}, N. and {Shang}, Z. and {Shannon}, R.~M. and {Shao}, L. and {Shara}, M.~M. and {Sharp}, R. and {Sokolowski}, M. and {Sollerman}, J. and {Staff}, J. and {Steele}, K. and {Sun}, T. and {Suntzeff}, N.~B. and {Tao}, C. and {Tingay}, S. and {Towner}, M.~C. and {Thierry}, P. and {Trott}, C. and {Tucker}, B.~E. and {V{\"a}is{\"a}nen}, P. and {Krishnan}, V. Venkatraman and {Walker}, M. and {Wang}, L. and {Wang}, X. and {Wayth}, R. and {Whiting}, M. and {Williams}, A. and {Williams}, T. and {Wolf}, C. and {Wu}, C. and {Wu}, X. and {Yang}, J. and {Yuan}, X. and {Zhang}, H. and {Zhou}, J. and {Zovaro}, H.},
        title = "{Follow Up of GW170817 and Its Electromagnetic Counterpart by Australian-Led Observing Programmes}",
      journal = {\pasa},
         year = 2017,
        month = dec,
       volume = {34},
          eid = {e069},
        pages = {e069},
          doi = {10.1017/pasa.2017.65},
archivePrefix = {arXiv},
       eprint = {1710.05846},
 primaryClass = {astro-ph.HE},
       adsurl = {https://ui.adsabs.harvard.edu/abs/2017PASA...34...69A}
}

@ARTICLE{Arcavi_2017,
       author = {{Arcavi}, Iair and {Hosseinzadeh}, Griffin and {Howell}, D. Andrew and {McCully}, Curtis and {Poznanski}, Dovi and {Kasen}, Daniel and {Barnes}, Jennifer and {Zaltzman}, Michael and {Vasylyev}, Sergiy and {Maoz}, Dan and {Valenti}, Stefano},
        title = "{Optical emission from a kilonova following a gravitational-wave-detected neutron-star merger}",
      journal = {\nat},
         year = 2017,
        month = nov,
       volume = {551},
       number = {7678},
        pages = {64-66},
          doi = {10.1038/nature24291},
archivePrefix = {arXiv},
       eprint = {1710.05843},
 primaryClass = {astro-ph.HE},
       adsurl = {https://ui.adsabs.harvard.edu/abs/2017Natur.551...64A}
}

@ARTICLE{Coulte_2017,
       author = {{Coulter}, D.~A. and {Foley}, R.~J. and {Kilpatrick}, C.~D. and {Drout}, M.~R. and {Piro}, A.~L. and {Shappee}, B.~J. and {Siebert}, M.~R. and {Simon}, J.~D. and {Ulloa}, N. and {Kasen}, D. and {Madore}, B.~F. and {Murguia-Berthier}, A. and {Pan}, Y. -C. and {Prochaska}, J.~X. and {Ramirez-Ruiz}, E. and {Rest}, A. and {Rojas-Bravo}, C.},
        title = "{Swope Supernova Survey 2017a (SSS17a), the optical counterpart to a gravitational wave source}",
      journal = {Science},
         year = 2017,
        month = dec,
       volume = {358},
       number = {6370},
        pages = {1556-1558},
          doi = {10.1126/science.aap9811},
archivePrefix = {arXiv},
       eprint = {1710.05452},
 primaryClass = {astro-ph.HE},
       adsurl = {https://ui.adsabs.harvard.edu/abs/2017Sci...358.1556C}
}

@ARTICLE{Covino_2017,
       author = {{Covino}, S. and {Wiersema}, K. and {Fan}, Y.~Z. and {Toma}, K. and {Higgins}, A.~B. and {Melandri}, A. and {D'Avanzo}, P. and {Mundell}, C.~G. and {Palazzi}, E. and {Tanvir}, N.~R. and {Bernardini}, M.~G. and {Branchesi}, M. and {Brocato}, E. and {Campana}, S. and {di Serego Alighieri}, S. and {G{\"o}tz}, D. and {Fynbo}, J.~P.~U. and {Gao}, W. and {Gomboc}, A. and {Gompertz}, B. and {Greiner}, J. and {Hjorth}, J. and {Jin}, Z.~P. and {Kaper}, L. and {Klose}, S. and {Kobayashi}, S. and {Kopac}, D. and {Kouveliotou}, C. and {Levan}, A.~J. and {Mao}, J. and {Malesani}, D. and {Pian}, E. and {Rossi}, A. and {Salvaterra}, R. and {Starling}, R.~L.~C. and {Steele}, I. and {Tagliaferri}, G. and {Troja}, E. and {van der Horst}, A.~J. and {Wijers}, R.~A.~M.~J.},
        title = "{The unpolarized macronova associated with the gravitational wave event GW 170817}",
      journal = {Nature Astronomy},
         year = 2017,
        month = oct,
       volume = {1},
        pages = {791-794},
          doi = {10.1038/s41550-017-0285-z},
archivePrefix = {arXiv},
       eprint = {1710.05849},
 primaryClass = {astro-ph.HE},
       adsurl = {https://ui.adsabs.harvard.edu/abs/2017NatAs...1..791C}
}

@ARTICLE{Cowperthwaite_2017,
       author = {{Cowperthwaite}, P.~S. and {Berger}, E. and {Villar}, V.~A. and {Metzger}, B.~D. and {Nicholl}, M. and {Chornock}, R. and {Blanchard}, P.~K. and {Fong}, W. and {Margutti}, R. and {Soares-Santos}, M. and {Alexander}, K.~D. and {Allam}, S. and {Annis}, J. and {Brout}, D. and {Brown}, D.~A. and {Butler}, R.~E. and {Chen}, H. -Y. and {Diehl}, H.~T. and {Doctor}, Z. and {Drout}, M.~R. and {Eftekhari}, T. and {Farr}, B. and {Finley}, D.~A. and {Foley}, R.~J. and {Frieman}, J.~A. and {Fryer}, C.~L. and {Garc{\'\i}a-Bellido}, J. and {Gill}, M.~S.~S. and {Guillochon}, J. and {Herner}, K. and {Holz}, D.~E. and {Kasen}, D. and {Kessler}, R. and {Marriner}, J. and {Matheson}, T. and {Neilsen}, E.~H., Jr. and {Quataert}, E. and {Palmese}, A. and {Rest}, A. and {Sako}, M. and {Scolnic}, D.~M. and {Smith}, N. and {Tucker}, D.~L. and {Williams}, P.~K.~G. and {Balbinot}, E. and {Carlin}, J.~L. and {Cook}, E.~R. and {Durret}, F. and {Li}, T.~S. and {Lopes}, P.~A.~A. and {Louren{\c{c}}o}, A.~C.~C. and {Marshall}, J.~L. and {Medina}, G.~E. and {Muir}, J. and {Mu{\~n}oz}, R.~R. and {Sauseda}, M. and {Schlegel}, D.~J. and {Secco}, L.~F. and {Vivas}, A.~K. and {Wester}, W. and {Zenteno}, A. and {Zhang}, Y. and {Abbott}, T.~M.~C. and {Banerji}, M. and {Bechtol}, K. and {Benoit-L{\'e}vy}, A. and {Bertin}, E. and {Buckley-Geer}, E. and {Burke}, D.~L. and {Capozzi}, D. and {Carnero Rosell}, A. and {Carrasco Kind}, M. and {Castander}, F.~J. and {Crocce}, M. and {Cunha}, C.~E. and {D'Andrea}, C.~B. and {da Costa}, L.~N. and {Davis}, C. and {DePoy}, D.~L. and {Desai}, S. and {Dietrich}, J.~P. and {Drlica-Wagner}, A. and {Eifler}, T.~F. and {Evrard}, A.~E. and {Fernandez}, E. and {Flaugher}, B. and {Fosalba}, P. and {Gaztanaga}, E. and {Gerdes}, D.~W. and {Giannantonio}, T. and {Goldstein}, D.~A. and {Gruen}, D. and {Gruendl}, R.~A. and {Gutierrez}, G. and {Honscheid}, K. and {Jain}, B. and {James}, D.~J. and {Jeltema}, T. and {Johnson}, M.~W.~G. and {Johnson}, M.~D. and {Kent}, S. and {Krause}, E. and {Kron}, R. and {Kuehn}, K. and {Nuropatkin}, N. and {Lahav}, O. and {Lima}, M. and {Lin}, H. and {Maia}, M.~A.~G. and {March}, M. and {Martini}, P. and {McMahon}, R.~G. and {Menanteau}, F. and {Miller}, C.~J. and {Miquel}, R. and {Mohr}, J.~J. and {Neilsen}, E. and {Nichol}, R.~C. and {Ogando}, R.~L.~C. and {Plazas}, A.~A. and {Roe}, N. and {Romer}, A.~K. and {Roodman}, A. and {Rykoff}, E.~S. and {Sanchez}, E. and {Scarpine}, V. and {Schindler}, R. and {Schubnell}, M. and {Sevilla-Noarbe}, I. and {Smith}, M. and {Smith}, R.~C. and {Sobreira}, F. and {Suchyta}, E. and {Swanson}, M.~E.~C. and {Tarle}, G. and {Thomas}, D. and {Thomas}, R.~C. and {Troxel}, M.~A. and {Vikram}, V. and {Walker}, A.~R. and {Wechsler}, R.~H. and {Weller}, J. and {Yanny}, B. and {Zuntz}, J.},
        title = "{The Electromagnetic Counterpart of the Binary Neutron Star Merger LIGO/Virgo GW170817. II. UV, Optical, and Near-infrared Light Curves and Comparison to Kilonova Models}",
      journal = {\apjl},
         year = 2017,
        month = oct,
       volume = {848},
       number = {2},
          eid = {L17},
        pages = {L17},
          doi = {10.3847/2041-8213/aa8fc7},
archivePrefix = {arXiv},
       eprint = {1710.05840},
 primaryClass = {astro-ph.HE},
       adsurl = {https://ui.adsabs.harvard.edu/abs/2017ApJ...848L..17C}
}

@ARTICLE{Drout_2017,
       author = {{Drout}, M.~R. and {Piro}, A.~L. and {Shappee}, B.~J. and {Kilpatrick}, C.~D. and {Simon}, J.~D. and {Contreras}, C. and {Coulter}, D.~A. and {Foley}, R.~J. and {Siebert}, M.~R. and {Morrell}, N. and {Boutsia}, K. and {Di Mille}, F. and {Holoien}, T.~W. -S. and {Kasen}, D. and {Kollmeier}, J.~A. and {Madore}, B.~F. and {Monson}, A.~J. and {Murguia-Berthier}, A. and {Pan}, Y. -C. and {Prochaska}, J.~X. and {Ramirez-Ruiz}, E. and {Rest}, A. and {Adams}, C. and {Alatalo}, K. and {Ba{\~n}ados}, E. and {Baughman}, J. and {Beers}, T.~C. and {Bernstein}, R.~A. and {Bitsakis}, T. and {Campillay}, A. and {Hansen}, T.~T. and {Higgs}, C.~R. and {Ji}, A.~P. and {Maravelias}, G. and {Marshall}, J.~L. and {Moni Bidin}, C. and {Prieto}, J.~L. and {Rasmussen}, K.~C. and {Rojas-Bravo}, C. and {Strom}, A.~L. and {Ulloa}, N. and {Vargas-Gonz{\'a}lez}, J. and {Wan}, Z. and {Whitten}, D.~D.},
        title = "{Light curves of the neutron star merger GW170817/SSS17a: Implications for r-process nucleosynthesis}",
      journal = {Science},
         year = 2017,
        month = dec,
       volume = {358},
       number = {6370},
        pages = {1570-1574},
          doi = {10.1126/science.aaq0049},
archivePrefix = {arXiv},
       eprint = {1710.05443},
 primaryClass = {astro-ph.HE},
       adsurl = {https://ui.adsabs.harvard.edu/abs/2017Sci...358.1570D}
}
\bibliographystyle{aasjournalv7}

%\onecolumngrid 



\end{document}